\pdfoutput=1
\documentclass[fleqn]{jfm}
\usepackage[]{graphicx}
\usepackage[utf8]{inputenc}
\usepackage[T1]{fontenc}
\usepackage[english]{babel}
\usepackage[usenames,dvipsnames,table,svgnames,x11names]{xcolor}
\usepackage{longtable}
\usepackage{amsfonts, amsmath}
\usepackage[squaren,Gray,cdot]{SIunits}
\usepackage{hhline}
\usepackage[normalem]{ulem}
\usepackage{notation}
\usepackage{appendix}
\usepackage{natbib}
\usepackage{multirow}
\usepackage{hyperref}
\usepackage{verbatim}
\hypersetup{
    colorlinks = true,
    urlcolor   = blue,
    citecolor  = SkyBlue3,
}
\usepackage{newtxtext}
\usepackage{newtxmath}

\renewcommand{\vec}[1]{\mathbf{#1}}

\DeclareMathOperator*{\argmax}{arg\,max}
\title{Fingering convection in a spherical shell}

\author{Th\'eo Tassin\aff{1},
  Thomas Gastine\aff{1}
    \corresp{\email{gastine@ipgp.fr}},
 \and Alexandre Fournier\aff{1}}

\affiliation{\aff{1}Universit\'e Paris Cité, Institut de physique du globe de Paris, CNRS, F-75005 Paris, France}

\begin{document}
\maketitle

\begin{abstract}
We use $\nsims$ three dimensional direct numerical simulations to study 
fingering convection in non-rotating spherical shells. We investigate the 
scaling behaviour of the flow lengthscale, the non-dimensional heat 
and compositional fluxes $Nu$ and $Sh$ and the mean convective velocity
over the fingering convection instability domain defined by $1 \leq 
R_\rho < Le$, $R_\rho$ being the ratio of density perturbations of thermal 
and compositional origins and $Le$ the Lewis number. 
We show that the chemical boundary layers are marginally unstable and 
adhere to the laminar Prandtl-Blasius model, hence explaining the asymmetry 
between the inner and outer spherical shell boundary layers.
We develop scaling laws for two asymptotic regimes close to the two edges of the 
instability domain, namely $R_\rho \lesssim Le$ and $R_\rho \gtrsim 1$.
For the former, we develop novel power laws of a small parameter 
$\epsilon$ measuring the distance to onset, which differ from 
theoretical laws published to date in Cartesian geometry. For the 
latter, we find that the Sherwood number $Sh$ gradually approaches a scaling 
$Sh\sim \rac^{1/3}$ when $\rac \gg 1$; 
and that the P\'eclet number accordingly follows $Pe \sim \rac^{2/3} 
|\rat|^{-1/4}$, $\rac$ being the chemical Rayleigh number.
When the Reynolds number exceeds a few tens, we report on a secondary 
instability which takes the form of large-scale toroidal jets which 
span the entire spherical domain. Jets distort 
the fingers resulting in Reynolds stress correlations, which in turn feed the 
jet growth until saturation. This nonlinear phenomenon can yield relaxation 
oscillation cycles.

\end{abstract}


\section{Introduction}

Double-diffusive effects are ubiquitous in fluid envelopes of planetary 
interiors. For instance, in the liquid iron core of terrestrial planets, 
the convective motions are driven by thermal and compositional perturbations
which originate from the secular cooling and the inner-core 
growth \citep[e.g.][]{Roberts2013}. Several internal evolution models coupled 
with \textit{ab-initio} computations and high-pressure experiments for core 
conductivity are suggestive of a thermally-stratified layer underneath the 
core-mantle boundary of Mercury \citep{Hauck2004} and the Earth 
\citep[e.g.][]{Pozzo2012,Ohta2016}. Such a fluid layer could harbor
fingering convection, where thermal stratification is stable whilst
compositional stratification is unstable \citep{Stern1960}. 
In contrast, in the deep interior of the gas giants Jupiter and Saturn, 
the internal models by \citet{Mankovich2020} suggest the presence of a 
stabilising Helium gradient opposed to the destabilising thermal stratification, 
a physical configuration prone to semi-convective instabilities 
\citep{Veronis1965}.

The linear stability analysis of double-diffusive systems, 
as carried out by e.g. \citet{Stern1960}, \cite{Veronis1965}, 
and \cite{Baines1969}, shows
that fingering convection occurs in planar geometry when the density ratio 
$R_\rho=|\alpha_T \Delta T|/\alpha_\xi \Delta \xi$ belongs to the interval 
\begin{equation}
 1 \leq R_\rho < Le,
 \label{eq:fingering_domain}
\end{equation}
where $\alpha_T$ ($\alpha_\xi$) are the thermal (compositional) expansion 
coefficient and $\Delta T$ ($\Delta \xi$) are the perturbations of thermal 
(compositional) origin. In the above expression, $Le$ is the Lewis number, 
defined by the ratio of the thermal and solutal diffusivities.  
Note that the upper bound of the instability domain, $Le$,
ignores a correcting term that is a function of the structure of the
unstable mode, and that becomes negligible in the limit of large
thermal and compositional Rayleigh numbers
to be defined below \citep[see e.g.][\S~8.1.2]{Turner1973}. 
Close to onset, 
the instability takes the form of vertically-elongated fingers, whose typical 
horizontal size $\mathcal{L}_h$ results from a balance between buoyancy and viscosity 
 and follows $\mathcal{L}_h = |\rat|^{-1/4}\,d$, where $\rat$ is the thermal 
Rayleigh number and $d$ is the vertical extent of the fluid domain 
\citep[e.g.][]{Schmitt1979,Taylor1989,Smyth2007}. Developed salt fingers 
frequently give rise to secondary instabilities, such as the collective 
instability \citep{Stern1969}, thermohaline staircase formation 
\citep{Radko2003} or jets \citep{Holyer1984}. The saturated state of the 
instability is therefore frequently made up of a mixture of small-scale fingers 
and large-scale structures commensurate with the size of the fluid domain.
A mean-field formalism can then be successfully employed to describe 
the secondary instabilities \citep[e.g.][]{Traxler2011a,Radko2013}.

Following \citet{Radko2012}, \citet{Brown2013} hypothesize that fingering 
convection saturates once the growth rate of the secondary instability is 
comparable to the one of the primary fingers. In the context of low-Prandtl 
numbers ($Pr \ll 1$) relevant to stellar interiors, they derive three branches 
of semi-analytical scaling laws for the transport of heat and chemical 
composition which depend on the value of $r_\rho = (R_\rho-1)/(Le-1)$.
These theoretical scaling laws accurately account for the behaviours observed 
in local 3-D numerical simulations \citep[see, 
e.g.][Fig.~2]{Garaud2018}. In the context of large Prandtl numbers, 
\citet{Radko2010} developed a weakly-nonlinear model for salt fingers which 
predicts that the scaling behaviours for the transport of heat and 
chemical composition should follow power laws of the distance to onset 
 \citep[see also][]{Stern1998,Radko2000,Xie2017}. 

Besides the growth of the celebrated thermohaline staircases 
\citep{Garaud2015a}, the fingering 
instability can also lead to the formation of large-scale horizontal flows or 
jets. By applying the methods of Floquet theory, \citet{Holyer1984} demonstrated 
that a non-oscillatory secondary instability could actually grow faster than 
the collective instability for fluids with $Pr \gtrsim 1$. This instability 
takes the form of a mean horizontally-invariant flow which distorts the fingers 
(see her Fig.~1). Cartesian numerical simulations in 2-D carried out by 
\citet{Shen1995} later revealed that the nonlinear evolution of this instability 
yields a strong shear that eventually disrupts the vertical coherence of the 
fingers (see his Fig.~2). Jets were also obtained in the 2-D simulations by 
\citet{Radko2010} for configurations with $Pr=10^{-2}$ and $Le=3$ and by 
\citet{Garaud2015} for $Pr=0.03$ and $Le\approx 33.3$. In addition to the direct 
numerical simulations, jets have also been observed in single-mode truncated 
models by \citet{Paparella1999} and \citet{Liu2022}.
The qualitative description of the instability developed by
\citet{Stern2005} underlines the analogy with tilting instabilities observed in 
Rayleigh-B\'enard convection \citep[e.g.][]{Goluskin2014}: jets shear apart  
 fingers, which in turn feed the growth of jets via Reynolds 
stress correlations. The nonlinear saturation of the instability can occur via 
the disruption of the fingers \citep{Shen1995}, but can also yield relaxation 
oscillations with a predator-prey like behaviour between jets and fingers
\citep{Garaud2015,Xie2019}. \citet{Garaud2015} suggest, however, that this 
instability may be confined to two-dimensional fluid domains, and hence question
the relevance of jet formation in 3-D when $Pr \ll 1$. The 3-D bounded 
planar models by \citet{Yang2016} for $Pr=7$ however exhibit jets for 
simulations with $R_\rho=1.6$. This raises the question of the physical 
phenomena which govern the instability domain of jets.
In addition, to date, jets have only been observed in local simulations 
containing
a few tens of fingers; their relevance in 3-D global domains therefore remains 
to be assessed.

The vast majority of the theoretical and numerical models discussed above 
adopt a local Cartesian approach. Two configurations are then considered. The 
most common, termed \emph{unbounded}, resorts to using a triply-periodic domain 
without boundary layers \citep[e.g.][]{Stellmach2011,Brown2013}. Conversely, in 
the \emph{bounded} configurations, the flow is maintained between two 
horizontal plates and boundary layers can form in the vicinity of the 
boundaries \citep[e.g.][]{Schmitt1979,Radko2000,Yang2015}. This latter 
configuration is also relevant to laboratory experiments in which 
thermal and/or chemical composition are imposed at the boundaries of the fluid 
domain \citep[e.g.][]{Taylor1989,Hage2010,Rosenthal2022}. 
One of the key issues of the bounded configuration is to express scaling 
laws that depend on the thermal $\Delta T$ and/or compositional $\Delta \xi$ 
contrasts imposed over the layer. Early experiments by 
\citet{Turner1967} and \citet{Schmitt1979b} suggest that the salinity flux 
grossly follows a $4/3$ power law on $\Delta \xi$, with additional secondary 
dependence on $R_\rho$, $Pr$ and $Le$ \citep[see][]{Taylor1996}.
Expressed in dimensionless quantities, this translates to $Sh \sim 
\rac^{1/3}$, $Sh$ being the Sherwood number and $\rac$ a 
composition-based Rayleigh number. This is the double-diffusive counterpart of 
the classical heat transport model for turbulent convection in which the heat 
flux is assumed to be depth-independent \citep{Priestley1954}. \citet{Yang2015} 
refined this scaling by extending the Grossmann-Lohse theory for classical 
Rayleigh-B\'enard convection \citep[hereafter GL, see][]{Grossmann2000} to the 
fingering configuration. Considering a suite of 3-D bounded Cartesian numerical 
simulations with $Pr=7$ and $Le=100$, \citet{Yang2016} then found
that the dependence of $Sh$  upon $\rac$ is well accounted for by the GL 
theory. 
Scaling laws for the Reynolds number $Re$ put forward by 
\citet{Yang2016} involve an extra dependence to the density 
ratio with an empirical exponent of the form $Re \sim R_\rho^{-1/4} 
\rac^{1/2}$. This latter scaling differs from the one obtained by 
\citet{Hage2010} --$Re 
\sim \rac |\rat|^{-1/2}$-- using experimental data with $Pr \approx 
9$ and $Le \approx 230$. Differences between the two could possibly result from 
the amount of data retained to derive the scalings. Both studies indeed also 
consider configurations with $R_\rho < 1$ where overturning convection can also
develop and modify the scaling behaviours.
The development of boundary layers makes the comparison between bounded
and unbounded configurations quite delicate, as it requires defining
effective quantities measured on the fluid bulk in the bounded 
configuration \citep[see][]{Radko2000,Yang2020a}.

In planetary interiors, fingering convection operates in a quasi spherical fluid 
gap enclosed between two rigid boundaries in the case of terrestrial bodies. 
For terrestrial planets possessing a metallic iron core, the Prandtl 
number $Pr$ is $\mathcal{O}(10^{-1})$, while the Lewis number $Le$ is 
$\mathcal{O}(10^3)$ \citep[e.g.][]{Roberts2013}.
The depth of the fingering convection region is more uncertain. In any event, 
it would correspond to a thin shell in the case of Earth, with a
ratio between the inner radius $r_i$ and the outer radius $r_o$
larger than $r_i/r_o=0.9$ \citep[e.g.][and references therein]{Labrosse2015}, 
while a deeper shell configuration is more likely for Mercury, since the 
corresponding fluid layer may be as thick as $880$~km \citep{Wardinski2021}, 
yielding a radius ratio close to $0.6$.
One of the main goals of this study is to assess the applicability of the 
results derived in local planar geometry to global spherical shells.
Because of curvature and the radial changes of the buoyancy force, 
spherical-shell convection differs from the plane layer configuration and 
yields asymmetrical boundary layers between both boundaries
\citep[e.g.][]{Gastine2015}. Only a handful of studies have considered 
fingering convection in spherical geometry and they all incorporate the effects 
of global rotation, which makes the comparison to planar models difficult.
\citet{Silva2019} and 
\citet{Monville2019} computed the onset of rotating double-diffusive 
convection in both the fingering and semi-convection regimes.
\citet{Monville2019}  and \citet{Guervilly2022} also considered nonlinear 
models with $Pr=0.3$ and $Le=10$ and observed 
the formation of large scale jets. \citet{Guervilly2022} also analysed the 
scaling behaviour of the horizontal size of the fingers and their mean
velocity. At a given rotation speed, the finger size $\mathcal{L}_h$ gradually transits from a 
large  horizontal scale close to onset to decreasing lengthscales on increasing 
supercriticality. When rotation becomes less influential, $\mathcal{L}_h$ 
tends to conform with \citeauthor{Stern1960}'s scaling. The mean fingering velocity was 
found to loosely follow the scalings by \citet{Brown2013} (see 
\citeauthor{Guervilly2022}'s
Fig.~10(\textit{b})), deviations to the theory likely occurring because of the 
imprint of rotation on the dynamics. 

While our long-term 
objective is to conduct global simulations of fingering convection 
in the presence of global rotation, we opt for an incremental approach.
Our immediate goal with the present work is twofold: (\textit{i}) 
to assess the salient differences (if any) between fingering convection 
in global, non-rotating spherical domains and fingering 
convection in local, non-rotating Cartesian domains;  
(\textit{ii}) to examine the occurrence of jet formation in 
spherical shell fingering convection. To do so, we conduct a systematic 
parameter survey of $\nsims$ direct numerical simulations in spherical 
geometry, varying the Prandtl number between $0.03$ and $7$, the 
Lewis number between $3$ and $33.3$ and the vigour of the forcing up to $\rac 
= 5\times 10^{11}$. The aspect ratio of the spherical shell is
the same for all simulations. While its value of $0.35$ is admittedly 
smaller than current estimates relevant to planetary interiors (see above), it
is meant to exacerbate
the differences that may exist between Cartesian and spherical setups; such
a deep shell is 
also less demanding on the angular resolution required to resolve fingers, 
and makes a systematic analysis possible.

The paper is organised as follows: in \S~\ref{sec:model}, we present the 
governing equations and the numerical model; in \S~\ref{sec:fingers} we focus 
on deriving scaling laws for fingering convection in spherical shells; in 
\S~\ref{sec:jets}, we analyse the formation of jets before concluding in 
\S~\ref{sec:conclu}.

\section{Model and methods}
\label{sec:model}

\subsection{Hydrodynamical model}

We consider a non-rotating spherical shell of inner radius 
$r_i$ and outer radius $r_o$ with $r_i/r_o=0.35$ filled with a Newtonian 
Boussinesq fluid of background density $\rho_c$. The spherical shell 
boundaries are assumed to be impermeable, no slip and held at constant 
temperature and chemical composition.
The kinematic viscosity 
$\nu$, the thermal diffusivity $\kappa_T$ and the diffusivity of chemical 
composition $\kappa_\xi$ are assumed to be constant. We adopt the following 
linear equation of state

\begin{equation}
  \rho = \rho_c [1-\alpha_{T}(T - T_c) -\alpha_{\xi}(\xi - \xi_c)],
\end{equation}
which ascribes changes in density ($\rho$) to fluctuations of temperature 
($T$) and chemical composition ($\xi$). In the above expression, $T_c$ and 
$\xi_c$ denote the reference temperature and composition, while 
$\alpha_T$ and $\alpha_\xi$ are the corresponding constant expansion 
coefficients. In the following, we adopt a dimensionless formulation using the 
spherical shell gap $d=r_o-r_i$ as the reference lengthscale and the viscous 
diffusion time $d^2/\nu$ as the reference timescale. Temperature and 
composition are respectively expressed in units of $\Delta 
T=T_{\text{top}}-T_{\text{bot}}$
and $\Delta \xi=\xi_{\text{bot}}-\xi_{\text{top}}$, their imposed contrasts over 
the shell. The dimensionless equations which govern the 
time evolution of the velocity $\vec{u}$, the pressure $p$, the temperature $T$ 
and the composition $\xi$ are expressed by

\begin{equation}
  \grad \cdot \vec{u} = 0,
  \label{eq:divu}
\end{equation}
\begin{equation}
\ddd{t}{\vec{u}} + \vec{u} \cdot\grad\vec{u}  = - \grad p
+\dfrac{1}{Pr}\left(-\rat T + \dfrac{\rac}{Le} \xi \right) g \er  
+ \grad^2 \vec{u},
\label{eq:momentum}
\end{equation}
\begin{equation}
 \ddd{t}{T} + \vec{u} \cdot \grad T = \dfrac{1}{Pr}\laplacien T,
\label{eq:temp}
\end{equation}
\begin{equation}
\ddd{t}{\xi}  + \vec{u} \cdot \grad \xi = \dfrac{1}{Le Pr} \laplacien \xi,
\label{eq:chem}
\end{equation}
where $\vec{e}_r$ is the unit vector in the radial direction and $g=r/r_o$ is 
the dimensionless gravity. The set of equations 
(\ref{eq:divu}-\ref{eq:chem}) is governed by four dimensionless numbers

\begin{equation}
\rat = -\dfrac{\alpha_T g_0 d^3 \Delta T}{\nu \kappa_T}, \quad 
\rac = \dfrac{\alpha_\xi g_0 d^3 \Delta \xi}{\nu \kappa_\xi}, \quad
Pr = \dfrac{\nu}{\kappa_T}, \quad
Le = \dfrac{\kappa_T}{\kappa_\xi}\,,
\end{equation}
termed the thermal Rayleigh, chemical Rayleigh, Prandtl, and Lewis numbers, respectively. 
Our definition of $\rat$ makes it negative, in order to stress the 
 stabilising effect of the thermal background. 
For the sake of clarity,  we will also make use of the Schmidt number 
$Sc=\nu/\kappa_\xi=Le\,Pr$ in the following. 
The stability of the convective system depends on the value of 
the density ratio $\Rp$ \citep{Stern1960} defined by

\begin{equation}
 \Rp = \dfrac{\alpha_T \Delta T}{\alpha_\xi \Delta \xi}\,,
\end{equation}
which relates to the above control parameters via $\Rp = Le |\rat|/\rac$. In 
order to compare models with different Lewis numbers $Le$, it is convenient to 
also introduce a normalised density ratio $\rrho$ following 
\citet{Traxler2011}
\begin{equation}
 \rrho = \dfrac{\Rp - 1}{Le - 1}\,.
\label{eq:r_density}
\end{equation}
This maps the instability domain for fingering convection, $1\leq \Rp < Le$,  to 
$0\leq \rrho<1$.

\subsection{Numerical technique}

We consider numerical simulations computed using the pseudo-spectral 
open-source code \texttt{MagIC}\footnote{Freely available at 
\url{https://github.com/magic-sph/magic}} \citep{Wicht2002}. \texttt{MagIC} has 
been validated against a benchmark for rotating double-diffusive convection in 
spherical shells initiated by 
\citet{Breuer2010}. The system of equations (\ref{eq:divu}-\ref{eq:chem}) is 
solved in spherical coordinates $(r,\theta,\phi)$, expressing the solenoidal 
velocity field in terms of poloidal ($W$) and toroidal ($Z$) potentials
\[
 \vec{u} = \vec{u}_P+\vec{u}_T = \grad \times( \grad \times 
W\,\vec{e}_r) + \grad \times Z\,\vec{e}_r\,.
\]
The unknowns $W$, $Z$, $p$, $T$ and $\xi$ are then expanded in spherical 
harmonics up to the maximum degree $\ell_\text{max}$ and order $m_\text{max}=\ell_\text{max}$ in the angular 
directions. Discretisation in the radial direction involves a 
Chebyshev collocation technique with $N_r$ collocation grid points 
\citep{Boyd2001}. The equations are advanced in time 
using the third order implicit-explicit Runge-Kutta schemes ARS343 
\citep{Ascher1997} and BPR353 \citep{Boscarino2013} which handle the linear 
terms of (\ref{eq:divu}-\ref{eq:chem}) implicitly. At each iteration, the 
nonlinear terms are calculated on the physical grid space 
and transformed back to spectral representation using the open-source 
\texttt{SHTns} library\footnote{Freely available at
\url{https://gricad-gitlab.univ-grenoble-alpes.fr/schaeffn/shtns}} 
\citep{Schaeffer2013}. The seminal work by \citet{Glatzmaier1984} and the more 
recent review by \citet{Christensen2015} provide additional insights on the 
algorithm \citep[the interested reader may also consult][with regard to its parallel implementation]{Lago2021}.

\subsection{Diagnostics}

We introduce here several diagnostics that will be used to characterise the 
convective flow and the thermal and chemical transports. We hence 
adopt the following notations to denote temporal and spatial averaging of
a field $f$:

\[
 \overline{f} = \dfrac{1}{t_\text{avg}}\int_{t_0}^{t_0+t_{\text{avg}}} f 
\mathrm{d}t,\  \langle f \rangle_V = \dfrac{1}{V}\int_V f\,\mathrm{d}V,
\ \langle f \rangle_S = \dfrac{1}{4\pi}\int_0^{2\pi}\int_0^\pi f\,\sin\theta 
\mathrm{d}\theta\mathrm{d}\phi,
\]
where 
time-averaging runs over the
interval $[t_0,t_0+t_{\text{avg}}]$, 
$V$ is the spherical shell volume and $S$ is the unit sphere. 
Time-averaged poloidal and toroidal energies are defined by

\begin{equation}
  \ekpol = \dfrac{1}{2}\overline{\langle \vec{u}_P\cdot\vec{u}_P \rangle_V} = 
  \sum_{\ell=1}^{\lmax} \ekpoll, 
\quad
  \ektor = \dfrac{1}{2}\overline{\langle\vec{u}_T\cdot\vec{u}_T  \rangle_V} = 
  \sum_{\ell=1}^{\lmax} \ektorl 
  \,,
\end{equation}
noting that both can be expressed as the sum of contributions from each spherical 
harmonic degree $\ell$, $\ekpoll$ and $\ektorl$.
The corresponding Reynolds numbers are then expressed by

\begin{equation}
\rep = \sqrt{2\ekpol}, \quad \ret=\sqrt{2\ektor}\,.
\end{equation}
In the following we also employ the chemical P\'eclet number, $Pe$,
to quantify the vertical velocity. It relates to the Reynolds number of the 
poloidal flow via
\begin{equation}
Pe = \rep\,Sc. 
\label{eq:pe_def}
\end{equation}

We introduce the notation $\Theta$ and $\Xi$ to define the time and 
horizontally averaged radial profiles of temperature and chemical composition
\begin{equation}
 \Theta = \overline{\langle T \rangle_S},\quad
 \Xi = \overline{\langle \xi \rangle_S}\,.
\end{equation}
Heat and chemical composition transports are defined at all radii by the 
Nusselt $Nu$ and Sherwood $Sh$ numbers
\begin{equation}
Nu = \dfrac{\overline{\langle u_r T \rangle_S}-
\dfrac{1}{Pr}\ddn{r}{\Theta}}{-\dfrac{1}{Pr}\ddn{r}{T_c}},
\quad Sh = \dfrac{\overline{\langle u_r \xi \rangle_S}-\dfrac{1}{Sc}
\ddn{r}{\Xi}}{-\dfrac{1}{Sc}\ddn{r}{\xi_c}}.
\label{eq:nu_sh}
\end{equation}
where $\mathrm{d}T_c /\mathrm{d} r = r_i r_o/r^2$ and $\mathrm{d}\xi_c 
/\mathrm{d} r = -r_i r_o/r^2$ are the gradients of the diffusive states.
In the numerical computations, those quantities are practically evaluated at 
the outer boundary, $r=r_o$, where the convective fluxes vanish.
Heat sinks and sources in fingering convection are provided by buoyancy power 
of thermal and chemical origins $\mathcal{P}_T$ and $\mathcal{P}_\xi$, 
expressed by
\begin{equation}
 \mathcal{P}_T = \dfrac{|Ra_T|}{Pr}\overline{\langle g u_r T \rangle_V},\quad
  \mathcal{P}_\xi = \dfrac{Ra_\xi}{Sc}\overline{\langle g u_r \xi 
\rangle_V}\,.
\end{equation}
On time average, the sum of these two buoyancy sources balances the viscous 
dissipation $D_\nu$, according
to 
\begin{equation}
\mathcal{P}_T  + \mathcal{P}_\xi  = D_\nu, 
\label{eq:DeltaP}
\end{equation}
where $D_\nu=\overline{\langle (\grad\times \vec{u})^2 \rangle_V}$. 

As shown in Appendix~\ref{sec:app1}, the thermal and compositional 
 buoyancy powers can be approximated in spherical 
geometry by 
\begin{equation}
 \mathcal{P}_T \approx - \dfrac{4 \pi r_i r_m}{V} 
\dfrac{|Ra_T|}{Pr^2}(Nu-1),\quad
  \mathcal{P}_\xi \approx \dfrac{4 \pi r_i r_m}{V} \dfrac{Ra_\xi}{Sc^2}(Sh-1)\,,
\end{equation}
where $r_m = (r_o+r_i)/2$ is the mid-shell radius. The turbulent flux ratio 
\citep{Traxler2011} is defined by
\begin{equation}
\gamma = \Rp \dfrac{|\langle u_r T \rangle_V|}{\langle u_r \xi \rangle_V}
\approx \Rp Le \dfrac{Nu-1}{Sh-1}.
 \label{eq:def_gamma}
\end{equation}

The typical flow lengthscale is estimated using an integral lengthscale 
\citep[see ][\S3.6.3]{Backus1996}

\begin{equation}
 \mathcal{L}_h = \dfrac{\pi r_m}{\sqrt{\ell_h(\ell_h+1)}} \approx 
\dfrac{\pi r_m}{\ell_h}, \quad
\end{equation}
in which the average spherical harmonic degree $\ell_h$ is defined according to
\begin{equation}
 \ell_h = \dfrac{\sum_{\ell,m} \ell\mathcal{E}_{\ell}^m}{\sum_{\ell,m} 
\mathcal{E}_{\ell}^m}, 
\label{eq:int_scale}
\end{equation}
where $\mathcal{E}_{\ell}^m$ denotes the time and radially
averaged poloidal kinetic energy at degree $\ell$ and order $m$. 
Note that this global $\ell_h$ is what uniquely defines the flow lengthscale, 
and is routinely used as a diagnostic in global spherical simulations 
\citep[e.g.][]{Christensen2006}. One could define a radially-varying 
$\ell_h$, by considering the radial profiles of the spherical harmonic expansion 
of the kinetic poloidal energy, and get accordingly 
 $\mathcal{L}_h(r)$. Preliminary inspections revealed
no sizeable changes of $\mathcal{L}_h(r)$ in the fluid bulk, and led
us to stick to the integral definition given above for the sake
of parcimony.

\subsection{Parameter coverage}
\label{sec:explo2}

\renewcommand{\arraystretch}{1.4}
\begin{table}
    \caption{Name, notation, definition, and range covered by the 
dimensionless control parameters employed in this
	 study.}
    \label{tab:dimensionless2}
    \begin{center}
    \begin{tabular}{l l l l}\hline
     \textbf{Name} & \textbf{Notation} & \textbf{Definition} 
     & \textbf{This study}\\\hhline{====}
       Thermal Rayleigh& $|\rat|$  & $|\alpha_T g_o \lscale^3 \Delta 
T/\nu 
\kappa_T|$ & $7.334 \times 10^4 - 1.83 \times 10^{11}$\\
       Chemical Rayleigh& $\rac$ & $\alpha_\xi g_o \lscale^3 \Delta 
\xi/\nu \kappa_\xi$ & $2 \times 10^{5} - 5 \times 10^{11}$\\
       Prandtl & $Pr$ &$\nu/\kappa_T$ & $0.03 - 7$\\
         Schmidt & $Sc$ &$\nu/\kappa_\xi$ & $1 - 30$\\
       Lewis & $Le$ &$\kappa_T/\kappa_\xi$ & $3 - 33.3$\\
       Density ratio  & $\Rp$ & $Le\left|\rat\right|/\rac$ & $1 - Le$ \\ \hline
    \end{tabular}
    \end{center}
\end{table}

Table~\ref{tab:dimensionless2} summarises the parameter space covered by this 
study. In order to compare our results against
previous numerical studies by e.g. \citet{Stellmach2011,Traxler2011,Brown2013,Yang2016},
the Prandtl number $Pr$ spans two orders of magnitude, from $0.03$ to $7$,
 while the Lewis number $Le$ varies from $3$ to $33.3$.
The thermal and chemical Rayleigh numbers  
$\rat$ and $\rac$ are comprised between $-1.83 \times 10^{11}$ and  
$-7.33 \times 10^4$, 
and  $2 \times 10^5$ and $5 \times 10^{11}$, respectively, thereby 
permitting an extensive description of the primary instability region
 for each value of the Lewis number $Le$ (recall 
Eq.~\ref{eq:fingering_domain}). In 
practice, this coverage leads to a grand total of 
$\nsims$ simulations. 

\begin{figure}
\includegraphics[width=\linewidth]{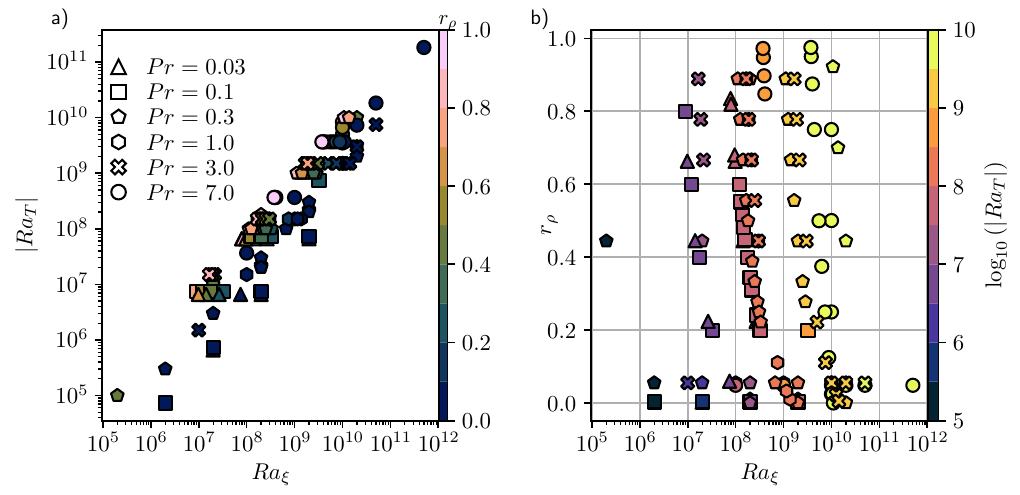}
\caption{
(\textit{a}) Normalised density ratio $\rrho$ in the 
thermal Rayleigh number~$|\rat|$ - chemical Rayleigh
number~$\rac$ plane for the $\nsims$~simulations performed for this study. 
(\textit{b}) $\log_{10}(|\rat|)$ in the $\rac$-$\rrho$~plane for the 
same simulations. Markers reflect the value of~$Pr$.}
\label{fig:exploration}
\end{figure}
Figure~\ref{fig:exploration} illustrates the exploration of the parameter 
space. Subsets of simulations were designed according to three different 
strategies. First by varying $(\rat, \rac)$, hence the buoyancy input power, 
while keeping the triplet $(Pr,Le,\Rp)$ constant. 
This gives rise to horizontal lines in 
Fig.~\ref{fig:exploration}(\textit{b}). 
Second by varying $\rac$, and consequently $\Rp$, at constant $(Pr,Le,\rat)$. 
These series are located on horizontal lines in 
Fig.~\ref{fig:exploration}(\textit{a}), and along branches of the same color in 
Fig.~\ref{fig:exploration}(\textit{b}), see e.g. the yellow circles with 
$\rac \lesssim 10^{10}$. This subset is meant at easing the comparison with previous 
local studies in periodic Cartesian boxes, where the box size is set 
according to the number of fingers it contains, which amounts to an implicit 
prescription of $\rat$. Third, we performed a series by varying $\rat$, keeping 
$(Pr,Le,\rac)$ constant. It shows as a vertical line of $5$~circles in both panels 
of Fig.~\ref{fig:exploration}, noting that $\rac$ is set to $10^{10}$ for that 
series. 

On the technical side, spatial resolution ranges from 
$(N_r,\ell_\text{max})=(41,85)$ to $(769,938)$ in the catalogue of simulations, 
notwithstanding a single simulation that used finite-differences in radius with 
$1536$ grid points in conjunction with $\ell_\text{max}=
m_\text{max}=1536$. 
Moderately supercritical cases were initialised by a random thermo-chemical 
perturbation applied to a motionless fluid. The integration of strongly 
supercritical cases started from snapshots of equilibrated solutions subject 
to weaker forcings, in order to shorten the duration of the 
transient regime. Numerical convergence was in most cases assessed by checking 
that the average power balance defined by Eq.~(\ref{eq:DeltaP}) 
was satisfied within $2$\% \citep{King2012}. In some instances, though, the 
emergence and growth of jets caused the solution to evolve over several viscous 
time scales $\tau_\nu$ without reaching a statistically converged state in the 
power balance sense. The convergence criterion we used instead was that of a 
stable cumulative temporal average of $\ektor$, which we assessed by 
visual inspection. For five strongly-driven
simulations, jets did not reach their saturated state because of 
a too costly numerical integration; this may result in  
an underestimated value of $\ektor$. These five cases feature a ``NS'' label
in the leftmost column of Tab.~\ref{tab:simu_tab2}, where the main 
properties of the $\nsims$~simulations are listed. The total computation 
time required for this study represents 20 million single core hours, executed 
for the most part on AMD Rome processors.

\subsection{Definition of boundary layers}

\label{subsec:def_BL}
\begin{figure}
  \centering
\includegraphics[width=\linewidth]{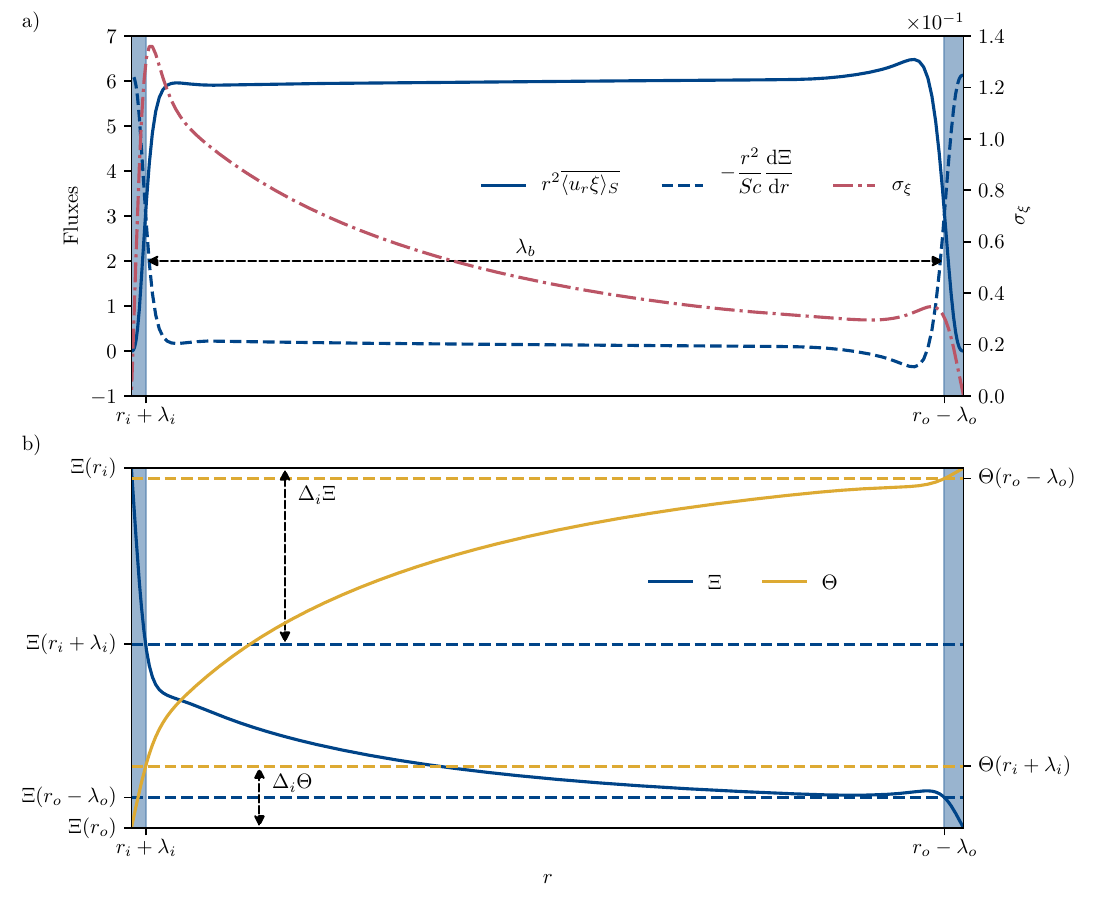}
\caption{%
(\textit{a}) Time-averaged radial profiles of convective and diffusive
 compositional fluxes, expressed by $r^2 \overline{\langle u_r \xi \rangle_S}$ 
and $-r^2 Sc^{-1} \mathrm{d}\Xi/\mathrm{d}r$ respectively (see 
Eq.~\ref{eq:nu_sh}), for a simulation with  
  $|\rat| = 3.66 \times 10^7$, $\Rp = 1.1$, $Pr = 7$, and $Le = 3$ 
(simulation 123 in Tab.~\ref{tab:simu_tab2}). 
The radial profile of the variance of composition $\sigma_\xi$ 
is represented by a dash-dotted line. 
 (\textit{b}) Time-averaged radial 
profiles of composition (solid blue line) and temperature (solid mustard line). 
Horizontal dashed lines highlight the values of composition and temperature
 at the edges of the convective bulk. $\dXii$ ($\dTi$) is the composition 
(temperature)
drop across the inner boundary layer. The blue-shaded 
areas highlight the inner and outer chemical boundary layers, of 
thickness $\lambda_i$ and $\lambda_o$, respectively, while
the convective bulk has thickness $\lambda_b$.}
\label{fig:def_BL}
\end{figure}

We now turn our attention to the practical characterisation of the boundary 
layers that emerge in our bounded setup, whose properties 
are governed by the least diffusive field, i.e. the chemical composition 
\citep[e.g.][]{Radko2000}. In the remainder of this study, $\lambda_i$ (resp. 
$\lambda_o$) will denote the thickness of the
inner (resp. outer) chemical boundary layer.
In contrast to planar configurations, boundary layer 
thicknesses at both boundaries differ ($\lambda_i\neq \lambda_o$) due to 
curvature and radial changes of gravitational acceleration 
\citep[e.g.][]{Vangelov1994,Gastine2015}.
Figure~\ref{fig:def_BL}(\textit{a}) shows the time-averaged radial profiles
of convective and diffusive chemical fluxes (recall~\ref{eq:nu_sh}),
alongside the variance of chemical composition, $\sigma_\xi(r)$, 
for a simulation with 
$|\rat| = 3.66 \times 10^7$, $\Rp = 1.1$, $Pr = 7$ and $Le = 3$ 
(simulation 123 in Tab.~\ref{tab:simu_tab2}).
Several methods have been introduced to assess the boundary layer thicknesses. 
An account of these approaches is given by \citet[\S~2.2]{Julien2012}, in the 
classical context of Rayleigh-B\'enard convection in planar geometry. In that 
setup, temperature is uniform within the convecting bulk, and a first approach 
consists of picking the depth at which 
the linear profile within the thermal boundary layer intersects
the temperature of the convecting core \citep[see 
also][their Fig.~3]{Belmonte1994}. 
A second possibility is to argue that the depth of the boundary layer is
that at which the standard deviation of temperature reaches 
a local maximum \citep[e.g.][their Fig.~4]{Tilgner1993}. \citet{Long2020} 
showed however that both approaches become questionable when convection 
operates 
under the influence of global rotation, in which case
the heterogeneous distribution of temperature causes
the linear intersection method to fail, or if Neumann 
boundary conditions are prescribed in place of Dirichlet conditions 
for the temperature field, in which case the maximum 
variance method is no longer adequate.  
A third option proposed by  
\citet{Julien2012}, and favoured by \citet{Long2020} in their study, 
is to define $\lambda_i$ and $\lambda_o$ at the locations where
convective and diffusive fluxes are equal.
Chemical boundary layers defined by this condition
appear as blue-shaded regions in Fig.~\ref{fig:def_BL}. 
They are thinner than those that may have been 
determined otherwise using either the linear intersection or the standard 
deviation approaches \citep[][]{Julien2012} and they are asymmetric, with 
$\lambda_i < \lambda_o$.
Figure~\ref{fig:def_BL}(\textit{b}) shows the time-averaged radial profiles of 
temperature and composition. We observe a pronounced 
asymmetry in both profiles with a larger temperature and composition drop 
accommodated across the inner boundary layer than across the outer one. 
Inspection of Fig.~\ref{fig:def_BL}(\textit{b}) also reveals 
that the profiles of composition
and temperature remain quite close to linear within the boundary layers 
determined with the flux equipartition method. 
In the following, we will adhere to this approach, and exploit 
the linearity of the profiles of temperature and composition in some of our 
derivations. The downside of this choice is that it does not incorporate the 
curvy part of the profiles at the edge of the boundary layers, hence possibly
overestimating the contrast of composition in the fluid bulk compared to other
boundary layer definitions.

Let $\lambda_b$ denote the thickness of the bulk of the fluid permeated by 
fingers, let $\dTb$ and $\dXib$ stand for the contrast of temperature and 
composition across this region, and let $\dXii$ $(\dXio)$ and $\dTi$ 
$(\dTo)$ be the composition and temperature drops across the inner (outer) 
chemical boundary layer. By choice, each contrast is a positive
quantity. We note for future usage that the following non-dimensional 
relationships hold
\begin{equation}
1 = \lambda_i+ \lambda_b + \lambda_o,\quad 
1 =  \dTi + \dTb +\dTo, \quad
1 = \dXii + \dXib + \dXio.
\label{eq:lambdab} 
\end{equation}

\section{Fingering convection}
\label{sec:fingers}

\subsection{Flow morphology}
\begin{figure}
\includegraphics[width=\textwidth]{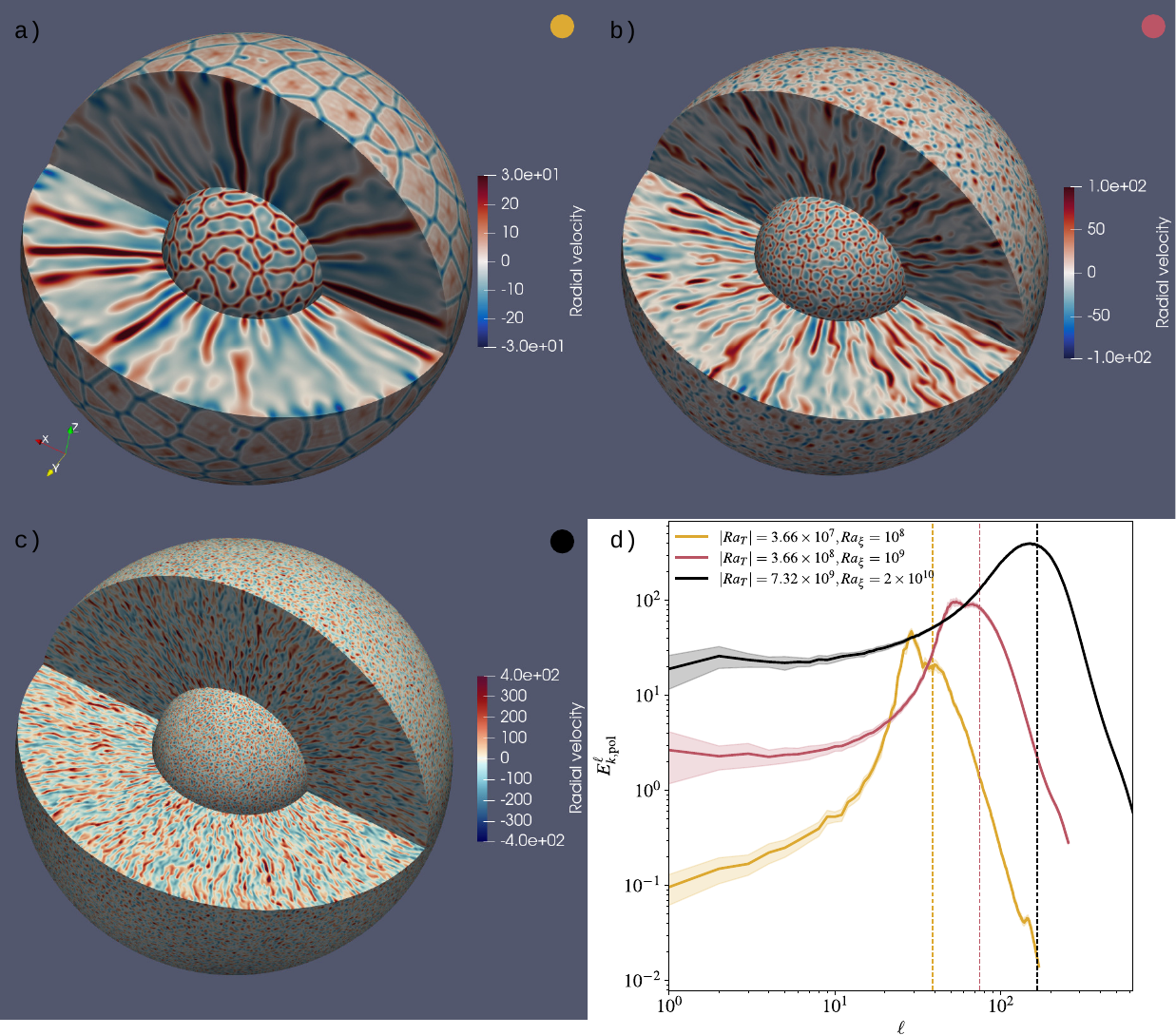}
\caption{(\textit{a}-\textit{c}) 
Three-dimensional renderings of the radial velocity 
for three simulations carried out using $\Rp = 1.1$, $Pr = 7$
and $Le = 3$,  and 
$\rac=10^8, 10^9,$ and $2\times 
10^{10}$ (simulations 123, 118 and 103 in Tab.~\ref{tab:simu_tab2}), 
for (\textit{a}), (\textit{b}) and (\textit{c}), 
respectively. Inner and outer spherical surfaces correspond to $r_i + 0.03$ 
and $r_o - 0.04$, respectively. (\textit{d}) Poloidal kinetic energy spectra as 
a function of spherical harmonic degree $\ell$ for these three 
simulations. Dashed vertical lines correspond to the average spherical harmonic 
degree $\ell_h$ (see Eq.~\ref{eq:int_scale}), of value $39$, $75$ and $166$, 
for $\rac=10^8, 10^9,$ and $2\times 10^{10}$.}
\label{fig:2D}
\end{figure}

We first focus on a series of three simulations to highlight the 
specificities of fingering convection in global spherical geometry. 
Figure~\ref{fig:2D} provides three-dimensional renderings of the fingers, along 
with the corresponding kinetic energy spectra, of three cases that share $\Rp = 
1.1$,  $Pr = 7$ and $Le = 3$ and differ by $\rac$, which increases from 
$10^8$ to $2\times 10^{10}$ (simulations 123, 118 and 103 in 
Tab.~\ref{tab:simu_tab2}). The convective power injected in the fluid 
is accordingly multiplied by $500$ between the simulation
closest to onset, illustrated in Fig.~\ref{fig:2D}(\textit{a}), and the 
most supercritical one, shown in Fig.~\ref{fig:2D}(\textit{c}). 
For the least forced simulation (Fig.~\ref{fig:2D}\textit{a}), the flow is 
dominated by coherent radial filaments that extend over the entire spherical 
shell. These particular structures are reminiscent of the
``elevator modes'' found in periodic planar models
\citep[e.g.][\S2.1]{Radko2013}. 
Inside each of these 
filaments $T$, $\xi$ and
$u_r$ can be considered to be quasi-uniform. The velocity field reaches
relatively small amplitudes, with a poloidal Reynolds number $\rep \approx 10$. 
Geometrical patterns link the fingers together over spherical surfaces, and
it appears that fingers emerge at the vertices of polygons in the vicinity
of the inner sphere. Fingers have an almost constant width in the 
bulk of the domain. The rather large polygonal patterns that 
appear on the outer surface of the three-dimensional rendering are due to the 
the weakening of the radial velocity in the vinicity of the outer boundary layer, that goes along with 
a widening of the finger as it penetrates the boundary layer. The spectrum of 
this simulation, displayed in Fig.~\ref{fig:2D}(\textit{d}), presents a marked 
maximum, with an average spherical harmonic degree $\ell_h$ (recall 
Eq.~\ref{eq:int_scale}) of $39$. 
The majority of the poloidal kinetic energy of the fluid is stored in degrees 
close to this peak. The quasi-homogeneous lateral thickness of fingers
in Fig.~\ref{fig:2D}(\textit{a}) illustrates this spectral concentration. 

Figure~\ref{fig:2D}(\textit{b}) reveals that a strengthening
of the forcing leads to an increase of the magnitude of the velocity, with 
$\rep = 28$, alongside a gradual loss of the coherent tubular structure 
of the fingers in favour of undulations and 
occasional branchings. They contract horizontally leading to a shift
of $\ell_h$ to a higher value of $75$. The geometrical patterns remain 
well-defined over the inner sphere but appear attenuated at the outer 
spherical surface, again the signature of the effect of the
outer boundary layer. Further increase of injected convective power 
causes the occasional fracture of fingers in the radial direction, see 
Fig.~\ref{fig:2D}(\textit{c}), 
as they assume the shape of flagella reminiscent of the modes of
\citet[][Fig.~1]{Holyer1984}. Though the fingers gradually lose their 
vertical coherence on increased convective forcings, they still present an 
anisotropic elongated structure in the radial direction.
Accordingly, the lateral thickness of fingers continues to decrease 
and $\ell_h$ now reaches a value of $166$. That amounts to finding
$\mathcal{O}(10^4)$ such elongated structures at any radius $r$
in the bulk of the flow.

\subsection{Mean profiles and compositional boundary layers}
\label{sec:instabilite_1}

\begin{figure}
\includegraphics[width=\textwidth]{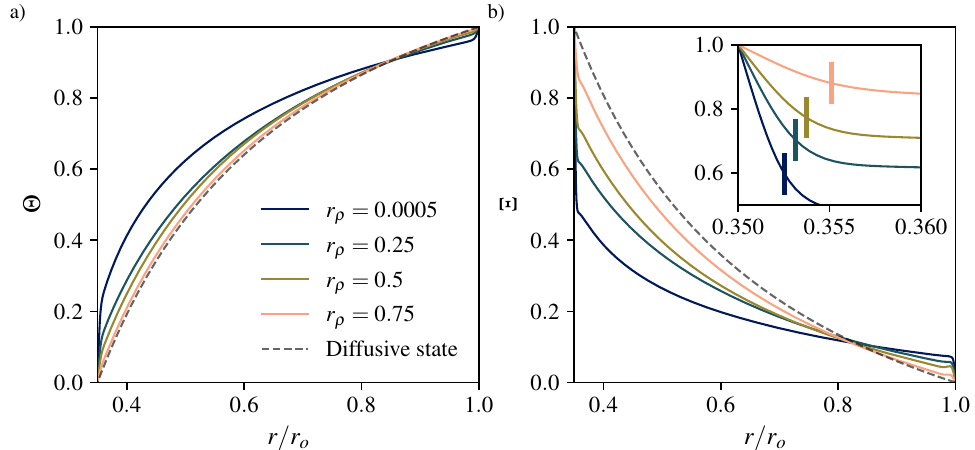}
\caption{%
(\textit{a}) Time-averaged radial profile of temperature, $\Theta$,  
for four simulations sharing  $|\rat| = 3.66 \times 10^{9}$, 
$Pr = 7$, and $Le = 3$, and increasing values of $\rrho$ 
(simulations 106, 110, 112 and 113 in Tab.~\ref{tab:simu_tab2}).
(\textit{b}) Time-averaged radial profile of composition, $\Xi$,   
for the same simulations. 
The inset shows a magnification of the inner boundary layer, whose
edge is shown with a thick vertical segment. On both panels, the dashed lines 
correspond to the diffusive reference profile.}
\label{fig:profil}
\end{figure}

We now  assess the impact of fingers on the average temperature
and composition profiles within the  spherical shell. 
Figures~\ref{fig:profil}(\textit{a})-(\textit{b}) show the 
time-averaged  radial profiles of temperature and composition of four 
simulations that share $|\rat| =3.66 \times 10^9$, $Pr = 7$, $Le = 3$, and 
differ by the prescribed $\rac$, whose value goes from
$4.392 \times 10^9$ to $1.1 \times 10^{10}$, with
a concomitant decrease of $\rrho$ from $0.75$ down to $5 \times 10^{-4}$.  

The increase of $\rac$ impacts composition more 
than temperature, as it leads to steeper boundary layers and flatter bulk 
profiles of $\xi$ that substantially deviate from
their diffusive reference, displayed with a dashed line in 
Fig.~\ref{fig:profil}(\textit{b}). Inspection of 
Fig.~\ref{fig:profil}(\textit{a}) shows
that this trend exists but is less marked for temperature. 
Accordingly, the bulk temperature
and composition drops $\dTb$ and $\dXib$ introduced above decrease as 
$\rac$ increases, in a much more noticeable manner for composition 
than for temperature. Boundary layer asymmetry inherent in curvilinear 
geometries \citep{Jarvis1993} is clearer with increasing $Ra_\xi$: for the 
largest forcing considered here about $40\%$ ($10\%$) of the contrast of 
composition is accommodated at the inner (outer) boundary layer.

To enable comparison with results from unbounded studies,
we seek scaling laws for the effective contrasts $\dTb$ and $\dXib$ that 
develop in the fluid bulk. To that end, and in line with the 
characterisation of boundary layers we introduced above, we first make use of 
the heat and composition flux conservation  
 over spherical 
surfaces. Assuming that temperature and composition vary linearly within 
boundary layers yields

\begin{equation}
  \left(\dfrac{r_i}{r_o}\right)^2 \dfrac{\dTi}{\dTo} = 
\dfrac{\lambda_i}{\lambda_o}, \quad \left(\dfrac{r_i}{r_o}\right)^2 
\dfrac{\dXii}{\dXio} =
        \dfrac{\lambda_i}{\lambda_o}  
        \label{eq:BL_relation1}, 
\end{equation}
where the $r_i/r_o$ factors emphasise the asymmetry of boundary layer
properties caused by spherical geometry. To derive scaling laws for the ratio 
of temperature and composition drops at both boundary layers, one must invoke a 
second hypothesis. In classical Rayleigh-B\'enard convection in an annulus, 
\citet{Jarvis1993} made the additional assumption that the boundary layers are 
marginally unstable \citep{Malkus1954}. In other words, a local Rayleigh number 
defined on the 
boundary layer thickness should be close to the critical value to trigger 
convection. Later numerical simulations in 3-D by \citet{Deschamps2010} in 
the context of infinite Prandtl number convection and by \citet{Gastine2015} 
for $Pr=1$ however showed that this hypothesis failed to correctly account for 
the actual boundary layer asymmetry observed in spherical geometry. For 
fingering 
convection in bounded domains, \citet{Radko2000} nonetheless showed that the 
marginal stability argument provides a reasonable description of the boundary 
layers for composition. We  here test this hypothesis by introducing
the local thermohaline Rayleigh numbers $Ra^{\lambda_i}$ and
$Ra^{\lambda_o}$ defined over the extent of the inner and outer boundary layers
\begin{equation}
Ra^{\lambda_{i}} = g_{i} \lambda_{i}^3 \left( \rac \dXii   - |\rat|
        \dTi \right), \quad
Ra^{\lambda_{o}} = g_{o} \lambda_{o}^3 \left( \rac \dXio   - |\rat|
        \dTo \right), 
\end{equation}
where $g_i$ and $g_o$ denote the acceleration of gravity at 
the inner and outer boundaries, with $g_i/g_o = r_i/r_o$. We note in 
passing that gravity increasing linearly with radius is the second factor 
responsible for the boundary layer asymmetry. 
Equating $Ra^{\lambda_{i}}$ and $Ra^{\lambda_{o}}$ to a critical value $Ra^c$ 
gives, in light of Eq.~\eqref{eq:BL_relation1}, 
\[
\left(\dfrac{\lambda_i}{\lambda_o}\right)^3 = \dfrac{r_o}{r_i} 
\dfrac{\dXio}{\dXii}. 
\]
Further use of Eq.~\eqref{eq:BL_relation1} yields
 \begin{equation}
 \dfrac{\lambda_o}{\lambda_i} = \left(\dfrac{r_i}{r_o}\right)^{-1/4}
\label{eq:dLi}
\end{equation}
and
\begin{equation}
\dfrac{\dXio}{\dXii} = \left(\dfrac{r_i}{r_o}\right)^{7/4}
\label{eq:dXi}, 
\end{equation}
both of which being a function of the sole radius ratio $r_i/r_o$.
\begin{figure}
  \includegraphics[width=\linewidth]{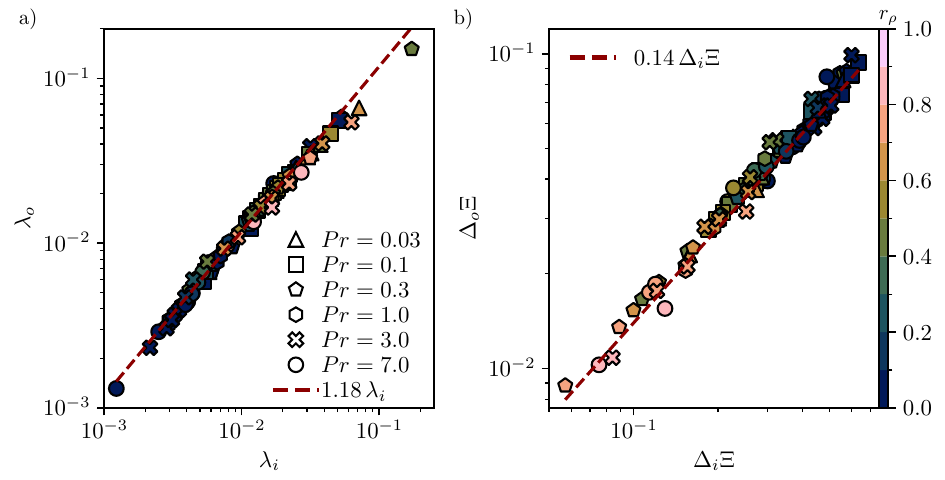}
\caption{(\textit{a}) 
Thickness of the outer boundary layer, $\lambda_o$, 
as a function of the thickness of the inner boundary layer, 
$\lambda_i$. (\textit{b}) Composition contrast across the outer boundary 
layer, $\dXio$, as a function of its inner counterpart, $\dXii$.
In both panels, the dashed lines correspond to a linear fit to 
data retaining only those simulations with $\lambda_i < 0.02$.} 
\label{fig:LXi_dXi}
\end{figure}

Figure~\ref{fig:LXi_dXi} shows the conformity of these two laws
with the numerical dataset, which has a constant radius ratio $r_i/r_o=0.35$. 
In Fig.~\ref{fig:LXi_dXi}(\textit{a}), we observe a close to linear
increase of $\lambda_o$ with $\lambda_i$ over two orders
of magnitude of variations. A least-squares fit performed for those simulations
with $\lambda_i < 0.02$ provides $\lambda_o=1.18\,\lambda_i$ instead
of the expected $\lambda_o=1.30\,\lambda_i$ from Eq.~\eqref{eq:dLi}. 
Simulations
causing this departure are those close to onset with 
thick boundary layers within which the linearity assumption 
may not hold. Likewise, we see in Fig.~\ref{fig:LXi_dXi}(\textit{b})
that Eq.~(\ref{eq:dXi}) captures the relative
ratio $\dXio/\dXii$  within the dataset. We find
$\dXio = 0.14 \dXii$ instead of the predicted $\dXio = 0.16 \dXii$, 
and  note that the dispersion about a linear law is
maximum for strongly-driven simulations (those with smaller
$\rrho$ in Fig.~\ref{fig:LXi_dXi}\textit{b}). These results indicate that in 
contrast with Rayleigh-Bénard convection, the marginal 
stability hypothesis provides a good way to characterise the boundary layer 
asymmetry in spherical shell fingering convection, with the caveat
that confirmation of this finding should be sought for radius ratios other
than the one considered in this study.

\begin{figure}
    \includegraphics[width=\linewidth]{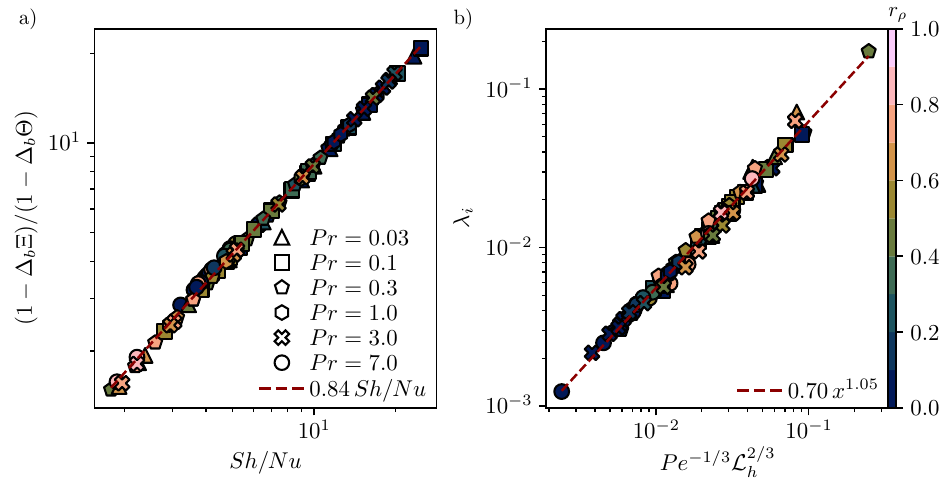}
\caption{(\textit{a}) $(1 - \dXib)/(1
- \dTb T)$ as a function of  $Sh/Nu$. (\textit{b}) Thickness of 
the inner boundary layer as a function of $Pe^{-1/3}\Lh^{2/3}$ derived in 
Eq.~\eqref{eq:scaling_BL}. In each panel, the result of the
linear regression is shown with a dashed line.}
\label{fig:BL_Nu_Sh}
\end{figure}

In order to obtain a relationship between the contrasts across the bulk, 
$\dXib$ and $\dTb$, recall 
that the Sherwood and Nusselt numbers $Sh$ and $Nu$
are expected to read at $r=r_o$
\begin{equation}
  Sh = \dfrac{r_o}{r_i} \dfrac{\dXio}{\lambda_o}, \quad Nu =
       \dfrac{r_o}{r_i} \dfrac{\dTo}{\lambda_o}\,,
       \label{eq:ShNu_BL}
\end{equation}
if the linearity assumption for composition and temperature 
holds within the chemical boundary layer.
Combining these definitions with 
Eqs.~\eqref{eq:lambdab} and (\ref{eq:dLi}-\ref{eq:dXi})
yields the following relationship between $\dTb$, $\dXib$, $Nu$, and $Sh$, 
\begin{equation}
  \dfrac{1 - \dXib}{1 - \dTb} = \dfrac{Sh}{Nu}.
  \label{eq:dXibdTb}
\end{equation}
Figure~\ref{fig:BL_Nu_Sh}(\textit{a}) shows that there is convincing 
evidence for
a linear relationship between $(1 - \dXib)/(1 - \dTb)$ and  $Sh/Nu$ 
across the parameter space sampled by the dataset; the slope
is equal to $0.84$ instead of the expected value of one. This 
might come from the uncertainty 
related to the departure from the linearity assumption for the chemical 
boundary layer,  possibly leading to underestimate the
true value of the Sherwood number.

For a better understanding of Eq.~\eqref{eq:dXibdTb}, 
it is insightful to split the mean radial profiles into the sum of a reference 
conducting state and fluctuations denoted by primed quantities
\[
 \Xi=\xi_c+\Xi',\ \Theta=T_c+\Theta'\,.
\]
Using the definition of the flux ratio~\eqref{eq:def_gamma}, 
Eq.~\eqref{eq:dXibdTb} can be rewritten as
\begin{equation}
 \dfrac{\dXib'}{\dTb'} \approx \Rp Le\gamma^{-1}\,,
 \label{eq:fraction_fluct}
\end{equation}
where we have assumed that $\Delta_b \xi_c = \Delta_b T_c \approx 1$.

Following \citet{Yang2015}, we now evaluate whether results coming from 
classical Rayleigh-B\'enard convection models regarding the nature of the 
boundary layers are still applicable to bounded double-diffusive convection. In 
that regard, the cornerstone of the GL theory is that 
the kinetic and thermal boundary layer thicknesses adhere to the 
Prandtl-Blasius model \citep[e.g.][\S9.2]{Schlichting18}. In this 
model, the boundary layer corresponds 
to a laminar flow over an horizontal plate and the temperature or chemical 
composition are treated as passive scalars. For fingering convection with $Le>1$, this 
translates to 
\begin{equation}
 \dfrac{\lambda_\xi}{\Lh} \sim Sc^{-1/2}\,Re_h^{-1/2} (Sc < 1), \quad 
 \dfrac{\lambda_\xi}{\Lh} \sim Sc^{-1/3}\,Re_h^{-1/2} (Sc > 1),
 \label{eq:PB}
\end{equation}
where $\lambda_\xi$ denotes the thickness of the inner or outer 
chemical boundary layer, 
and $Re_h = u_h \Lh/\nu$ is a Reynolds number defined using the lengthscale 
$\Lh$ and the near-boundary horizontal flow velocity $u_h$. 
For tubular fingers, mass conservation between the 
finger core of typical diameter $\lambda_\xi$ \citep[][their Figure~10]{Yang2016} 
and the horizontal flow that converge towards the finger in the boundary 
layers demands
\begin{equation}
u_h \lambda_\xi \pi \Lh \approx u_r \dfrac{\pi}{4} \lambda_\xi^2 (Sc < 
1), 
\quad \dfrac{\lambda_\xi}{\lambda_U}u_h \lambda \pi \Lh \approx u_r \dfrac{\pi}{4} 
\lambda_\xi^2\ (Sc > 1) \,, 
\end{equation}
where $\lambda_U$ denotes the thickness of the velocity boundary layer.
The factor $\lambda_\xi/\lambda_U$ in the equation on the right reflects the fact 
that the velocity boundary layer is nested in the compositional one when 
$Sc > 1$. As such, and assuming linear boundary layers, the relevant horizontal 
flow velocity has to be rescaled by the relative thickness of both boundary 
layers. Using Eq.~\eqref{eq:PB}, this then leads to the unique form
\begin{equation}
 \lambda_\xi \sim Pe^{-1/3}{\Lh}^{2/3},
 \label{eq:scaling_BL}
\end{equation}
for both Schmidt number end-members. Figure~\ref{fig:BL_Nu_Sh}(\textit{b}) 
shows the thickness of the inner compositional boundary layer ($\lambda_\xi = \lambda_i$) as a function of 
this theoretical scaling for all the simulations where the boundary layers 
could be evaluated. The reduced scatter of the data as well as the slope of the 
best fit power law being close to one indicate an excellent agreement 
with the Prandtl-Blasius model combined with the hypothesis of tubular fingers.

\subsection{Effective density ratio}

In order to compare the dynamics with unbounded domains,
the common practice in bounded planar geometry \citep{Schmitt1979, Radko2000, 
Yang2020a} consists in introducing an effective density ratio within the bulk 
of the domain expressed by

\[
\Rpe = \Rp \dfrac{\dTb}{\dXib}.
\]
This measure is appropriate to fingering convection in Cartesian domains with $\Rp \ll Le$, 
given the piecewise-linear nature of the composition profile 
\citep[see e.g. Fig.~2 of][]{Yang2020}. This estimate is however not suitable 
close to onset as boundary layer definition becomes ill-posed. An extra 
complication arises in spherical geometry since our definition of boundary 
layers only retains the linear part of the drop of composition, which tends 
to overestimate $\dXib$ (recall Fig.~\ref{fig:def_BL} and the discussion 
 in \S~\ref{subsec:def_BL}). As such, it appears more reliable to introduce 
an effective density ratio (and its normalised counterpart) based on the 
gradients of temperature and  composition at mid-depth:

\begin{equation}
\Rpe = -\Rp 
\left. \dfrac{\mathrm{d}
\Theta}{\mathrm{d}r}\right|_{r_m}
\left(\left.\dfrac{\mathrm{d}
\Xi}{\mathrm{d}r}\right|_{r_m}\right)^{-1}\,,\quad
\rpe = \dfrac{\Rpe - 1}{Le -1}.
\end{equation}
We saw above that the contrast of composition across the bulk (or fingering region)
is systematically lower than that of temperature. It hence follows 
that $\Rpe > \Rp$ and $\rpe > \rrho$.

\subsection{Finger width}

We now derive scaling laws for fingering convection based on our $\nsims$
bounded simulations, beginning with the typical lateral extent $\Lh$ of a 
finger, as defined in 
Eq.~\eqref{eq:int_scale}. 
\begin{figure}
\includegraphics[width=\linewidth]{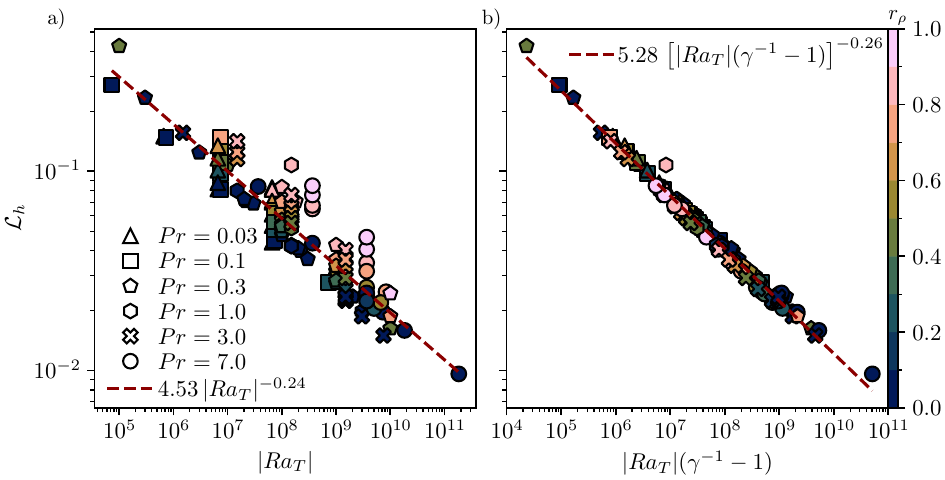}
\caption{%
(\textit{a}) Horizontal lengthscale $\mathcal{L}_h$ (Eq.~\ref{eq:int_scale}) as 
a function of $|\rat|$. (\textit{b}) $\mathcal{L}_h$ as a 
function of $|\rat|(\gamma^{-1}-1)$.  In both 
panels, the dashed lines  correspond to power law fits.}
\label{fig:lPolPeak}
\end{figure}
In his seminal contribution, \cite{Stern1960} predicts a scaling of the form 
\begin{equation}
        \Lh \propto |\rat|^{-1/4}
        \label{eq:Lh1}
\end{equation}
for an unbounded Boussinesq fluid subjected to uniform thermal and chemical background gradients 
when $Le \gg \Rp \gg 1$. A least-squares fit of our dataset yields
\begin{equation}
        \Lh = 4.23\,|\rat|^{-0.23}, 
        \label{eq:Lh2}
\end{equation}
where the exponent found is rather close to the expected value of $-0.25$, for 
values of $|\rat|$ spanning $6$ orders of magnitude. Yet, inspection 
of Fig.~\ref{fig:lPolPeak}(\textit{a}) reveals that this scaling fails to capture 
a second dependency to $\rrho$. At a given 
value of $|\rat|$, we observe indeed that $\Lh$ is an increasing function of 
$\rrho$. In order to make progress, let us assume that the finger width  
is controlled by a balance between buoyancy and viscous forces. 
In addition, we resort to the {\it tall finger} hypothesis 
introduced by \citet[][p.~192]{Stern1975} and christened by \cite{Smyth2007}, which 
consists of neglecting along-finger derivatives in favor of cross-finger 
derivatives. Accordingly, the time and volume average 
of the radial component 
of Eq.~\eqref{eq:momentum} becomes 
\begin{equation}
     \overline{\avg{\left(-\dfrac{\rat}{Pr}T + \dfrac{\rac}{LePr} \xi\right) 
\dfrac{r}{r_o}}{V}} \sim \overline{\avg{\laplacien \vec{u} \cdot \er}{V}}
 \sim \dfrac{\rep}{\Lh^2}. 
 \label{eq:mom_Lh}
\end{equation}
Likewise, the average of the heat equation \eqref{eq:temp} leads to
\begin{equation}
\rep \dfrac{\dTb}{\lambda_b}
\sim \dfrac{\overline{\avg{T}{V}}}{Pr\Lh^2}.
\end{equation}
Finally, a relationship between $\overline{\avg{T}{V}}$ and 
$\overline{\avg{\xi}{V}}$ is expressed by means
of the flux ratio $\gamma$, 
\begin{equation}
\overline{\avg{\xi}{V}} \sim \dfrac{Le|\rat|}{\gamma\rac} 
\overline{\avg{T}{V}}, 
\label{eq:xi_Lh}
\end{equation}
which, in light of Eq.~\eqref{eq:def_gamma}, assumes implicitly that the radial velocity correlates well with 
thermal and chemical fluctuations. Upon combining 
Eqs.~(\ref{eq:mom_Lh}-\ref{eq:xi_Lh}), we obtain
\begin{equation}
\Lh^{-4} \sim (\gamma^{-1}-1) |\rat|
\dfrac{\dTb}{\lambda_b}. 
\label{eq:Lh_tot}        
\end{equation}
The bulk temperature gradient $\dTb/\lambda_b$ aside, this expression is 
equivalent to that 
proposed by \citet{Taylor1989} in the discussion of their experimental results.  
Figure~\ref{fig:lPolPeak}(\textit{b}) shows $\Lh$ as a function of 
$|\rat|(\gamma^{-1}-1)$.  
The extra factor $(\gamma^{-1}-1)$ removes the dispersion
observed in Fig.~\ref{fig:lPolPeak}(\textit{a}). A least-squares fit 
of $\log_{10}(\Lh)$ versus 
$\log_{10}\left[|\rat|(\gamma^{-1}-1)\right]$ leads to
\begin{equation}
    \Lh = 5.28\left[(\gamma^{-1} -1)|\rat|\right]^{-0.26}.
    \label{eq:Lh_fit}
\end{equation}
The exponent of $|\rat|$ remains close to the value of $-0.25$ proposed by 
\citet{Stern1960}. In order to assess the effect of  
the $\dTb/\lambda_b$ term, we introduce the misfit 
\begin{equation}
    \chi_y^2 = \sqrt{\dfrac{1}{N} \sum_{i = 1}^N \left|\dfrac{\tilde{y}_i - 
y_i}{y_i}\right|^2},
\end{equation}
where $N$ is the number of simulations, $y_i$ is the $i$-th measured value of 
$\log_{10}(\Lh)$ and $\tilde{y}_i$ its prediction
 by the least-squares fit of interest. In the absence of a correction factor, 
recall Eq.~\eqref{eq:Lh2}, 
we obtain $\chi_y^2=0.091$. 
With the full correction~(\ref{eq:Lh_tot}), $\chi_y^2=0.023$, which amounts to 
a fourfold reduction. The  misfit increases by a modest amount to $0.025$ if we
omit the factor $\dTb/\lambda_b$ in Eq.~\eqref{eq:Lh_tot}. It thus 
appears 
reasonable to ignore that factor for the sake of parsimony. 
In the remainder of this study, we will therefore adhere to 
\begin{equation}
    \Lh^4 \sim \dfrac{\gamma}{1 - \gamma} |\rat|^{-1}.
    \label{eq:Lh}
\end{equation}

\begin{figure}
\centering
\includegraphics[width=\linewidth]{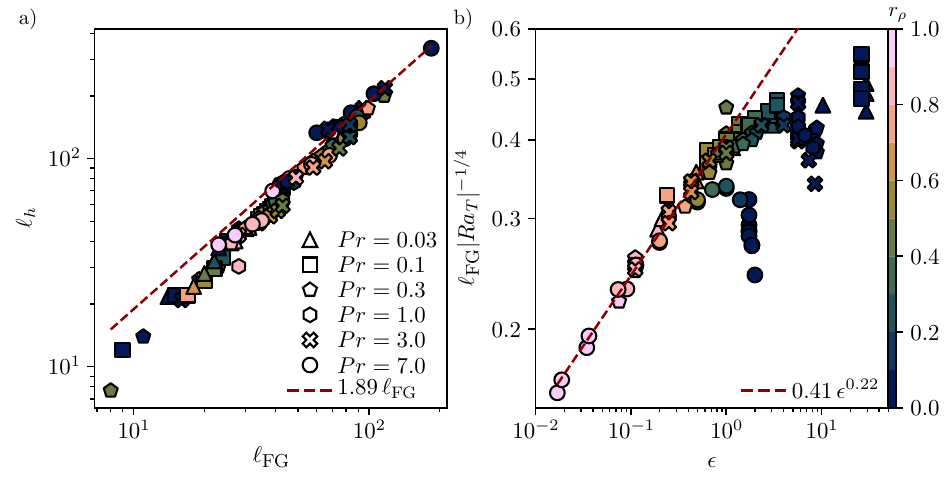}
\caption{%
(\textit{a}) Mean spherical harmonic degree $\lh$ as a function of 
the degree of the fastest growing mode $\lc$; the dashed line shows the 
result of the linear 
regression. (\textit{b}) Most linearly unstable degree $\lc|\rat|^{-1/4}$ 
as a function of the supercriticality parameter $\epsilon$. The dashed line in 
panel (\textit{a}) corresponds to a linear fit forced over the entire dataset, yielding
a prefactor equal to $1.89$, while 
the best fit in panel (\textit{b}) was obtained from the simulations with $\epsilon < 0.5$.}
\label{fig:threshold}
\end{figure}

Following the idea of \citet{Schmitt1979}, we now examine whether the average 
spherical harmonic degree $\ell_h$ of fingers in developed convection relates to 
the degree of the fastest growing mode, $\ell_\text{FG}$. 
Linearisation of the system (\ref{eq:divu}-\ref{eq:chem}) is conducted by 
considering small perturbations of the poloidal scalar $W$, temperature $T$ 
and composition $\xi$. Performing a spherical harmonic expansion of these 
variables leads to equations decoupled in harmonic degree $\ell$  and 
independent of the spherical harmonic order $m$. We use the ansatz 
\begin{equation}
[W_\ell, T_\ell, \xi_\ell](t, r) = [\hat{W_\ell}, \hat{T_\ell}, 
\hat{\xi_\ell}](r) 
\exp\left(\tau_\ell t\right), 
\label{eq:uTXi}
\end{equation}
where $[\hat{W_\ell}, \hat{T_\ell}, \hat{\xi_\ell}]$ are the radial shape 
functions of the perturbation of degree $\ell$. Focusing on the 
real-valued eigenvalues $\tau_\ell$ relevant to the fingering instability,
we obtain the generalised eigenvalue problem 
\begin{equation}
\mathcal{A}_\ell \vec{X}_\ell = \tau_\ell \mathcal{B}_\ell \vec{X}_\ell,
\end{equation}
where $\mathcal{A}_\ell$ and $\mathcal{B}_\ell$ are real dense matrices, 
$\vec{X}_\ell\equiv[W_\ell, T_\ell, \xi_\ell]^T$ is the state vector,  
and we understand that each eigenvalue depends on  $\ell$, $\rat$,
$\rac$, $Pr$ and $Le$. We resort to the \texttt{Linear Solver Builder} 
(\texttt{LSB}) package developed by \citet{Valdettaro2007} to determine the 
harmonic degree $\lc$ of the fastest 
growing  mode which corresponds to $\lc = \argmax_\ell \tau_\ell$ for  any 
given setup of the numerical dataset.

Figure~\ref{fig:threshold}(\textit{a}) shows $\lh$ as a function of $\lc$. To 
first order, $\lh$ grows almost linearly with $\lc$ and 
the proportionality coefficient linking the two harmonic degrees seems to
weakly depend on the input parameters. We find that $\lh$ is 
consistently greater than $\lc$. The linear fit $\lh = 1.89\,\lc$
is shown as a dashed line in Fig.~\ref{fig:threshold}(\textit{a}), and is 
overall in agreement with the simulations. Significant departures occur for  
$\lc < 30$, as the least turbulent simulations tend towards verifying $\lh = 
\lc$ (see e.g. the pentagon at the bottom left of 
Fig.~\ref{fig:threshold}\textit{a}). Far from onset, a 
large number of modes are excited and the broadening of the spectra
noticeable in Fig.~\ref{fig:2D}(\textit{d}) reflects the nonlinear interaction 
of theses modes.

The fastest growing modes can be expanded in powers of a supercriticality 
parameter $\epsilon$ which quantifies the distance to onset of fingering 
convection
\begin{equation}
 \epsilon = \dfrac{\rac-|\rat|}{|\rat|}=\dfrac{Le}{\Rp}-1\,.
 \label{eq:def_eps}
\end{equation}
For finite Prandtl numbers, the first order 
contribution reads \citep[e.g.][]{Schmitt1979,Stern1998,Radko2010}
\begin{equation}
 \lc^4 \sim 
\epsilon |\rat|,
\label{eq:onset_eps}
\end{equation}
in the limit of vanishing $\epsilon$.
Figure~\ref{fig:threshold}(\textit{b}) shows
the spherical harmonic degree of the fastest growing mode $\lc 
|\rat|^{-1/4}$ as a function of the distance to onset $\epsilon$. The best fit 
to the data reveals a good agreement with the expected theoretical scaling
whenever $\epsilon \ll 1$. Significant departure appear beyond $\epsilon 
\approx 0.5$ indicating the limit of validity of the scaling 
\eqref{eq:onset_eps}.

We hence retain from Fig.~\ref{fig:threshold} that the hypothesis put 
forward by \cite{Schmitt1979} --i.e. that the finger width relates to the 
horizontal size of the fastest growing mode-- better works for weakly nonlinear 
models, i.e. when $\epsilon \ll 1$. This observation will later be used to 
derive scaling laws for weakly nonlinear fingering convection.

\subsection{Scaling laws for transport}

We now aim at deriving integral scaling laws for the transport 
of composition and heat, and for the mean vertical velocity.

\begin{figure}
\includegraphics[width=\linewidth]{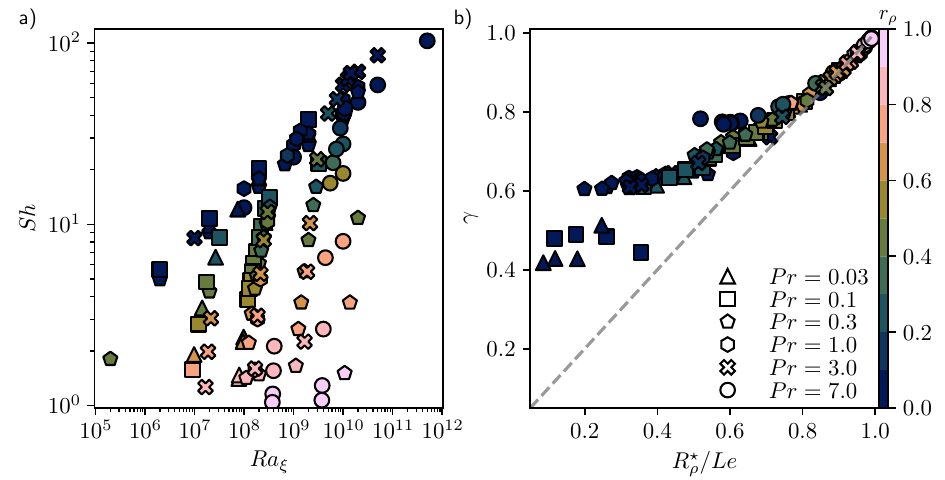}
\caption{(\textit{a}) Sherwood number $Sh$ as a function of $Ra_\xi$.   
(\textit{b}) Flux ratio $\gamma$ as a function of $\Rpe/Le$. The 
dashed line corresponds to the first bisector $\gamma=\Rpe/Le$.}
\label{fig:Sh_ga}
\end{figure}

Figure~\ref{fig:Sh_ga}(\textit{a}) shows $Sh$ as a function 
of $Ra_\xi$, and two trends emerge: first, at any given $\rac$,  numerical 
models with $\rrho < 0.2$ (dark blue symbols) provide 
an effective upper bound for the transport of chemical composition. This upper
bound of the Sherwood number $Sh$ grows with $\rac$ according to a power law 
that  appears almost independent of $Pr$. Second, close to onset ($\rrho \simeq 
1$), simulations are organised along branches representative 
of a dependency of $Sh$ on $\rac$ much steeper than the one of the effective upper bound. 
Each branch corresponds to fixed values of $\rat$, $Pr$, and $Le$, and 
gradually tapers off to the $\rrho \ll 1 $ trend as the value of $\rrho$ 
decreases along the branch. This prompts us to analyse the regimes $\rrho \simeq 
1$ and $\rrho \ll 1$ separately.
 
It is also informative to inspect the variations of the 
turbulent flux ratio $\gamma$, as defined by 
Eq.~\eqref{eq:def_gamma}, in both limits. Figure~\ref{fig:Sh_ga}(\textit{b}) 
shows $\gamma$ against $\Rpe/Le$.
When $\rrho \ll 1$, $\gamma$ substantially deviates from $\Rpe/Le$ and 
gradually saturates around values which decrease upon increasing the Lewis number:
simulations with $Le=3, Pr=7$ (circles) saturate at $\gamma \approx
0.8$, while those with $Pr\leq1$ (pentagons, hexagons, squares and triangles) 
gradually tapper off around values of $0.6$ and $0.4$ for $Le=10$ and $Le 
\ge30$, respectively. The series of simulations
with $Pr =7$ (circles) seems to suggest that the $\gamma \approx 0.8$ 
 plateau may actually precede an increase
at even lower values of $\Rpe/Le$, a trend predicted by 
\citet{Kunze1987} in the limit of $Pr, Le \gg 1$. 
In contrast, close to onset,  the flux ratio
is well described by $\Rpe/Le$, 
\begin{equation}
  \gamma \approx \dfrac{\Rpe}{Le}, 
  \label{eq:gamma_Rpe}
\end{equation}
a behaviour consistent with the arguments put forward by 
\cite{Schmitt1979}.

\subsubsection{The $\rrho \simeq 1$ regime}
\label{sec:wnl_scalings}

We shall start our analysis by investigating the weakly-nonlinear regime 
characterized by small values of the supercriticality parameter $\epsilon$ 
defined above in Eq.~\eqref{eq:def_eps}. In the limit where $\gamma \approx 
\Rpe/Le$, the scaling law obtained for the finger width 
\eqref{eq:Lh_tot} can be rewritten by breaking down the mean radial profiles 
into a sum of the conducting state and fluctuations. This yields
\[
 \Lh^{-4} \sim 
|\rat|\left[\dfrac{Le}{\Rp}\left(1+\dfrac{\dXib'}
{\lambda_b }\right)-1-\dfrac{\dTb'}{\lambda_b}\right]\,,
\]
where we have 
 assumed that $|\mathrm{d}\xi_c /\mathrm{d} r| = 
\mathrm{d} T_c /\mathrm{d} r  \approx 1$. Now, based on our previous findings 
that the finger width is well accounted for by the horizontal size of the 
fastest growing mode whenever $\epsilon \ll 1$ (recall 
Fig.~\ref{fig:threshold}), we have
\[
|\rat|\left[\dfrac{Le}{\Rp}\left(1+\dfrac{\dXib'}
{\lambda_b }\right)-1-\dfrac{\dTb'}{\lambda_b}\right]
\sim |\rat|\left(\dfrac{Le}{\Rp}-1\right)\,,
\]
which leads to
\begin{equation}
 \dfrac{\dTb'}{\lambda_b} \sim \dfrac{\epsilon}{Le^2 \gamma^{-1}-1}\,,
 \label{eq:dTb_prime}
\end{equation}
in which the ratio of $\dXib'/\dTb'$ has been estimated using 
Eq.~\eqref{eq:fraction_fluct}.

Making use of the definitions \eqref{eq:lambdab} then allows 
to relate the ratio $\dTb'/\lambda_b$ to its boundary layer counterpart, namely
\begin{equation}
 \dfrac{\dTb'}{\lambda_b} = 
\dfrac{\dTo'}{\lambda_o}\dfrac{\lambda_o+\lambda_i(\dTi'/\dTo')}{1-\lambda_i-
\lambda_o } \approx \dfrac{\dTo'}{\lambda_o} \lambda_o[1+(r_i/r_o)^{-3/2}]\,,
\end{equation}
where the latter equality has been derived using our previous findings 
regarding the boundary layer asymmetry (Eqs.~\ref{eq:dLi}-\ref{eq:dXi}).
Noting that $Nu-1\approx (r_o/r_i) \dTo'/\lambda_o$ and substituting the above 
expression 
in Eq.~\eqref{eq:dTb_prime} yields
\begin{equation}
 Nu - 1 \sim f(r_i/r_o)\lambda_o^{-1} \dfrac{\epsilon}{Le^2\gamma^{-1}-1},
 \label{eq:radko2000}
\end{equation}
where $f(r_i/r_o) = (r_i/r_o)^{-1}[1+(r_i/r_o)^{-3/2}]^{-1}$ accounts for 
dependence to the radius ratio $r_i/r_o$ inherent in 
spherical shell geometry. Besides these geometrical factors,
this expression is strictly equivalent to Eq.~(5.10) derived by 
\cite{Radko2000} in bounded Cartesian domains under the assumption of tall 
laminar fingers in the fluid bulk.

The end of the derivation however differs from \citet{Radko2000}, in that 
they consider a boundary layer model where the buoyancy term is retained, while 
we have shown 
earlier that the compositional boundary layers rather adhere to the 
Prandtl-Blasius model (see Eq.~\ref{eq:scaling_BL} and 
Fig.~\ref{fig:BL_Nu_Sh}).

To eliminate the P\'eclet number $Pe$ in Eq.~\eqref{eq:scaling_BL}, one 
can simply assume that viscous dissipation is well approximated by
$D_\nu \sim (\rep/\Lh)^2$. The power balance (\ref{eq:DeltaP}) (see also 
Appendix~\ref{sec:app1}) then directly yields
\begin{equation}
Pe \sim \Lh |\rat|^{1/2} Le (Nu -1)^{1/2} (\gamma^{-1}-1)^{1/2}\,.
\label{eq:Pe_gene}
\end{equation}

Hence combining the equations \eqref{eq:scaling_BL}, \eqref{eq:radko2000} and 
\eqref{eq:Pe_gene}, the scaling relation for $Nu-1$ reads:
\[
 Nu-1 \sim 
\dfrac{\epsilon^{6/5}|\rat|^{3/10}Le^{2/5}(\gamma^{-1}-1)^{3/10}}{
(\gamma^{-1}Le^2-1)^{6/5}}\,,
\]
where the order one geometrical factors have been omitted.
To end up with a scaling relation which solely depends on control quantities, 
one can further approximate the flux ratio by $\gamma \approx \Rp/Le$, or 
equivalently assume that $\gamma^{-1}-1 \approx \epsilon$. Noting that this 
latter assumption is more restrictive than Eq.~\eqref{eq:gamma_Rpe} and that 
$\Rp/Le=1-\epsilon$, the first order contribution leads to
\begin{equation}
 Nu-1 \sim \epsilon^{3/2}|\rat|^{3/10}Le^{-2}\,,
 \label{eq:nu_eps}
\end{equation}
while the corresponding scaling behaviour for $Sh-1$ then 
reads
\begin{equation}
 Sh-1\approx Le^2 (Nu-1) \sim \epsilon^{3/2}|\rat|^{3/10}\,.
 \label{eq:sh_eps}
\end{equation}
The scaling for the vertical velocity \eqref{eq:Pe_gene} 
then becomes
\begin{equation}
 Pe \sim \epsilon |\rat|^{2/5}\,.
 \label{eq:pe_eps}
\end{equation}

\begin{figure}
\centering
\includegraphics[width=\textwidth]{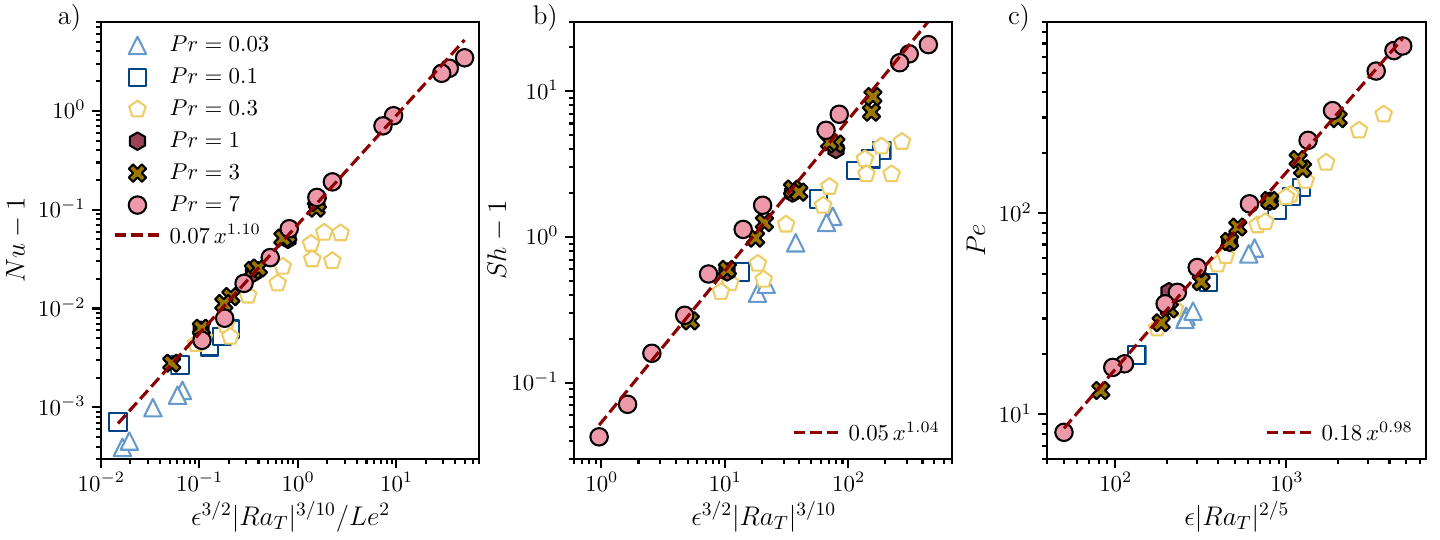}
\caption{(\textit{a}) Vertical convective transport $Nu-1$
as a function of the theoretical weakly-nonlinear scaling 
\eqref{eq:nu_eps} for all the simulations with $\epsilon < 1$.
(\textit{b}) Vertical convective transport of composition
as a function of the  scaling \eqref{eq:sh_eps}.
(\textit{c}) Convective velocity as a function of the scaling 
\eqref{eq:pe_eps}. The dashed lines correspond to the best fits obtained for 
the numerical simulations with $Pr \geq 1$ and $\epsilon < 0.5$.}
 \label{fig:transpi}
\end{figure}

Equations~(\ref{eq:nu_eps}-\ref{eq:pe_eps}) form the weakly nonlinear 
scaling behaviour for bounded fingering convection. The scaling exponents on 
the supercriticality parameter $\epsilon$ differ from those derived by 
\citet{Radko2000}, due to different boundary layer models. 
Figure~\ref{fig:transpi} shows an evaluation of these predictive scaling laws 
for the $47$ simulations with $\epsilon < 1$. Among those, the $25$
simulations with $Pr \geq 1$ are reasonably well accounted for by the weakly 
nonlinear laws. Best-fits using the $22$ simulations with $\epsilon < 
0.5$ yield scaling exponents of $\approx 1.1$  for the heat and composition 
transport, 
moderately larger than the expected slope of one, while the vertical velocity 
follows an exponent closer to unity. In contrast, the remaining $22$ 
simulations with $Pr < 1$
are more scattered, they significantly depart from the asymptotic laws and seem to 
demand an extra dependence on the Prandtl number. This is 
particularly obvious for the scaling of the Sherwood number shown in 
Fig.~\ref{fig:transpi}(\textit{b}).

The previous derivation rests on several assumptions that become
questionable in the limit of small Prandtl numbers. Among the most likely 
shortcomings, one can think of: (\textit{i}) the approximation of the 
flux ratio by $\gamma \approx \Rp/Le$ getting bolder on decreasing 
$Pr$; (\textit{ii}) the relation between the finger width and the size of the 
fastest growing mode breaking down at low $Pr$ or involving a more intricate 
dependence on $\epsilon$ \citep{Schmitt1979,Radko2010}; (\textit{iii}) inertia 
becoming 
sizeable thus invalidating the laminar tall finger hypothesis. All 
in all, these additional hurdles hamper the derivation of predictive scaling 
laws for bounded domains that would hold in the small Prandtl number limit.

For the purpose of comparison with the local unbounded computations by 
\citet{Brown2013} when $Pr < 1$, we nevertheless show in 
Appendix~\ref{sec:heuristic}
that adjusted diagnostics which incorporate the modification of the background 
profiles into account can be described by simple polynomials fits on 
$Pr$ and $\epsilon^\star/Le$, where $\epsilon^\star=Le/\Rpe-1$.

\subsubsection{The $\rrho \ll 1$ regime}
\label{sec:smallrrho}

Turning our attention to the second limit, we retain arbitrarily those 41
simulations with $\rrho < 0.1$, which cover all values 
of $Pr$ from $0.03$ to $7$.  

\begin{figure}
\centerline{\includegraphics[width=\textwidth]{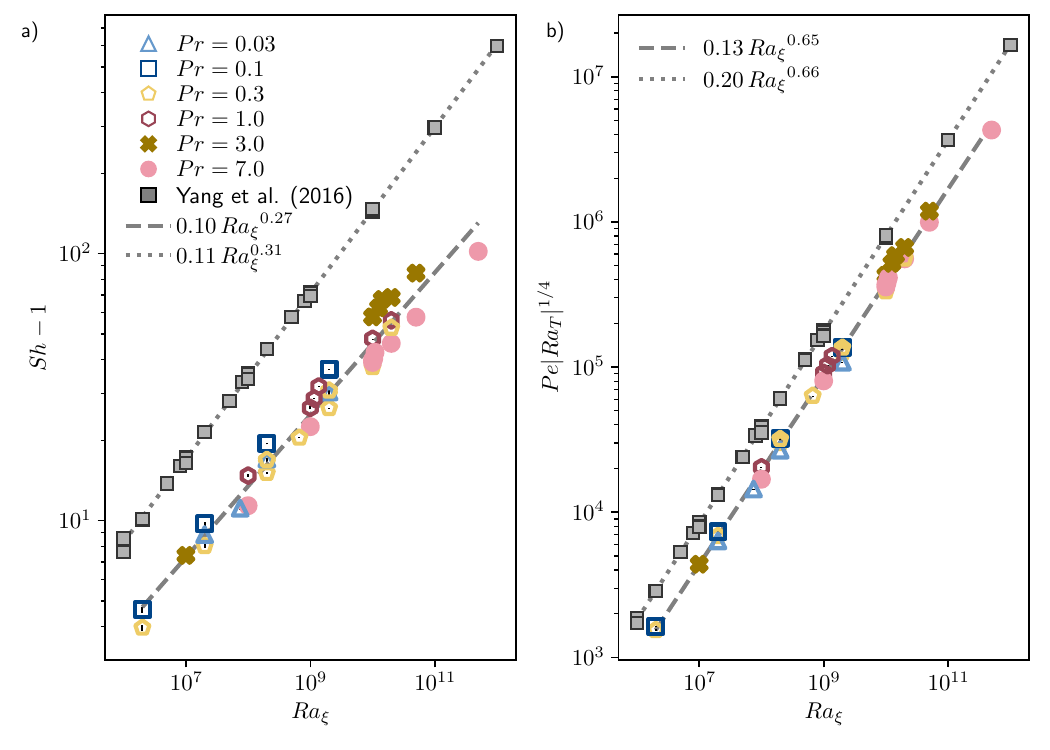}}
\caption{(\textit{a}) $Sh-1$ as a function of $\rac$ for our numerical 
simulations with $\rrho < 0.1$ alongside the data from \citet{Yang2016} that 
fulfill $\Rp > 1$. (\textit{b}) $Pe|Ra_T|^{1/4}$ as a function of $Ra_\xi$. In 
both panels, solid and dashed lines correspond to least-square fits to our 
data and to those of \citet{Yang2016}, respectively.}
\label{fig:lowrrho}
\end{figure}

Figure~\ref{fig:lowrrho}(\textit{a}) shows $Sh-1$ versus $\rac$ for these 
simulations, alongside the $31$ bounded Cartesian simulations of 
\citet{Yang2016} that satisfy $\rrho > 0$. 
A least-squares fit of $Sh-1$ as a function of $\rac$ yields 
$Sh-1 \sim \rac^{0.27}$ for our dataset and $Sh-1 \sim \rac^{0.31}$ for that of 
\citet{Yang2016}. 
The extension of the theory of \citet{Grossmann2000} for Rayleigh-B\'enard
convection to fingering convection by \citet{Yang2015} predicts $Sh \sim 
\rac^{1/3}$ in the $\rac \gg 1$ limit when dissipation occurs in the fluid bulk. 
However, for the range of $\rac$ covered in this study, 
a non-negligible fraction of dissipation is expected to happen within the 
boundary layers. In addition, our mixture of Prandtl numbers above and below 
unity may result in superimposed transport regimes. 
As such, the smaller value of the scaling exponent, as well as the 
larger spread of the data, compared with those of \citet{Yang2016} over a 
comparable range of $\rac$,  are not surprising: \citet{Yang2016} consider
a single combination of parameters $(Pr=7,Le=100)$ that makes it possible 
to reach values of $\rrho$ smaller than ours. 
We only retained the simulations by \citet{Yang2016} that satisfied the 
criterion $\rrho>0$, viz. $\Rp>1$; yet fingers in their case remain stable down 
to $\Rp \approx 0.1$, where the scaling $Sh-1 \sim \rac^{0.31}$  still holds 
(not shown). Finally, we note that the chemical transport for the $Pr < 1$ 
data shown in Fig.~\ref{fig:lowrrho}(\textit{a}) also follows the $\rac^{0.27}$ 
law, in stark contrast with the predictions of a constant $Sh$ for a fixed $(Le, Pr)$ pair 
derived by \citet{Brown2013} in unbounded planar models with $Pr\ll 1$.

Using Eq.~\eqref{eq:Pe_gene}, the definition of the flux ratio 
\eqref{eq:def_gamma} and the scaling for $Sh-1$ with $\rac$ we just 
discussed, and assuming that $[\gamma (1-\gamma)]^{1/2}$ can be considered 
constant, we expect
\begin{equation}
 Pe \sim \rac^{2/3}|\rat|^{-1/4}\,,
 \label{eq:Pe_ultimate}
\end{equation}
in the asymptotic $Sh \sim \rac^{1/3}$ 
regime. Figure~\ref{fig:lowrrho}(\textit{b}) shows $Pe  | \rat |^{1/4} $ as a 
function of $\rac$ for our dataset and
that of \citet{Yang2016}. Least-squares fits yield
scaling exponents that are remarkably close to $2/3$ for both subsets, and a 
spread of our data along the best-fit line less pronounced than in 
Fig.~\ref{fig:lowrrho}(\textit{a}). The difference in the prefactors of the 
best-fit lines describing our dataset and that of \cite{Yang2016} can be 
ascribed to the differences in model setups (Cartesian versus spherical 
geometry and constant gravity versus gravity increasing with $r$).

A few comments are in order  with regard to Eq.~\eqref{eq:Pe_ultimate}: 
operating at fixed $Le=100$, 
\citet{Yang2016} propose a  scaling  for the  vertical velocity in terms
of the Reynolds number $Re$, 
\[
 Re \propto \Rp^{-1/4}  \rac^{1/2} \propto   \rac^{3/4} | \rat |^{-1/4}    
Le^{-1/4},
\]
which, in light of their Fig.~4(b), does not yield as good an agreement
with their data than the scaling  proposed  here for $Pe$, and shown  
in Fig.~\ref{fig:lowrrho}(\textit{b}). Expressing  our proposed scaling in terms of 
$Re$ gives 
\begin{equation}
   Re \sim \rac^{2/3}    | \rat |^{-1/4} Sc^{-1}, 
\end{equation}
which exhibits a slightly smaller  dependency to $\rac$ than the one 
 put forward by  \citet{Yang2016}, and does a better job
of fitting their data (not shown). \citet{Yang2016}
acknowledged that additional 
dependency on $Pr$ and $Sc$ (or $Le$)  might occur since only one combination 
is considered in their study, in particular when discussing  the
scaling $Re \sim \rac | \rat |^{-1/2}$ proposed by 
\citet{Hage2010} based on their experimental data with $Pr \approx 9$ and 
$Sc\approx 2200$.
The exponents found by \citet{Hage2010} are markedly different
than the ones inferred from our analysis. It should be noted, 
however, that their experimental data cover a region of parameter space where the density 
ratio is mostly smaller than unity, in which fingers  can be  gradually 
replaced by large-scale convection. Under those  circumstances, the hypothesis 
that dissipation can be expressed by $D_\nu \sim (\rep/\Lh)^2$, with $\Lh$ the 
typical finger width, breaks down.

\section{Toroidal jets}
\label{sec:jets}

In a substantial subset of simulations, a secondary 
instability develops on top of the radially-oriented fingers, in the form of 
large-scale horizontal flows. Jets formation has been observed in 
two-dimensional unbounded simulations by \citet{Radko2010} and 
\citet{Garaud2015} for $Pr < 1$ fluids. More recently, \citet{Yang2016} 
reported the formation of alternating zonal jets in three-dimensional bounded 
geometry with $Pr=7$ and $Le=100$ when $Ra_\xi > 10^{10}$. The purpose of this 
section is to characterise the spatial and temporal distribution of jets 
forming in spherical shell fingering convection.

\label{sec:instabilite_2} 
\subsection{Flow morphology} 

\begin{figure}
\centering 
\includegraphics[width=\linewidth]{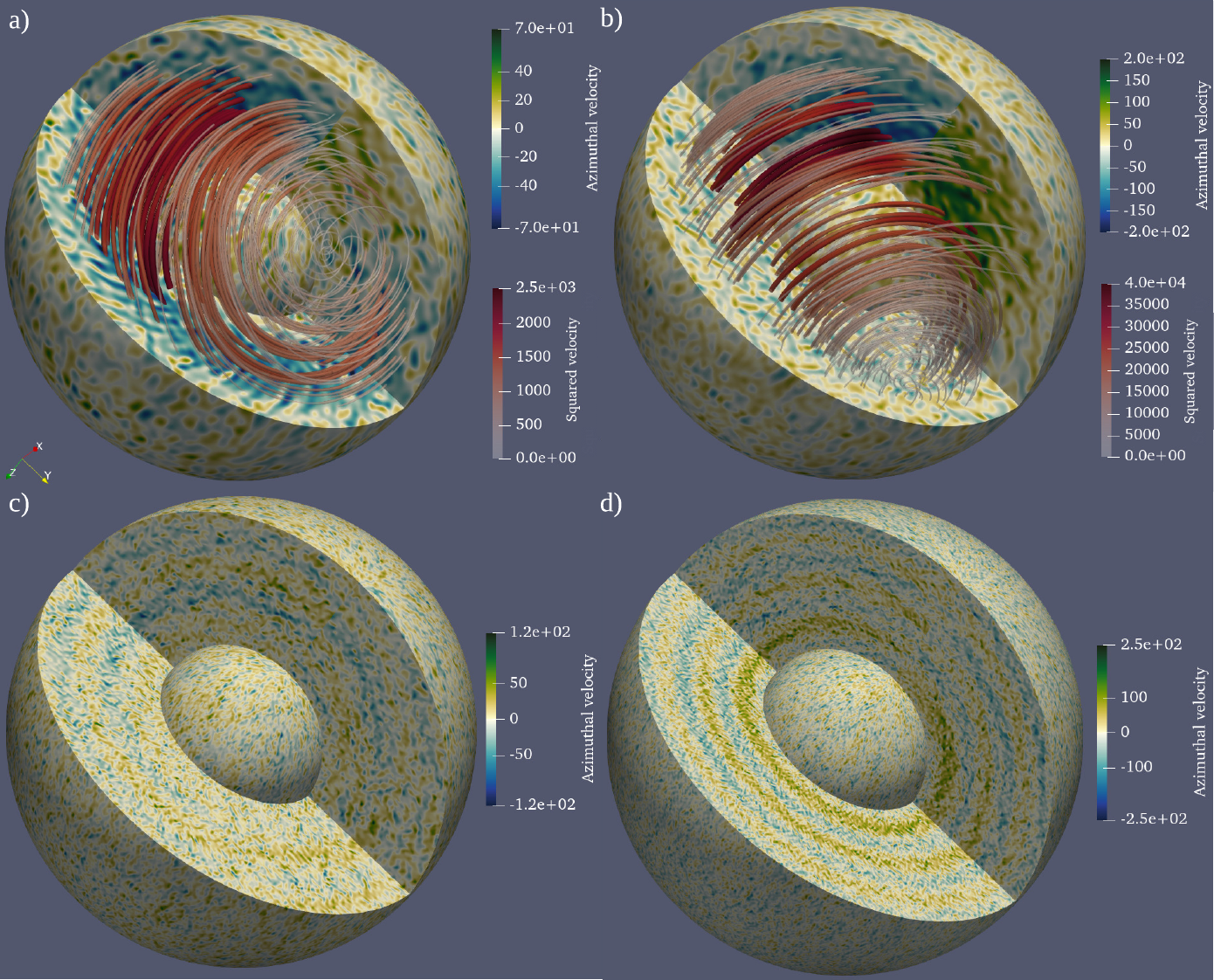}
\caption{Three-dimensional renderings of the instantaneous 
longitudinal velocity $u_\varphi$ for four simulations. 
The inner and outer spherical surfaces correspond to
$r_i  + 0.03$ and $r_o - 0.04$. (\textit{a}-\textit{b}) The two simulations 
share $Pr=0.3$, $Le=10$, and $\rat=-10^8$ (simulations 57 and 53 in 
Tab.~\ref{tab:simu_tab2}). $\Rp$ is equal to $6$ in (\textit{a})  and 
$4$ in (\textit{b}). Streamlines correspond to the large-scale component 
of the flow, truncated at a spherical harmonic degree $\ell = 7$. The width 
and the color of the streamlines vary according to the the local squared 
velocity. The two simulations on the bottom row 
(\textit{c}-\textit{d}) share $Pr=3$,  $Le=10$ and $\Rp=1.5$ (simulations 
84 and 80). The simulation 
shown in (\textit{c}) has $(\rat,\rac)=(-1.5 \times 10^9, 10^{10})$, while that 
shown in (\textit{d}) has $(\rat,\rac)=(-7.5 \times 10^9, 5 \times 10^{10})$.}
\label{fig:Vp_2D}
\end{figure}

Figure~\ref{fig:Vp_2D} shows three-dimensional renderings of snapshots of the 
azimuthal velocity for four selected numerical simulations. Upper panels 
(\textit{a}-\textit{b}) correspond to simulations which share the same 
parameters $Pr=0.3$, $|\rat|=10^8$, $Le=10$ and only differ in the values of the
density ratio $\Rp$, $\Rp=6$ in Fig.~\ref{fig:Vp_2D}(\textit{a}) and $\Rp=5$ in 
Fig.~\ref{fig:Vp_2D}(\textit{b}). To highlight the spatial distribution 
of the toroidal jet, Fig.~\ref{fig:Vp_2D}(\textit{a}-\textit{b}) also shows 
streamlines of the large-scale component of the flow truncated at spherical harmonic
degree $\ell =7$. In both simulations, the large-scale component of the flow takes the form 
of one single jet which reaches its maximum amplitude around mid-shell (red 
thick streamline tubes). We note in passing that 
the 22 simulations with $Pr<1$ of our dataset which develop toroidal jets systematically 
feature one singly-oriented jet that span most of the spherical shell volume.
The absence of background rotation precludes the existence of a preferential direction in the domain, 
and the axis of symmetry of the jet has the freedom to evolve
over time.
In panel (\textit{a}), the toroidal Reynolds number reaches $24$, comparable to 
the velocity of the fingers $\rep=41$.
The fivefold increase of the buoyancy power between 
Fig.~\ref{fig:Vp_2D}(\textit{a}) and Fig.~\ref{fig:Vp_2D}(\textit{b}) results in a stronger jet
which now reaches $\ret=87$, a value slightly larger than the finger velocity 
$\rep=77$. Interestingly, a further decrease of $\Rp$ to $3.25$ (not shown) 
results in the decrease of $\ret$ and the eventual demise of toroidal jets 
for lower $\Rp$. This is suggestive of a minimum threshold value of $\Rp$ 
favourable to trigger jet formation. 

The numerical models shown in the lower panels (\textit{c}) and (\textit{d}) of 
Fig.~\ref{fig:Vp_2D} share the same values of $Pr=3$, $Le=10$ and $\Rp=1.5$ but 
differ in their values of $\rac$ and $\rat$. In the case shown in 
Fig.~\ref{fig:Vp_2D}(\textit{c}) with $\rac=10^{10}$, faint multiple 
jets of alternated directions develop. 
Their amplitudes remain however weak compared to the finger 
velocity with $\ret=14$ and $\rep=72$. As can be seen in the equatorial plane 
(colatitude $\theta=\pi/2$), 
jets do not exhibit a perfectly coherent concentric nature over the entire 
fluid volume but rather adopt a spiralling structure with significant amplitude 
variations. As shown in Fig.~\ref{fig:Vp_2D}(\textit{d}), an increase of the 
convective driving to $\rac = 5 \times 10^{10}$ goes along with the formation 
of a stack of $6$ alternated jets. Though their amplitude remains smaller than 
the fingering velocity ($\ret=45$ and $\rep=134$), toroidal jets now adopt a 
quasi-concentric structure with well-defined boundaries. This latter 
configuration is reminiscent to the simulations of \citet{Yang2016} who also 
report the formation of multiple jets in local Cartesian numerical models with 
$Pr =7$ and $\Rp=1.6$ when $\rac \geq 10^{11}$. 
Similarly to 
\citet{Yang2016}, we also observe that toroidal jets develop in configurations 
with $Pr \geq 1$ for small values of $\Rp$ and 
sufficiently large values of $\rac$. Critical values of those parameters 
required to trigger jet formation are further discussed below.

\subsection{Time evolution}
\label{sec:dynamique}

\begin{figure}
\centering
\includegraphics[width=\linewidth]{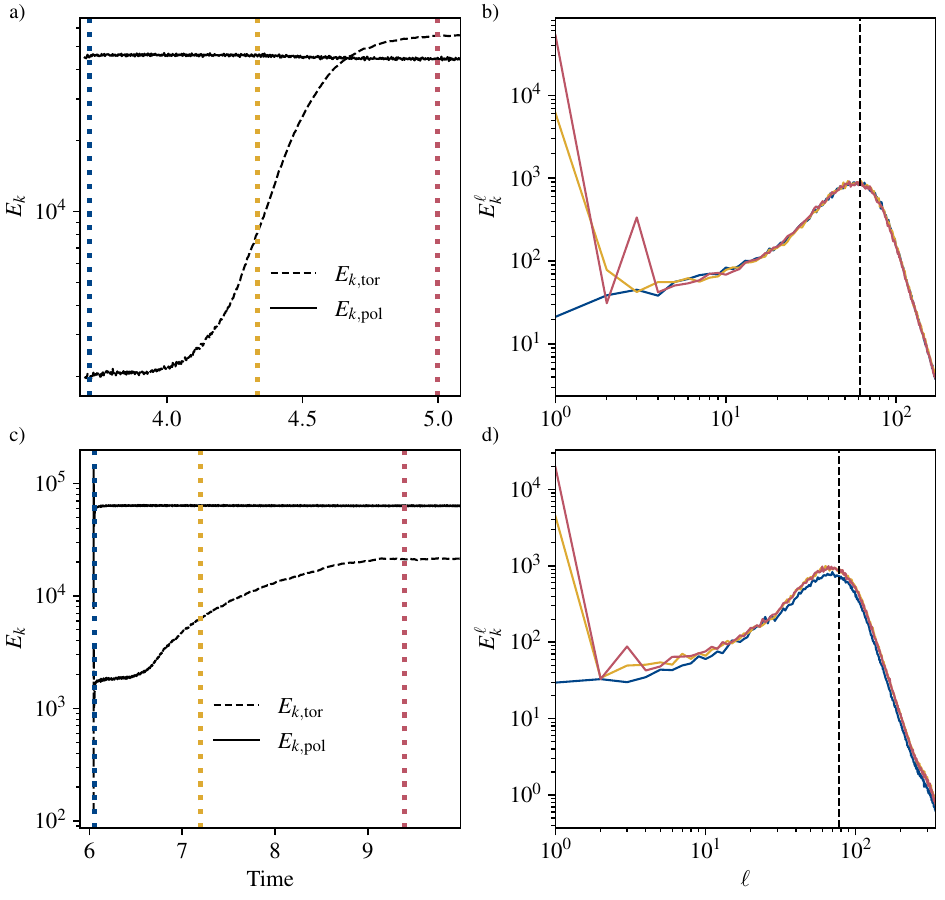}
\caption{%
Left panels: time evolution of the poloidal (toroidal) kinetic 
energy $\ekpol$ ($\ektor$). Right panels: kinetic energy spectra as a function 
of the spherical harmonic degree $\ell$ shown at three distinct times which 
correspond to the vertical dotted lines in the left panels. Panels 
(\textit{a}) and (\textit{b}) correspond to a simulation with $\rat = -10^8$, 
$Pr = 0.3$, $Le = 10$, and $\Rp = 4$ (simulation 53 in 
Tab.~\ref{tab:simu_tab2}), while panels (\textit{c}) and 
(\textit{d}) correspond to a simulation with $\rat = -1.5\times 10^8$, $Pr =
1$, $Le = 10$, and $\Rp = 1.3$ (simulation 72).  In panels (\textit{b}) and 
(\textit{d}), the vertical lines correspond to  $\ell_h$.}
\label{fig:ekin_spec}
\end{figure} 

To illustrate the growth of toroidal jets, we show in
Fig.~\ref{fig:ekin_spec} 
the time evolution of the poloidal and toroidal kinetic energy alongside 
kinetic energy spectra for two illustrative numerical simulations
which feature jets with $\rat =
-10^8$, $Pr = 0.3$, $Le = 10$ and $\Rp = 4$ (upper panels, simulation 53 
in Tab.~\ref{tab:simu_tab2}) and $\rat = -1.5 
\times 10^8$, $Pr = 1$, $Le = 10$ and $\Rp = 1.3$ (lower panels, 
simulation 72).
For the case with $Pr=0.3$ (Fig.~\ref{fig:ekin_spec}\textit{a}), $\ektor$ is 
initially $25$ times weaker than $\ekpol$. Beyond $t\approx4.2$, $\ektor$ grows 
exponentially over approximately one viscous diffusion time and saturates at a 
value which exceeds $\ekpol$. In contrast, the case with $Pr=1$ 
(Fig.~\ref{fig:ekin_spec}\textit{c}) exhibits a much slower growth of the 
toroidal energy: $\ektor$ gains one order of magnitude in more than $3$ viscous 
diffusion times. At the saturation of the instability around $t\approx 9$, 
$\ektor$ remains a factor $3$ smaller than $\ekpol$ in this case.
The growth of the toroidal energy goes along with the formation of one or 
several large scale jets (see Fig.~\ref{fig:Vp_2D}), which are clearly visible 
in kinetic energy spectra.  
Figure~\ref{fig:ekin_spec}(\textit{b}-\textit{d}) show the kinetic energy 
spectra as a function of the spherical harmonic degree at three different 
times: before the start of the instability (blue lines), during the 
exponential growth of the jets (yellow lines) and at the saturation of the 
instability (red lines). In both cases, the initial spectral distribution of 
energy are typical of fingering convection  with a well-defined maximum around 
the 
average spherical harmonic degree $\ell_h$ which corresponds to the mean horizontal size of 
the fingers (recall Fig.~\ref{fig:2D}\textit{d}). The growth of $\ektor$ 
manifests itself by an increase of several orders of magnitude of the energy at 
the largest scale $\ell=1$. In the saturated state, the kinetic energy spectra 
now reach their maxima at $\ell=1$ and feature a secondary peak of smaller 
amplitude at $\ell=3$. The toroidal energy at degree $\ell=1$ 
$\ektor^{1}$ is hence a good measure of the energy contained in the jets.
Beyond $\ell \gtrsim 6$, the spectra remain quite 
similar to their distribution prior to the onset of jet formation. This indicates a limited feedback of the 
growth of the jets on the horizontal size of the fingers.

\begin{figure}
\centering \includegraphics[width=\linewidth]{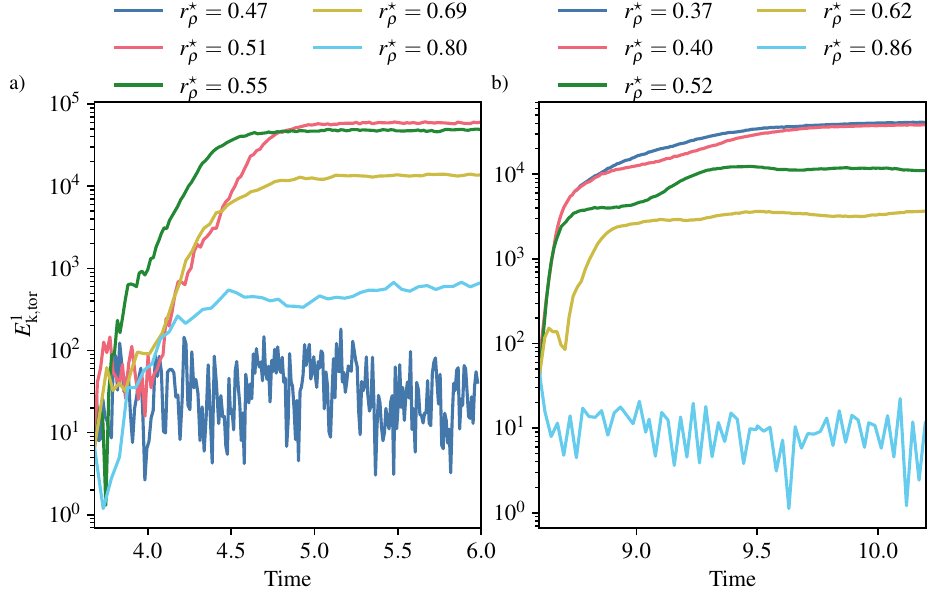}
\caption{%
(\textit{a}) Time evolution of the toroidal kinetic energy 
contained in the $\ell=1$ harmonic degree at a given 
radius $r$ for simulations with 
$Ra_T = -10^8, Pr = 0.3$, $Le = 10$, and a varying 
$\rpe$ (simulations 50, 52, 53, 56 and 58 in 
Tab.~\ref{tab:simu_tab2}). For all of them but one, $r = 0.793\,r_o$. The 
simulation 
with $\rpe=0.40$ is shown for $r=0.858\,ro$. (\textit{b}) Same for simulations 
with $Ra_T = -3.66 \times 10^9, Pr = 7$, $Le = 3$ 
(simulations 106, 107, 109, 110 and 112) and  $r = 0.785\,r_o$.}
\label{fig:ektor_VS}
\end{figure} 

To further characterise the physical parameters propitious to the formation of 
toroidal jets, Fig.~\ref{fig:ektor_VS} shows the time evolution of the toroidal 
kinetic energy contained in the 
$\ell=1$ degree at a given radius around $\approx 0.8 \,r_o$  for two series of 
simulations. Figure~\ref{fig:ektor_VS}(\textit{a}) shows simulations with $Ra_T 
= -10^8, Pr = 0.3$, $Le = 10$ and increasing $\rpe$. For the case with the 
smallest $\rpe=0.47$, jets do not form since the toroidal energy at $\ell=1$ 
oscillates but does not grow over time. For $\rpe \geq 0.5$, $\ektor^1$ 
grows exponentially over one viscous diffusion time before reaching saturation. 
The amplitude of $\ektor^1$ reaches its maximum for the cases with $\rpe\approx 
0.5-0.55$ and then decreases for larger values. The growth rate of the 
instability 
remains however markedly similar over the range of $\rpe$ considered here. 
The simulation with $\rpe=0.80$ is the last one of the series that features 
jets. For the models with $Pr=0.3$, jets hence only develop on a bounded 
interval of $\rpe$.
Figure~\ref{fig:ektor_VS}(\textit{b}) shows simulations with $Ra_T = -3.66 
\times 10^9$, $Pr = 7$ and $Le=3$. All the numerical models with $\rpe < 0.86$ 
present an exponential growth of $\ektor^1$ with once again comparable growth 
rates. 
In contrast to the $Pr < 1$ cases, jets appear 
to form below a threshold value of $\rpe$, while displaying a monotonic
trend: the lower $\rpe$ below the threshold,  the stronger the jets.

\begin{figure}
\centering \includegraphics[width=.8\linewidth]{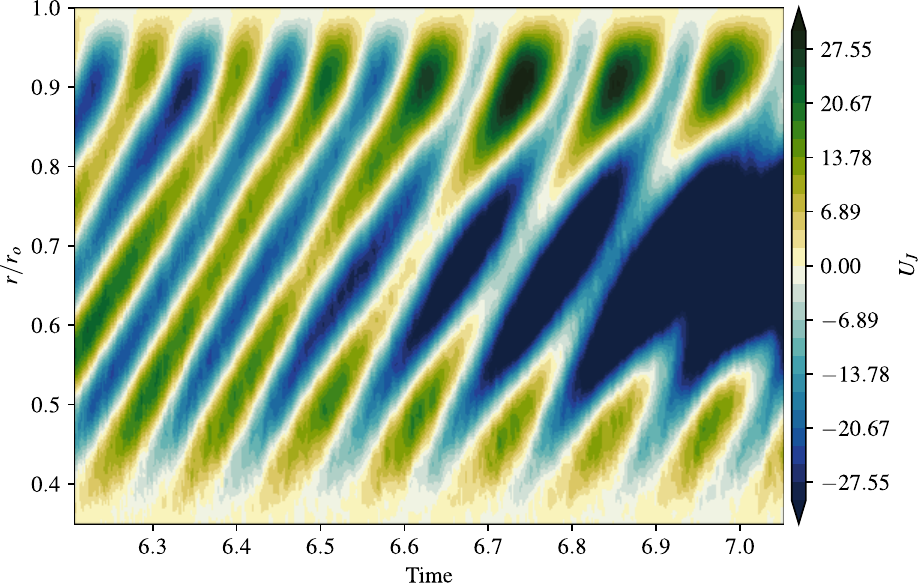}
\caption{%
Longitudinal average of the  longitudinal velocity 
 in the equatorial plane (colatitude $\theta=\pi/2$), $U_J$, 
as a function of radius $r$ and time $t$. Control parameters
for this simulation are $Pr = 3$, $Le = 10$, $\rac = 2 \times 10^{10}$, and 
$\Rp = 1.5$ (simulation 81 in Tab.~\ref{tab:simu_tab2}).}
\label{fig:hammer}
\end{figure} 

Once the toroidal jets have saturated, $Pr< 1$ and $Pr \geq 1$ simulations exhibit
distinct time evolutions: cases with $Pr < 1$ are dominated by one single jet 
with a well-defined rotational symmetry with no preferred axis
 (see Fig.~\ref{fig:Vp_2D}\textit{a},\textit{b}), while $Pr \geq 1$ 
cases usually feature a more complex stack of multiple alternated jets with 
different axes of symmetry. The latter are also more prone to time variations 
than the former. To illustrate this phenomenon, Fig.~\ref{fig:hammer} shows the 
time evolution of the longitudinal average of $u_\phi$ in the equatorial plane 
for a simulation with $Pr=3$, $\Rp=1.5$, $Le=10$ and $\rac=2\times 10^{10}$.
For this simulation, the $\ell=1$ toroidal energy is mostly axisymmetric, 
and the inspection of the azimuthally-averaged velocity thus provides a good 
insight into jet dynamics. The zonal flow pattern in the first half of the 
time series consists of three pairs of alternated jets. Jets are nucleated 
in the vicinity of the bottom boundary before slowly drifting outwards with a 
constant speed until reaching the outer boundary after $\approx 0.3$ viscous 
diffusion time. In between, another jet carrying the opposite direction 
has emerged forming a quasi-periodic behaviour. This oscillatory phenomenon is 
gradually interrupted beyond $t\approx6.6$. The mid-shell westward jet then
strengthens and widens, while the surrounding eastward jets vanish. The 
multiple drifting jets therefore transit to a single jet configuration within less 
than one viscous diffusion time. The long-term stability of the multiple jets 
configuration when $Pr \geq 1$ is hence in question.
Since the jets merging seems to occur on timescales commensurate to the 
viscous timescale, a systematic survey of the stability of the multiple jets 
configuration is numerically daunting. 
We 
decided to instead focus on a selected subset of multiple jets simulations with 
$Pr\geq 1$ which were 
integrated longer to examine the merging phenomenon. 
As stated above,  the $5$ multiple jets cases whose integration is too short to 
assess a possible merging feature an additional  ``NS''  in the last 
column of Table~\ref{tab:simu_tab2}. It is however striking to note that all 
the simulations that have been pursued longer eventually evolve into a single 
jet configuration; the merging time taking up to twice the viscous diffusion 
time. This phenomenon is reminiscent to the 2-D numerical models by 
\citet{Xie2019} who also report jets merging over time scales $4$ orders 
of magnitude larger than the thermal diffusion time at the scale of a finger.
For comparison purposes, the time integration of the case shown in 
Fig.~\ref{fig:hammer} corresponds to roughly $7000$ thermal diffusion times at 
the finger scale.

\subsection{Instability domain}

\begin{figure}
\centering
\includegraphics[width=\linewidth]{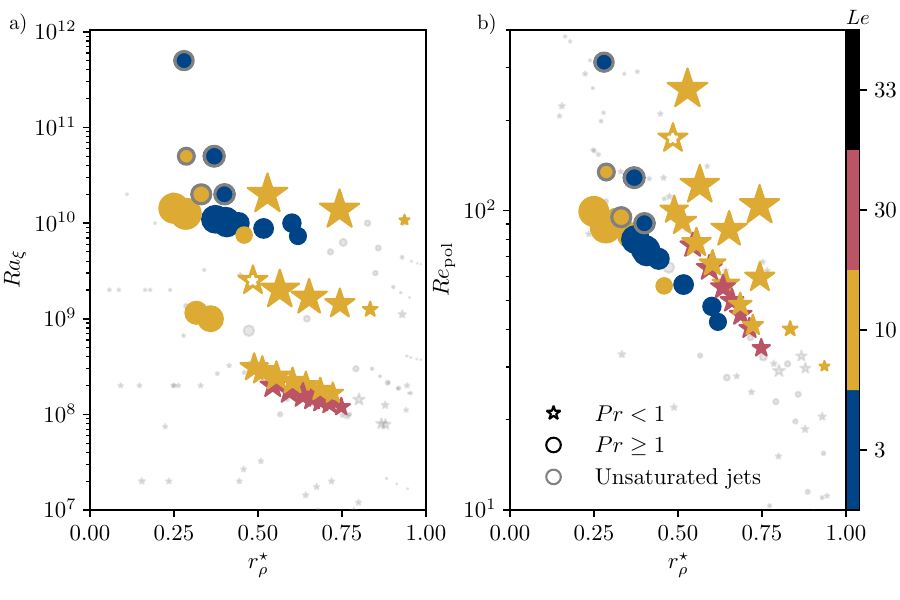}
\caption{(\textit{a}) Location of the $\nsims$~simulations computed in this 
study in the $(\rpe,\rac)$ plane. (\textit{b}) Same in the $(\rpe,\rep)$ plane. 
Stars and discs denote $Pr<1$ and $Pr\ge1$, respectively.  
Grey symbols correspond to models without jets. Conversely, 
symbols corresponding to simulations with single or multiple jets according to 
the criterion \eqref{eq:crit_jets} are colored according to the value of 
the Lewis number $Le$. Symbol size is proportional to $\ret^1/\rep$, where 
$\ret^1$ is the Reynolds number constructed from the spherical harmonic degree 
$\ell=1$ component of the toroidal flow. The $5$ simulations with jets whose 
symbol has a grey rim were not integrated for a long enough time 
to witness the possible merging of their multiple jets, which 
may lead to an underestimation of $\ret^1/\rep$. The white star with a yellow 
rim corresponds to the simulation further discussed in Fig.~\ref{fig:Rp4}.}
\label{fig:jet_space} 
\end{figure}

Jet formation systematically leads to the growth of the toroidal kinetic energy 
content at the largest scales (recall Fig.~\ref{fig:ekin_spec}). To 
distinguish jet-producing simulations from the others, 
we introduce the following empirical criterion 

\begin{equation}
    \zeta = \left(1 - \dfrac{\sum_{\ell=1}^{\ell = 5} E_{k, 
\mathrm{tor}}^\ell}
{\sum_{\ell=1}^{\ell = 11} E_{k, \mathrm{tor}}^\ell}\right)^{-1} > 20, 
\label{eq:crit_jets}
\end{equation}
whose purpose is to reveal the emergence of large-scale toroidal 
energy, for spherical harmonic degrees $\ell$ in the range $ \llbracket 1, 5 \rrbracket$, 
using a baseline defined by the spherical harmonic degrees $\ell$ in the range $ \llbracket 6, 11 \rrbracket$, 
whose level is immune to jet formation, see Fig.~\ref{fig:ekin_spec}(\textit{b})
and Fig.~\ref{fig:ekin_spec}(\textit{d}). The threshold of $20$ is arbitrary 
and enables a clear separation to be made between those simulations producing
jets and the others. 

Figure~\ref{fig:jet_space} shows the location of the $\nsims$ simulations in 
the $(\rpe,\rac)$ (left panel) and $(\rpe,\rep)$ (right panel) planes. In 
order to highlight the models which develop jets, the coloured symbols 
in Fig.~\ref{fig:jet_space} correspond to cases with toroidal jets and the 
symbol size scales with the relative energy content in $\ektor^1$. 
Configurations prone to jet formation are only observed beyond $\rac \approx 
10^8$ for $Pr < 1$ and beyond $\rac \approx 10^9$ for $Pr \geq 1$, which 
indicates that a minimum level of convective forcing is required to 
trigger the onset
of jet formation. This is in line with the findings of 
\citet{Yang2016} who also found jets forming beyond $\rac\geq 10^{11}$ in their 
models with $Pr=7$, $Le=100$ and $\Rp=1.6$.

For the numerical models with $Pr < 1$ (yellow and red stars), jets form over a 
bounded domain of $\rpe$. The instability domain 
grossly spans $\rpe \in [0.4, 0.8]$ but also features an additional 
dependence on $\rat$. The three quasi-horizontal branches of yellow stars 
correspond to numerical simulations with $|\rat|=10^8$, $10^9$ and $10^{10}$, 
respectively. For each value of $|\rat|$, the largest relative energy content 
in the large scale jets is attained close to the lower boundary of the 
instability domain. Increasing $|\rat|$ goes along with a gradual shift of the 
instability domain towards larger values of $\rpe$. Simulations with $Pr=0.1$ 
and $Le=30$ (red stars) feature a smaller instability domain than $Pr=0.3$, 
$Le=10$ (yellow stars). For the lowest $Pr=0.03$ configurations considered 
here, not a single model satisfies the criterion employed to detect jets.
We note, however, that two simulations with $Pr=0.03$ and $\rpe \approx 0.8$ 
feature a sizeable increase in 
$\ektor^{1}$, such that $\ektor^{1} \approx 10 \ektor^{2}$, 
 yet insufficient to fulfill the criterion. This indicates that the 
instability domain for jet formation 
 shrinks upon decreasing $Pr$, with a concomitant decrease of jet amplitude. 
 Jets are hence unlikely to form in the $Pr \ll 1$ 
regime \citep{Garaud2015}.

For the models with $Pr \ge 1$ (disks in Fig.~\ref{fig:jet_space}), jets 
develop for $\rac \geq 10^9$ for $Pr=3$ (yellow disks) and for $\rac \geq 
10^{10}$ for $Pr=7$ (blue disks) for a range of $\rpe$ which roughly spans $[0, 
0.5]$. At a given convective forcing $\rac$, the largest relative jet amplitude 
seems to correspond to the smallest values
of $\rpe$. Because of the slow merging of the multiple jets 
configuration when $Pr \ge 1$, their final amplitude
is however hard to assess for those 5 simulations 
that have not reached saturation and are 
represented 
by symbols with a grey rim in Fig.~\ref{fig:jet_space}. 

It is clear from Fig.~\ref{fig:jet_space}(\textit{a}) that the sole value of $\rac$ 
does not provide a reliable criterion for jet formation, since its critical 
value depends on both $Pr$ and $Le$. As an attempt to devise a more generic 
criterion, Fig.~\ref{fig:jet_space}(\textit{b}) shows the distribution 
of simulations in the $(\rpe,\rep)$ plane. Series of simulations with a fixed 
value of $|\rat|$ now define inclined branches, along which 
 the amplitude of $\rep$ increases upon decreasing $\rpe$. All 
the configurations prone to jet formation satisfy $\rep > 30$. This is of 
course not a sufficient condition, since, as discussed earlier, the $Pr 
\lesssim 1$ configurations also demand $\rpe\in[0.4, 0.8]$, while the $Pr \geq 
1$ require $\rpe < 0.5$ to develop jets. Overall, this stresses the need of a 
sufficiently vigorous background of fingering convection to trigger the 
secondary 
instability leading to jet formation.
 
 \begin{figure}
  \centering
  \includegraphics[width=\linewidth]{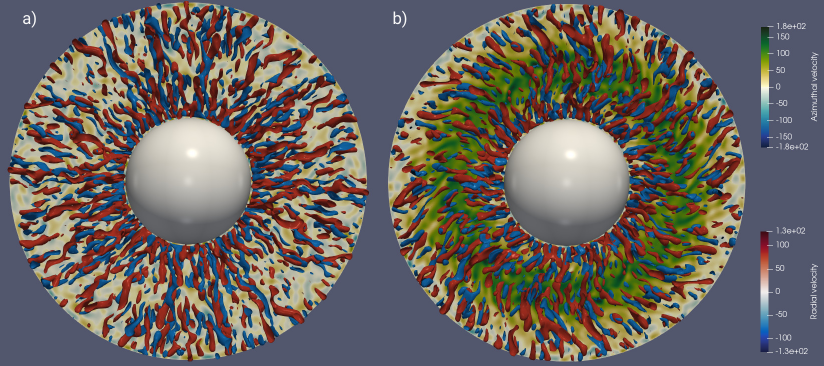}
  \caption{Renderings of the velocity for a numerical model with $Pr=0.3$, 
$Le=10$, $\rac=2\times 10^8$ and $\Rp=5$ (simulation 55 in 
Tab.~\ref{tab:simu_tab2}) at $t=3.98$ (\textit{a}) and at 
$t=5.68$ (\textit{b}). The equatorial plane shows the azimuthal component of 
the velocity $u_\phi$, while the red and blue surfaces correspond to the 
iso-levels of the radial velocity $u_r=\pm 80$ in the vicinity of the 
equatorial plane.}
\label{fig:tilting}
 \end{figure}

To better characterise the mechanism of jet formation, we now focus on one 
particular model with $Pr=0.3$, $Le=10$, $\rac=2\times 10^8$ and $\Rp=5$ where 
only one single jet develops (simulation 55 in 
Tab.~\ref{tab:simu_tab2}). To ease the analysis, the axis of symmetry of 
the jet has been enforced to perfectly align with the $z$-axis by imposing a 
twofold azimuthal symmetry. 
Figure~\ref{fig:tilting} shows two snapshots of the convective flow close to the 
equatorial plane prior to the jet formation and at the saturation of the 
instability. Before the jets start to grow (Fig.~\ref{fig:tilting}\textit{a}), 
fingers present an almost tubular structure. After almost two viscous 
diffusion times, a strong jet aligned with the $z$-axis (colatitude 
$\theta=0$) has developed and reaches a velocity which exceeds the radial flow 
(Fig.~\ref{fig:tilting}\textit{b}). Fingers have lost their vertical structure 
and are now distorted in the direction of the background shear. This is 
reminiscent to the analysis by \citet{Holyer1984} who showed that fingering 
convection is prone to develop secondary instabilities that can be either 
oscillatory or non-oscillatory. The latter takes the form of horizontal motions 
perpendicular to the axis of the fingers (see her Fig.~1).
This secondary instability shears the initially tubular fingers, 
while distorted fingers yield a correlation between the 
convective flow components that can in turn feed the shear by Reynolds stress 
\citep[see][\S~3.1]{Stern2005}.
The 2-D numerical models by \citet{Shen1995} 
with $Pr=7$, $Le=100$ and $R_\rho=2$  showed that this secondary 
instability saturates once the fingers are disrupted by shear.

\begin{figure}
\centering
\includegraphics[width=.8\linewidth]{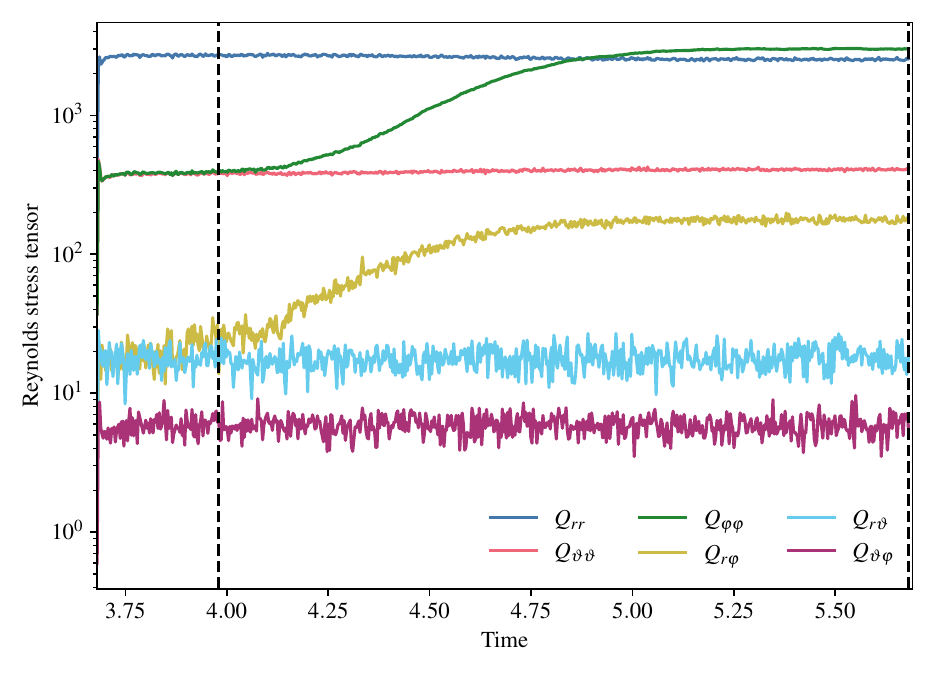}
\caption{%
Time evolution of the Reynolds stress tensor $Q_{ij}$ (Eq.~\ref{eq:qij}) for a 
simulation with $Pr = 0.3$, $Le = 10$, $\Rp = 
5$, and $\rac = 2 \times 10^8$ (simulation 55 in 
Tab.~\ref{tab:simu_tab2}), and subjected to the enforcement of a twofold 
azimuthal symmetry, such that the jet that forms has no component along
$\etheta$. The two vertical lines correspond to the snapshots shown in 
Fig.~\ref{fig:tilting}.}
\label{fig:reynolds_stress}
\end{figure}

In the case of $Pr < 1$ where only one single jet develops, it is always 
possible to define without loss of generality a local frame  $(\vec{e}_{r}, \vec{e}_\vartheta, 
\vec{e}_\varphi)$ in which the jets velocity can be expressed along 
$\vec{e}_\varphi$ only. The time and azimuthal average of the azimuthal 
component of the Navier-Stokes equations (\ref{eq:momentum}) expressed in this 
frame of reference then yields a balance between Reynolds and viscous stresses 
 given by

\begin{equation}
\dfrac{1}{r^2}\ddd{r}{}\left(r^2
\overline{\avg{u_r u_\varphi}{\varphi}}\right) + \dfrac{1}{r
\sin\vartheta}\ddd{\vartheta}{}\left(\sin\vartheta \overline{\avg{u_\vartheta
u_\varphi}{\varphi}} \right) = \nabla^2 \overline{U_J} -\dfrac{1}{r^2\sin^2\vartheta} \overline{U_J}
\label{eq:vzon}
\end{equation} 
where $\langle\cdots\rangle_\varphi$ denotes an azimuthal average and
$U_J={\avg{u_\varphi}{\varphi}}$ corresponds to the radial 
profile of the toroidal jets. Since the initial fingers are predominantly 
radial with $|u_r| \gg |u_\vartheta|$ and the jets are almost concentric 
with little $\vartheta$ dependence (see Fig.~\ref{fig:Vp_2D}), 
Eq.~\eqref{eq:vzon} can be simplified as follows

\begin{equation}
 \dfrac{1}{r^2}\ddd{r}{}\left(r^2
\overline{\avg{u_r u_\varphi}{\varphi}}\right) \sim \dfrac{1}{r^2} 
\dfrac{\partial}{\partial r}\left(r^2\dfrac{\partial \overline{U_J}}{\partial r}\right)\,.
\end{equation}
To examine this balance in more details, we compute the volume average of 
the Reynolds stress tensor via

\begin{equation}
 Q_{ij} = \dfrac{2\pi}{V}\int_{r_i}^{r_o} \left | \int_0^\pi \avg{u_i 
u_j}{\varphi} 
\sin\vartheta \mathrm{d}\vartheta \right|r^2  \mathrm{d}r,
\quad (i,j)\in\{r,\vartheta,\varphi\}^2\,.
\label{eq:qij}
\end{equation}
Figure~\ref{fig:reynolds_stress} shows the time evolution of the Reynolds 
stress tensor for the numerical model already shown in Fig.~\ref{fig:tilting}. 
Initially, the flow takes the form of elongated fingers with a strong radial 
coherence, $Q_{rr}$ being two orders of magnitude larger than 
$Q_{r\vartheta}$ and $Q_{r\varphi}$ and one order of magnitude larger than the 
horizontal components $Q_{\vartheta\vartheta}$ and $Q_{\varphi\varphi}$. When 
the secondary instability sets in around $t\approx 4$, horizontal jets develop 
and $Q_{\varphi\varphi}$ increases exponentially over one viscous diffusion time 
to finally exceed $Q_{rr}$ beyond $t\approx 5$. The correlation $Q_{r\phi}$ 
increases concomitantly, while the other Reynolds stress terms remain unchanged.
Similarly to the tilting instabilities observed in Rayleigh-Bénard convection 
\citep[e.g.][]{Goluskin2014}, the horizontal shear flow grows from the 
correlation between the radial and the azimuthal component of the background 
flow which here corresponds to the tilt of the fingers. Let us 
stress here that we observe that the same mechanism is at work for $Pr>1$ fluids, 
  leading to the
formation of multiple alternated jets, in agreement with the Cartesian analysis
of \citet[][her Fig.~1]{Holyer1984}. 


\begin{figure}
\centering
\includegraphics[width=\linewidth]{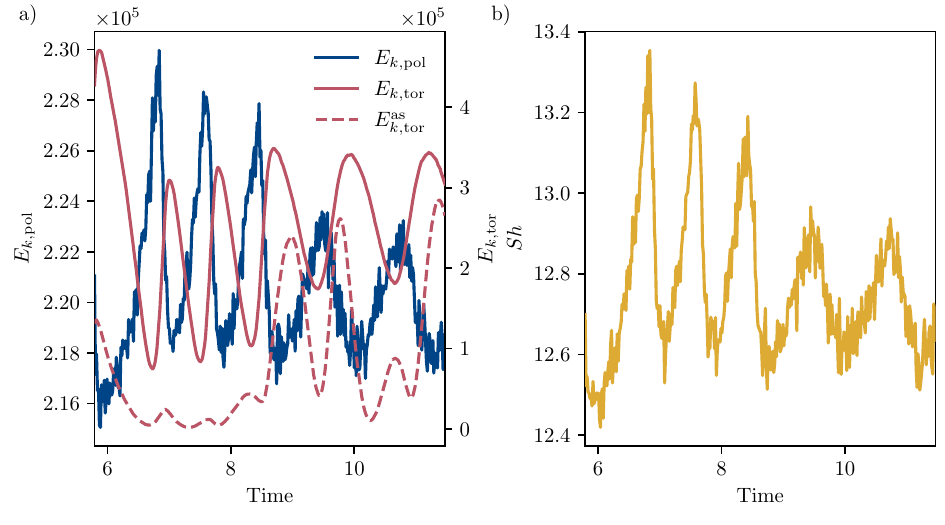}
\caption{(\textit{a}) Time evolution of the poloidal kinetic energy
(solid blue), the toroidal kinetic energy (solid red), and the axisymmetric 
toroidal energy (dashed red) for a simulation performed with $\rat = -10^9$, $Pr
= 0.3$, $Le = 10$, and  $\Rp = 4$ (simulation 39 in 
Tab.~\ref{tab:simu_tab2}). (\textit{b}) Time evolution of  $Sh$ for 
that simulation, that is represented with the white star with a yellow rim 
in Fig.~\ref{fig:jet_space}.}
\label{fig:Rp4}
\end{figure} 

Interestingly, close to the lower boundary of the instability domain for jets 
formation for $Pr < 1$ simulations (Fig.~\ref{fig:jet_space}), we found one 
numerical model (highlighted by a white star with a yellow rim) for which the 
interplay between shear and fingers via Reynolds stresses drives relaxation 
oscillations. Those are visible in Fig.~\ref{fig:Rp4} which shows the time 
evolution of the poloidal and toroidal kinetic energy (left panel) alongside 
that of the Sherwood number (right panel) for a numerical model 
with $\rat=-10^9$, $Pr=0.3$, $Le=10$ and $\Rp=4$.
It comes as no surprise that the temporal evolution of the poloidal energy 
$\ekpol$ is strongly correlated with that of the chemical transport $Sh$.
Relative changes of the two quantities are actually comparable, reaching 
approximately $5\%$ of their mean values.
This numerical model features a single jet, whose axis of symmetry changes over 
time. This manifests itself by the relative variations of the axisymmetric 
toroidal energy, which suddenly grows beyond $t\approx 9$ when the jet 
comes in better alignment with the $z$-axis. Toroidal and poloidal energy oscillate 
over time with a typical period close to the viscous diffusion time, 
a behaviour that resembles that 
of the 2-D models by \citet{Garaud2015}. Strong toroidal jets 
disrupt the fingers, hence reducing the efficiency of the radial flow and 
heat transport. This in turn is detrimental to feeding the jets via Reynolds 
stresses. The shear subsequently decays permitting a more efficient radial transport.
Repeating this cycle yields out-of-phase oscillations for $\ekpol$ and 
$\ektor$. In agreement with the findings by \citet{Xie2019}, relaxation 
oscillations disappear when increasing $\Rp$ while keeping all the other 
parameters constant. Conversely, we do not observe jets forming for lower 
values of $\Rp$. This indicates that relaxation oscillations likely
mark the boundary of the secondary instability domain.

\section{Summary and conclusion}
\label{sec:conclu}

We investigate the properties of fingering convection in a non-rotating 
spherical shell of radius ratio $r_i/r_o=0.35$ using  a catalogue of $\nsims$ 
three-dimensional simulations, one of our aims 
being to examine the extent to which predictions derived in local Cartesian 
domains would be adequate in a global spherical context. We focus on scaling 
laws for the transport of composition, heat, and momentum, and studied the 
possible occurrence of jet-forming secondary instabilities, over a broad range 
of Prandtl numbers, from $Pr = 0.03$ to $Pr=7$. 

In spherical shells, curvature and the linear increase of gravitational acceleration 
with radius yield asymmetric 
boundary layers. 
A dedicated analysis of the chemical boundary layers enables us to show
that (\textit{i}) the boundary layer asymmetry can be rationalised by 
assuming that the boundary layers are marginally-unstable; (\textit{ii}) the 
thickness of the boundary layers is well ascribed by the laminar 
Prandtl-Blasius model.

A single horizontal scale, defined in practice at mid-depth, suffices to
describe the lateral extent of fingers, which does not show substantial
variations with radius in the fluid bulk.
We show that the typical finger width
is controlled by a balance between 
buoyancy and viscous forces, and that its expression can be derived by making 
the classical ``tall finger'' assumption, whereby along-finger (radial) 
derivatives are neglected against cross-finger (horizontal) derivatives 
\citep[see e.g.][]{Taylor1989,Smyth2007}. 
 The excellent 
agreement between the prediction and the dataset seen in 
Fig.~\ref{fig:lPolPeak}(\textit{b}) stresses the crucial importance of the 
correcting factor involving the flux ratio $\gamma$, compared with the original 
$|\rat|^{-1/4}$ scaling proposed by \citet{Stern1960}, even though this 
correction implies the loss of predictivity.

\begin{table}
\caption{Comparison of the scaling laws reported in unbounded and bounded
Cartesian models with the ones obtained in this study. For an easier 
comparison with existing scaling laws, we define a Péclet number based on the 
finger width $Pe_{\Lh}=Pe\Lh$. The abbreviations B13
corresponds to \cite{Brown2013}, HT10 to \cite{Hage2010} RS00 to
\cite{Radko2000}, R10 to \cite{Radko2010}, RS12 to \cite{Radko2012}, S79 to
\cite{Schmitt1979}, T67 to \cite{Turner1967} and Y16 to \cite{Yang2016}. The 
$\approx$ signs correspond to numerical fits where no 
theoretical derivation could be provided. The question mark in the last row 
accounts for an uncertainty on the scaling for $\gamma$.}
\label{tab:scalings}
\begin{center}
\begin{tabular}{l l l l}
\hline
Quantity & Unbounded Cartesian & Bounded Cartesian & This study \\
\hline
\multicolumn{4}{c}{ Weakly nonlinear ($Pr > 1$)} \\
$\Lh$ & $\epsilon^{-1/4} |\rat|^{-1/4}$ (R10) & $\epsilon^{-1/4} |\rat|^{-1/4}$
(RS00) & $\epsilon^{-1/4} |\rat|^{-1/4}$ \\
$Nu-1$ & $\epsilon^2Le^{-2}$  (R10) &
$\epsilon^{13/10} |\rat|^{3/10}
Le^{-2}$ (RS00) &
$\epsilon^{3/2}|\rat|^{3/10}Le^{-2}$ \\
$Sh-1$ & $\epsilon^2$ (R10) & $f(\Rp,Le,Pr)\rac^{1/3}$ (T67) &
$\epsilon^{3/2}|\rat|^{3/10}$ \\
$Pe_{\Lh}$ & $\epsilon$ (R10)  & $\epsilon^{3/4}$ (RS00)
& $\epsilon |\rat|^{3/20}$ \\
$Pe$ & $\epsilon^{5/4}|\rat|^{1/4}$ (R10)  & $\epsilon |\rat|^{1/4}$ (RS00) & $ 
\epsilon |\rat|^{2/5}$ \\
$\gamma$ & $\Rp/Le$ (R10) & $\Rp/Le$ (S79) & $\Rp/Le$ \\
\hline
\multicolumn{4}{c}{ Weakly nonlinear ($Pr < 1$)} \\
$\Lh$ & $\epsilon^{-1/4} |\rat|^{-1/4}$ (B13) & \hfill - \hfill &
$[(\gamma^{-1}-1)|\rat|]^{-1/4}$ \\
$Nu^\star-1$ & $\epsilon^2Le^{-2}$ (B13) & \hfill - \hfill &
$\approx(\epsilon^\star Le^{-1} Pr^{1/2})^{1.4}$ \\
$Sh^\star-1$ & $\epsilon^2$ (B13) & \hfill - \hfill &
$\approx(\epsilon^\star Le^{-1} Pr^{1/2})^{1.4}Le^2$  \\
$Pe_{\Lh}$ & $\epsilon$ (B13) & \hfill - \hfill & $Pe\Lh$ \\
$Pe|\rat|^{-1/4}$ & $\epsilon^{5/4}$ (B13) & \hfill - \hfill &
$\approx {\epsilon^\star}^{0.95}Pr^{0.17}Le^{0.30}$
\\
$\gamma$ & $\Rp/Le$ (B13) & \hfill - \hfill & $\Rpe/Le$ \\
\hline
\multicolumn{4}{c}{ Strongly driven, $0 < \rrho \ll 1$} \\
$\lambda_\xi$  & Not relevant &
$\left\lbrace\begin{array}{l}
  Pe^{-1/3}\Lh^{2/3}\ \text{(Y16)} \\
  Pe^{-1/2}\Lh^{1/2}\ \text{(HT10)} 
 \end{array}\right.$
& $Pe^{-1/3}\Lh^{2/3}$ \\
$\Lh$ & $f(Le,Pr)|\rat|^{-1/4}$ (B13) & $\left\lbrace\begin{array}{l}
            \rac^{-1/4}\Rp^{-1/8}\ \text{(Y16)} \\
            |\rat|^{-1/3}\rac^{1/9}\ \text{(HT10)}
           \end{array}\right.$
& $[(\gamma^{-1}-1)|\rat|]^{-1/4}$ \\
 $Sh-1$ &
$\left\lbrace\begin{array}{l}
  f(Pr,Le)\ \text{(B13)} \\
 (\Rp-1)^{-1/2}\ \text{(RS12)}
 \end{array}\right.$
& $\left\lbrace\begin{array}{l}
  \rac^{1/3}\ \text{(Y16)} \\
  f(\Rp,Le,Pr)\rac^{1/3}\ \text{(T67)} \\
  \rac^{4/9}|\rat|^{-1/12}\ \text{(HT10)}
\end{array}\right.$ & $\rac^{1/3}$ \\
$Pe$ &
$\begin{array}{l}
  Le Pr^{1/2} \Lh^{-1} \ \text{(B13)} \\
 \end{array}$
& $\left\lbrace\begin{array}{l}
           \rac^{3/4}|\rat|^{-1/4}\frac{Sc}{Le^{1/4}}\ \text{(Y16)} \\
           \rac |\rat|^{-1/2}Sc\ \text{(HT10)}
          \end{array}\right.$ & $\rac^{2/3} |\rat|^{-1/4}$ \\
$\gamma$ & $\left\lbrace\begin{array}{l}
 \Rp(1+LePr)^{1/2}\ \text{(B13)} \\
 \approx\alpha_1e^{-\alpha_2\Rp}+\alpha_3\ \text{(RS12)}
 \end{array}\right.$ & $\approx f(\rac)\Rp^{-1/4} $ (Y16) & $f(Pr,Le)$?
\\
\end{tabular}
\end{center}
\end{table}

While the law expressing the finger width is adequate for all cases, 
establishing scaling laws for transport requires to distinguish between cases 
close to onset and cases that were strongly-driven. 
To stress the differences and similarities with planar 
models, Table~\ref{tab:scalings} summarises all the scaling laws established 
in \S~\ref{sec:fingers} alongside those reported in unbounded and bounded 
Cartesian models.

For the weakly nonlinear scalings corresponding to the two first sections 
of Tab.~\ref{tab:scalings}, we introduce a small parameter 
$\epsilon=\Rp/Le-1$ which measures the distance to onset.
We provide novel predictive theoretical scaling laws of the form 
$\epsilon^{\alpha_1}|\rat|^{\alpha_2}Le^{\alpha_3}$ for transport in 
bounded domains
by making use of two cornerstone assumptions: (\textit{i}) the 
finger width relates to the width of the fastest growing 
mode when $\epsilon \ll 1$ \citep{Schmitt1979}; (\textit{ii}) compositional 
boundary layers are well-described by the Prandtl-Blasius model. 
Our scaling laws differ --unsurprisingly-- from the theoretical 
derivations carried out in unbounded models but also from 
the ones derived in bounded models \citep{Radko2000}. This latter discrepancy 
originates from the different assumption retained to model the compositional 
boundary layer.
Our simulations with $Pr > 1$ and $\epsilon < 1$ are found to nicely adhere to 
these new theoretical scalings, while $Pr < 1$ models show less of an agreement.
For the latter, several assumptions entering the derivation, such as neglecting 
the effect of shear and inertia, becomes likely dubious on decreasing $Pr$.
We nonetheless show that adjusted diagnostics which incorporate the 
modification of the background profiles into account for the $Pr < 1$ spherical 
shell data favourably compare with the local unbounded models by 
\citet{Brown2013} and can be described by simple polynomial fits 
provided in the middle section of Tab.~\ref{tab:scalings}.
While those  numerical fits merely provide a heuristic description of the 
scaling behaviour of the transport of heat and chemical composition, they have 
the merit to better describing the $Pr < 1$ data and could prove beneficial for 
future studies.

For strongly-driven cases, corresponding to the last section of 
Tab.~\ref{tab:scalings}, we show that
the Sherwood number trends towards the asymptotic dependency of
$\rac^{1/3}$, regardless of 
the value of $Pr$, even if a direct numerical verification of this scaling is 
beyond our computational reach. Our strongly-driven cases are not  perfectly in 
the $\rac\gg 1$ regime, which implies that a non-negligible fraction of 
dissipation occurs inside boundary layers. In addition, the mixture of Prandtl 
numbers within the ensemble of simulations leads to variety of transport 
mechanisms, and consequently to some residual scatter in 
Fig.~\ref{fig:lowrrho}(\textit{a}). 
The analysis of the  power balance in conjunction with $Sh \sim 
\rac^{1/3}$ prompts us to propose a novel scaling law 
for the velocity $Pe$ that accounts 
extremely well for our data and those of \citet{Yang2016}. This law 
is adequate for regimes where fingers dominate the dynamics, regardless of the 
geometry. Conversely, this law may prove of limited value to account for data  
obtained for $\Rp < 1$ (or equivalently $\rrho<0$), i.e. when overturning 
convection can also occur. Among the most salient differences between 
the scaling behaviours shown in Tab.~\ref{tab:scalings} in this regime, we 
would like to stress that our simulations with $Pr\ll 1$ feature scaling 
behaviours for the chemical transport that grow as $\rac^{1/3}$ in stark 
contrast with the predictions from unbounded domains which predict a constant 
value for any given ($Pr,Le$) pair \citep{Brown2013}.

A secondary instability may develop in the form of large-scale toroidal jets. 
These features have been observed in both unbounded and bounded planar models 
which -by construction- deal with low aspect ratios, so it was far from 
clear that they could actually grow in global geometry. Here we report for the 
first time on large-scale toroidal jets that develop in non-rotating spherical 
shells.
If the properties and evolution of those jets depend on the Prandtl number
being larger or smaller than unity, their formation is in any event contingent 
upon a minimum level of driving. For low $Pr$ fluids, a single jet develops
and we find that the level of forcing leading to jet formation is bounded, 
meaning that jets might disappear in the $\rac\gg1$ limit, all other 
control parameters remaining fixed. In addition, our results indicate that the 
interval of forcing over which jet formation occurs shrinks as $Pr$ decreases. 
Consequently, we do not expect toroidal jets to form in spherical geometry in 
the $Pr\ll1$ limit, in agreement with the conclusion drawn by 
\citet{Garaud2015} in Cartesian geometry. 
 For $Pr$ above unity, multiple alternated jets form, and our 
numerical results suggest that there is no upper bound on the level of forcing 
that favours jet formation. We were not able to study the possible merging of 
those multiple jets in a systematic manner, due to the computational cost of 
such an investigation, as 
the merging and subsequent saturation may occur on a timescale commensurate 
to the viscous timescale. The analysis and characterisation of the 
merging 
process appears as a pending issue worth of future examination. As envisionned 
by \citet{Holyer1984} and \citet{Stern2005}, jets draw their energy from the 
Reynolds stress correlations that come from the sheared fingers, a mechanism 
akin to the tilting instabilities found in classical Rayleigh-Bénard convection 
\citep[e.g.][]{Goluskin2014}. The nonlinear saturation of this secondary 
instability can yield relaxation oscillations with a quasi-periodic exchange of 
energy between fingers and jets.
We finally note that a more detailed description of the region of parameter 
space where jets may develop would entail a linearisation to be performed
around the fingering convection state, a task which is not straightforward
in global spherical geometry.

Fingering convection can eventually lead to the formation of compositional 
staircases. Staircases were found by \citet{Stellmach2011} in a triply-periodic 
domain,  and more recently by \citet{Yang2020}, in a bounded Cartesian configuration, 
upon reaching $\rac \gtrsim 10^{12}$, slightly above the maximum value
considered in this work ($\rac=5\times 10^{11}$). Future work should seek 
confirmation of spontaneous layer formation in a spherical shell. 

Finally, let us recall that this study ignored the effect of background rotation 
from the outset, on the account of parsimony. In view of planetary 
applications, and in light of the studies by \citet{Monville2019} and
\citet{Guervilly2022}, a sensible next step is to add background rotation to the 
physical set-up and to analyse how it affects the understanding developed
here in the non-rotating case.

\backsection[Acknowledgements]{We thank the three anonymous referees for 
their thorough and insightful comments that helped improve the quality of 
manuscript. Figures were generated using {\tt matplotlib} \citep{Hunter2007}
and {\tt paraview} (\href{https://www.paraview.org}{https://www.paraview.org}) 
using the color schemes from \citet{Thyng2016}.}
\backsection[Funding]{Numerical computations were performed on GENCI resources (Grants A0090410095 and 
A0110410095) and on the S-CAPAD/DANTE platform at IPGP.}
\backsection[Declaration of interests]{The authors report no conflict of interest.}
\appendix
\section{Numerical database}
{\scriptsize
    \begin{longtable}{lrrrrrrrrrrrrrl}
\caption{Control parameters and output diagnotics for the $\nsims$
simulations performed for this study. Simulations are sorted according to
increasing $Pr$, then increasing $\rat$, and finally increasing
$\Rp$.
The quantity $\ektor^\%$ is the fraction of kinetic energy contained
in the toroidal flow, expressed in percent.
The single simulation resorting to finite differences in the
radial direction is indicated with the $\mathrm{f}$ superscript
next to the value of $N_r$. A star in the leftmost column
indicates simulations featuring jets, with an additional ``NS'' if
the jets had yet to reach saturation.} \\
    \label{tab:simu_tab2}
     \# & $|\rat|$ & $(N_r,\ell_\mathrm{max})$ & $\Rp$ & $\Rpe$ & $\gamma$
           & $\ektor^\%$ & $Nu$
           &$Sh$ & $(\lambda_i, \lambda_o)$ &
    $(\dXii, \dXio)$ & $\rep$ &$\ret^1$ &$\lh$ & \\
    & $(\times 10^7)$ & & & & & & & &($\times 10^{-1}$) & & & & &\\
    \endfirsthead

\caption{Control parameters and output diagnotics}\\
   \# & $|\rat|$ & $(N_r,\ell_\mathrm{max})$ & $\Rp$ & $\Rpe$ & $\gamma$
           & $\ektor^\%$ & $Nu$
           &$Sh$ & $(\lambda_i, \lambda_o)$ &
    $(\dXii, \dXio)$ & $\rep$ &$\ret^1$ &$\lh$ & \\
    & $(\times 10^7)$ & & & & & & & &($\times 10^{-1}$) & & & & &\\ \hline
    \endhead 
\hline
\multicolumn{14}{c}{$Pr = 0.03, Le = 33.3$} \\ 
\hline
$1$ & $6.6$ & $(513, 426)$ & $1.10$ & $2.86$ & $0.42$ & $22.00$ & $1.33$ & $31.31$ & $(0.06, 0.07)$ & $(0.43, 0.06)$ & $1201.4$ & $17.3$ & $74$ \\
$2$ & $6.6$ & $(241, 213)$ & $8.25$ & $13.26$ & $0.61$ & $12.48$ & $1.02$ & $10.23$ & $(0.12, 0.14)$ & $(0.29, 0.04)$ & $290.3$ & $5.0$ & $61$ \\
$3$ & $6.6$ & $(217, 213)$ & $15.40$ & $19.98$ & $0.71$ & $8.96$ & $1.01$ & $4.90$ & $(0.18, 0.21)$ & $(0.20, 0.03)$ & $140.4$ & $3.2$ & $54$ \\
$4$ & $6.6$ & $(193, 213)$ & $22.44$ & $25.32$ & $0.81$ & $5.84$ & $1.00$ & $2.39$ & $(0.30, 0.33)$ & $(0.16, 0.02)$ & $67.3$ & $2.1$ & $47$ \\
$5$ & $6.6$ & $(193, 213)$ & $23.00$ & $25.76$ & $0.81$ & $5.60$ & $1.00$ & $2.26$ & $(0.32, 0.35)$ & $(0.16, 0.02)$ & $62.9$ & $2.4$ & $46$ \\
$6$ & $6.6$ & $(193, 213)$ & $27.50$ & $29.04$ & $0.88$ & $4.89$ & $1.00$ & $1.47$ & $(0.00, 0.00)$ & $(0.00, 0.00)$ & $32.6$ & $4.6$ & $40$ \\
$7$ & $6.6$ & $(193, 213)$ & $28.00$ & $29.42$ & $0.89$ & $4.38$ & $1.00$ & $1.41$ & $(0.00, 0.00)$ & $(0.00, 0.00)$ & $29.7$ & $3.9$ & $39$ \\
$8$ & $0.66$ & $(201, 213)$ & $1.10$ & $3.95$ & $0.43$ & $20.23$ & $1.19$ & $18.03$ & $(0.12, 0.14)$ & $(0.50, 0.07)$ & $528.9$ & $13.6$ & $40$ \\
$9$ & $0.66$ & $(129, 128)$ & $2.95$ & $8.23$ & $0.51$ & $15.57$ & $1.06$ & $12.13$ & $(0.16, 0.19)$ & $(0.46, 0.07)$ & $285.5$ & $7.7$ & $37$ \\
$10$ & $0.66$ & $(129, 128)$ & $8.25$ & $15.78$ & $0.64$ & $11.00$ & $1.01$ & $6.59$ & $(0.25, 0.28)$ & $(0.37, 0.05)$ & $128.5$ & $4.3$ & $32$ \\
$11$ & $0.66$ & $(129, 128)$ & $15.40$ & $21.76$ & $0.73$ & $7.74$ & $1.00$ & $3.46$ & $(0.37, 0.40)$ & $(0.29, 0.04)$ & $63.0$ & $2.5$ & $28$ \\
$12$ & $0.66$ & $(129, 128)$ & $22.44$ & $26.39$ & $0.82$ & $4.89$ & $1.00$ & $1.91$ & $(0.72, 0.66)$ & $(0.27, 0.04)$ & $30.8$ & $1.4$ & $24$ \\
$13$ & $0.066$ & $(97, 133)$ & $1.10$ & $5.99$ & $0.43$ & $17.53$ & $1.09$ & $9.86$ & $(0.25, 0.29)$ & $(0.56, 0.08)$ & $223.0$ & $10.2$ & $22$ \\
\hline
\multicolumn{14}{c}{$Pr = 0.1, Le = 30$} \\ 
\hline
$14$ & $73.34$ & $(385, 426)$ & $6.80$ & $10.86$ & $0.61$ & $7.43$ & $1.06$ & $21.67$ & $(0.05, 0.07)$ & $(0.28, 0.04)$ & $285.8$ & $2.9$ & $118$ \\
$15$ & $7.334$ & $(321, 213)$ & $1.10$ & $3.49$ & $0.48$ & $13.29$ & $1.52$ & $37.90$ & $(0.05, 0.06)$ & $(0.47, 0.07)$ & $493.3$ & $6.2$ & $73$ \\
$16$ & $7.334$ & $(241, 213)$ & $6.80$ & $13.01$ & $0.63$ & $6.35$ & $1.04$ & $14.02$ & $(0.11, 0.13)$ & $(0.36, 0.05)$ & $127.6$ & $2.9$ & $64$ \\
$17$ & $7.334$ & $(217, 213)$ & $8.00$ & $14.31$ & $0.65$ & $5.79$ & $1.03$ & $12.19$ & $(0.12, 0.14)$ & $(0.33, 0.05)$ & $109.3$ & $2.2$ & $63$ \\
$18$ & $7.334$ & $(217, 213)$ & $10.00$ & $16.23$ & $0.68$ & $5.07$ & $1.02$ & $9.76$ & $(0.13, 0.16)$ & $(0.30, 0.05)$ & $86.3$ & $3.0$ & $60$ \\
$19^*$ & $7.334$ & $(241, 213)$ & $11.00$ & $16.77$ & $0.69$ & $38.62$ & $1.02$ & $8.45$ & $(0.14, 0.17)$ & $(0.28, 0.04)$ & $76.4$ & $58.0$ & $59$ \\
$20^*$ & $7.334$ & $(241, 213)$ & $12.60$ & $18.20$ & $0.71$ & $40.09$ & $1.01$ & $7.05$ & $(0.15, 0.18)$ & $(0.25, 0.04)$ & $64.2$ & $50.9$ & $57$ \\
$21^*$ & $7.334$ & $(217, 213)$ & $14.00$ & $19.38$ & $0.73$ & $36.43$ & $1.01$ & $6.07$ & $(0.16, 0.19)$ & $(0.23, 0.04)$ & $55.4$ & $40.4$ & $56$ \\
$22^*$ & $7.334$ & $(217, 213)$ & $15.00$ & $20.16$ & $0.75$ & $32.40$ & $1.01$ & $5.47$ & $(0.17, 0.20)$ & $(0.22, 0.03)$ & $49.9$ & $33.1$ & $55$ \\
$23^*$ & $7.334$ & $(217, 213)$ & $16.00$ & $20.91$ & $0.76$ & $27.57$ & $1.01$ & $4.93$ & $(0.18, 0.21)$ & $(0.21, 0.03)$ & $44.9$ & $27.1$ & $54$ \\
$24^*$ & $7.334$ & $(217, 213)$ & $17.00$ & $21.67$ & $0.78$ & $22.28$ & $1.01$ & $4.44$ & $(0.19, 0.23)$ & $(0.20, 0.03)$ & $40.4$ & $20.9$ & $53$ \\
$25^*$ & $7.334$ & $(217, 213)$ & $18.40$ & $22.69$ & $0.80$ & $13.88$ & $1.00$ & $3.85$ & $(0.21, 0.24)$ & $(0.18, 0.03)$ & $34.7$ & $13.4$ & $51$ \\
$26$ & $0.7334$ & $(129, 128)$ & $1.10$ & $5.27$ & $0.49$ & $10.99$ & $1.27$ & $20.43$ & $(0.12, 0.12)$ & $(0.55, 0.07)$ & $206.5$ & $4.6$ & $40$ \\
$27$ & $0.7334$ & $(129, 128)$ & $6.80$ & $15.74$ & $0.66$ & $5.13$ & $1.02$ & $8.46$ & $(0.22, 0.26)$ & $(0.43, 0.07)$ & $54.2$ & $1.7$ & $33$ \\
$28$ & $0.7334$ & $(129, 128)$ & $12.60$ & $20.56$ & $0.75$ & $3.19$ & $1.01$ & $4.80$ & $(0.31, 0.35)$ & $(0.34, 0.05)$ & $28.0$ & $1.1$ & $30$ \\
$29$ & $0.7334$ & $(129, 128)$ & $18.40$ & $24.17$ & $0.83$ & $1.71$ & $1.00$ & $2.82$ & $(0.44, 0.46)$ & $(0.27, 0.04)$ & $15.1$ & $0.6$ & $26$ \\
$30$ & $0.7334$ & $(129, 128)$ & $24.20$ & $27.07$ & $0.90$ & $0.40$ & $1.00$ & $1.57$ & $(0.00, 0.00)$ & $(0.00, 0.00)$ & $6.6$ & $0.2$ & $22$ \\
$31$ & $0.0733$ & $(49, 85)$ & $1.10$ & $7.81$ & $0.48$ & $8.35$ & $1.13$ & 
$10.73$ & $(0.25, 0.27)$ & $(0.60, 0.09)$ & $83.4$ & $3.2$ & $22$ \\
$32$ & $0.0073$ & $(41, 85)$ & $1.10$ & $10.66$ & $0.44$ & $6.26$ & $1.06$ & 
$5.64$ & $(0.52, 0.55)$ & $(0.64, 0.09)$ & $33.1$ & $2.0$ & $12$ \\
\hline
\multicolumn{14}{c}{$Pr = 0.3, Le = 10$} \\ 
\hline
$33^*$ & $1000$ & $(433, 682)$ & $5.00$ & $5.75$ & $0.72$ & $80.96$ & $1.14$ & $10.86$ & $(0.04, 0.05)$ & $(0.11, 0.02)$ & $253.6$ & $510.2$ & $200$ \\
$34^*$ & $1000$ & $(321, 597)$ & $7.30$ & $7.69$ & $0.82$ & $79.67$ & $1.03$ & $3.70$ & $(0.07, 0.08)$ & $(0.06, 0.01)$ & $103.7$ & $204.1$ & $174$ \\
$35^*$ & $1000$ & $(321, 597)$ & $9.30$ & $9.42$ & $0.95$ & $1.98$ & $1.01$ & $1.51$ & $(0.00, 0.00)$ & $(0.00, 0.00)$ & $30.2$ & $3.9$ & $134$ \\
$36$ & $202$ & $(541, 682)$ & $1.01$ & $1.99$ & $0.60$ & $12.23$ & $4.12$ & $53.74$ & $(0.03, 0.03)$ & $(0.38, 0.05)$ & $879.5$ & $5.9$ & $139$ \\
$37$ & $150$ & $(433, 512)$ & $1.50$ & $2.74$ & $0.62$ & $9.34$ & $2.54$ & $38.55$ & $(0.04, 0.04)$ & $(0.35, 0.05)$ & $563.1$ & $3.8$ & $134$ \\
$38$ & $100$ & $(289, 341)$ & $3.50$ & $5.03$ & $0.69$ & $5.51$ & $1.30$ & $16.10$ & $(0.06, 0.07)$ & $(0.23, 0.03)$ & $210.1$ & $6.0$ & $119$ \\
$39^*$ & $100$ & $(241, 341)$ & $4.00$ & $5.36$ & $0.70$ & $52.90$ & $1.21$ & $12.77$ & $(0.07, 0.08)$ & $(0.20, 0.03)$ & $173.9$ & $187.0$ & $115$ \\
$40^*$ & $100$ & $(241, 341)$ & $5.00$ & $6.08$ & $0.74$ & $77.61$ & $1.11$ & $8.18$ & $(0.08, 0.10)$ & $(0.16, 0.02)$ & $121.4$ & $229.9$ & $109$ \\
$41^*$ & $100$ & $(193, 341)$ & $6.00$ & $6.87$ & $0.77$ & $72.19$ & $1.06$ & $5.50$ & $(0.10, 0.12)$ & $(0.12, 0.02)$ & $86.3$ & $138.1$ & $103$ \\
$42^*$ & $100$ & $(193, 341)$ & $7.00$ & $7.69$ & $0.82$ & $54.91$ & $1.03$ & $3.70$ & $(0.12, 0.14)$ & $(0.10, 0.02)$ & $59.7$ & $65.2$ & $97$ \\
$43^*$ & $100$ & $(193, 341)$ & $8.00$ & $8.50$ & $0.87$ & $7.20$ & $1.02$ & $2.64$ & $(0.15, 0.18)$ & $(0.09, 0.01)$ & $40.2$ & $11.2$ & $90$ \\
$44$ & $100$ & $(193, 341)$ & $9.00$ & $9.36$ & $0.93$ & $1.10$ & $1.01$ & $1.66$ & $(0.00, 0.00)$ & $(0.00, 0.00)$ & $20.5$ & $1.5$ & $76$ \\
$45$ & $30$ & $(257, 341)$ & $1.50$ & $3.15$ & $0.62$ & $8.31$ & $2.08$ & $27.39$ & $(0.06, 0.07)$ & $(0.40, 0.06)$ & $316.9$ & $3.2$ & $90$ \\
$46$ & $22$ & $(257, 341)$ & $1.10$ & $2.61$ & $0.61$ & $9.69$ & $2.62$ & $30.83$ & $(0.06, 0.07)$ & $(0.44, 0.06)$ & $366.4$ & $4.1$ & $82$ \\
$47$ & $20.2$ & $(289, 341)$ & $1.01$ & $2.48$ & $0.60$ & $10.12$ & $2.80$ & $31.70$ & $(0.06, 0.07)$ & $(0.45, 0.06)$ & $380.6$ & $4.3$ & $79$ \\
$48$ & $18$ & $(129, 170)$ & $9.00$ & $9.49$ & $0.94$ & $0.42$ & $1.01$ & $1.48$ & $(0.00, 0.00)$ & $(0.00, 0.00)$ & $11.1$ & $0.3$ & $46$ \\
$49$ & $10$ & $(193, 213)$ & $1.50$ & $3.51$ & $0.63$ & $7.44$ & $1.84$ & $21.43$ & $(0.09, 0.09)$ & $(0.43, 0.06)$ & $211.7$ & $2.9$ & $68$ \\
$50$ & $10$ & $(161, 213)$ & $3.00$ & $5.27$ & $0.69$ & $5.31$ & $1.26$ & $12.51$ & $(0.11, 0.14)$ & $(0.33, 0.05)$ & $111.2$ & $5.0$ & $65$ \\
$51^*$ & $10$ & $(129, 170)$ & $3.25$ & $5.40$ & $0.70$ & $54.68$ & $1.21$ & $11.00$ & $(0.12, 0.14)$ & $(0.31, 0.05)$ & $100.2$ & $96.4$ & $64$ \\
$52^*$ & $10$ & $(129, 170)$ & $3.50$ & $5.62$ & $0.71$ & $56.27$ & $1.18$ & $10.07$ & $(0.12, 0.15)$ & $(0.30, 0.04)$ & $91.9$ & $89.3$ & $63$ \\
$53^*$ & $10$ & $(129, 170)$ & $4.00$ & $5.99$ & $0.72$ & $55.80$ & $1.13$ & $8.47$ & $(0.14, 0.16)$ & $(0.27, 0.04)$ & $77.9$ & $81.5$ & $62$ \\
$54^*$ & $10$ & $(129, 170)$ & $4.50$ & $6.43$ & $0.74$ & $52.16$ & $1.10$ & $7.16$ & $(0.15, 0.17)$ & $(0.25, 0.04)$ & $66.4$ & $57.1$ & $60$ \\
$55^*$ & $10$ & $(129, 170)$ & $5.00$ & $6.78$ & $0.76$ & $45.30$ & $1.08$ & $6.10$ & $(0.16, 0.19)$ & $(0.22, 0.03)$ & $56.8$ & $50.8$ & $59$ \\
$56^*$ & $10$ & $(129, 170)$ & $5.50$ & $7.17$ & $0.78$ & $36.75$ & $1.06$ & $5.18$ & $(0.17, 0.20)$ & $(0.21, 0.03)$ & $48.5$ & $31.5$ & $57$ \\
$57^*$ & $10$ & $(129, 170)$ & $6.00$ & $7.50$ & $0.80$ & $25.17$ & $1.05$ & $4.42$ & $(0.18, 0.22)$ & $(0.19, 0.03)$ & $41.3$ & $21.8$ & $56$ \\
$58$ & $10$ & $(129, 170)$ & $7.00$ & $8.20$ & $0.85$ & $5.04$ & $1.03$ & $3.21$ & $(0.22, 0.26)$ & $(0.16, 0.02)$ & $29.1$ & $5.3$ & $52$ \\
$59$ & $10$ & $(129, 170)$ & $8.00$ & $8.91$ & $0.89$ & $1.41$ & $1.01$ & $2.22$ & $(0.32, 0.33)$ & $(0.15, 0.02)$ & $18.6$ & $1.1$ & $46$ \\
$60$ & $10$ & $(129, 170)$ & $9.00$ & $9.47$ & $0.94$ & $0.32$ & $1.00$ & $1.42$ & $(0.00, 0.00)$ & $(0.00, 0.00)$ & $8.9$ & $0.2$ & $39$ \\
$61$ & $3$ & $(129, 170)$ & $1.50$ & $3.96$ & $0.64$ & $6.70$ & $1.62$ & $16.08$ & $(0.12, 0.14)$ & $(0.47, 0.07)$ & $134.7$ & $3.7$ & $47$ \\
$62$ & $2.2$ & $(129, 170)$ & $1.10$ & $3.37$ & $0.61$ & $7.69$ & $1.89$ & $17.54$ & $(0.12, 0.14)$ & $(0.50, 0.07)$ & $153.7$ & $3.1$ & $46$ \\
$63$ & $2.06$ & $(129, 170)$ & $1.03$ & $3.26$ & $0.61$ & $7.91$ & $1.95$ & $17.81$ & $(0.12, 0.14)$ & $(0.51, 0.07)$ & $157.9$ & $3.0$ & $45$ \\
$64$ & $2.02$ & $(129, 170)$ & $1.01$ & $3.22$ & $0.61$ & $7.99$ & $1.98$ & $17.89$ & $(0.12, 0.14)$ & $(0.51, 0.07)$ & $159.1$ & $3.1$ & $45$ \\
$65$ & $1$ & $(65, 128)$ & $5.00$ & $7.47$ & $0.79$ & $2.68$ & $1.05$ & $4.26$ & $(0.32, 0.36)$ & $(0.31, 0.05)$ & $24.7$ & $1.0$ & $30$ \\
$66$ & $0.3$ & $(65, 128)$ & $1.50$ & $5.00$ & $0.65$ & $5.20$ & $1.33$ & $9.03$ & $(0.25, 0.30)$ & $(0.52, 0.08)$ & $55.0$ & $1.8$ & $26$ \\
$67$ & $0.03$ & $(65, 128)$ & $1.50$ & $5.39$ & $0.64$ & $3.95$ & $1.16$ & $4.97$ & $(0.53, 0.58)$ & $(0.57, 0.09)$ & $22.0$ & $1.1$ & $14$ \\
$68$ & $0.01$ & $(65, 85)$ & $5.00$ & $7.86$ & $0.81$ & $0.58$ & $1.01$ & $1.81$ & $(1.73, 1.51)$ & $(0.56, 0.09)$ & $4.0$ & $0.2$ & $8$ \\
\hline
\multicolumn{14}{c}{$Pr = 1, Le = 10$} \\ 
\hline
$69$ & $300$ & $(601, 794)$ & $1.50$ & $3.22$ & $0.63$ & $3.63$ & $3.33$ & $57.20$ & $(0.03, 0.03)$ & $(0.41, 0.06)$ & $255.8$ & $1.6$ & $165$ \\
$70$ & $150$ & $(433, 629)$ & $1.50$ & $3.44$ & $0.63$ & $3.36$ & $3.00$ & $48.94$ & $(0.04, 0.04)$ & $(0.43, 0.06)$ & $198.5$ & $2.0$ & $139$ \\
$71$ & $15$ & $(257, 341)$ & $1.10$ & $3.57$ & $0.63$ & $3.19$ & $2.76$ & $32.79$ & $(0.07, 0.08)$ & $(0.51, 0.07)$ & $107.4$ & $2.1$ & $77$ \\
$72^*$ & $15$ & $(257, 341)$ & $1.30$ & $3.85$ & $0.63$ & $25.26$ & $2.35$ & $29.63$ & $(0.07, 0.08)$ & $(0.49, 0.07)$ & $93.1$ & $51.8$ & $77$ \\
$73^*$ & $15$ & $(257, 341)$ & $1.50$ & $4.23$ & $0.64$ & $30.21$ & $2.10$ & $27.42$ & $(0.07, 0.09)$ & $(0.48, 0.07)$ & $82.5$ & $58.6$ & $77$ \\
$74$ & $15$ & $(257, 341)$ & $2.00$ & $5.26$ & $0.68$ & $3.94$ & $1.77$ & $24.00$ & $(0.08, 0.10)$ & $(0.46, 0.07)$ & $64.5$ & $8.6$ & $77$ \\
$75$ & $15$ & $(257, 341)$ & $5.00$ & $8.12$ & $0.83$ & $0.80$ & $1.15$ & $10.34$ & $(0.12, 0.15)$ & $(0.29, 0.05)$ & $23.0$ & $0.8$ & $63$ \\
$76$ & $15$ & $(257, 341)$ & $7.00$ & $8.97$ & $0.89$ & $0.25$ & $1.05$ & $4.99$ & $(0.17, 0.19)$ & $(0.19, 0.03)$ & $11.5$ & $0.2$ & $54$ \\
$77$ & $15$ & $(257, 341)$ & $8.00$ & $9.27$ & $0.92$ & $0.09$ & $1.02$ & $3.00$ & $(0.22, 0.23)$ & $(0.15, 0.02)$ & $7.2$ & $0.1$ & $49$ \\
$78$ & $15$ & $(257, 341)$ & $9.00$ & $9.59$ & $0.95$ & $0.02$ & $1.01$ & $1.58$ & $(0.00, 0.00)$ & $(0.00, 0.00)$ & $4.1$ & $0.0$ & $30$ \\
$79$ & $1.5$ & $(121, 170)$ & $1.50$ & $6.09$ & $0.70$ & $1.57$ & $1.65$ & $15.74$ & $(0.15, 0.19)$ & $(0.54, 0.09)$ & $32.8$ & $0.8$ & $41$ \\
\hline
\multicolumn{14}{c}{$Pr = 3, Le = 10$} \\ 
\hline
$80^{*\text{NS}}$ & $750$ & $(769, 938)$ & $1.50$ & $3.58$ & $0.62$ & $10.62$ & 
$4.48$ & $85.48$ & $(0.02, 0.02)$ & $(0.44, 0.06)$ & $134.3$ & $39.2$ & $218$ \\
$81^{*\text{NS}}$ & $300$ & $(541, 682)$ & $1.50$ & $3.98$ & $0.63$ & $15.19$ 
& $3.85$ & $69.49$ & $(0.03, 0.03)$ & $(0.47, 0.06)$ & $94.9$ & $39.5$ & $174$ 
\\
$82^*$ & $150$ & $(541, 682)$ & $1.05$ & $3.25$ & $0.61$ & $48.63$ & $4.84$ & $68.32$ & $(0.03, 0.03)$ & $(0.51, 0.07)$ & $99.2$ & $95.6$ & $139$ \\
$83^*$ & $150$ & $(541, 682)$ & $1.20$ & $3.57$ & $0.61$ & $51.80$ & $4.15$ & $63.64$ & $(0.03, 0.04)$ & $(0.49, 0.07)$ & $87.7$ & $90.2$ & $142$ \\
$84^*$ & $150$ & $(433, 597)$ & $1.50$ & $4.40$ & $0.64$ & $3.65$ & $3.46$ & $58.92$ & $(0.04, 0.04)$ & $(0.49, 0.07)$ & $72.6$ & $9.6$ & $145$ \\
$85^*$ & $150$ & $(481, 629)$ & $2.00$ & $5.13$ & $0.67$ & $7.55$ & $2.59$ & $49.04$ & $(0.04, 0.05)$ & $(0.45, 0.07)$ & $56.0$ & $15.4$ & $145$ \\
$86$ & $150$ & $(481, 629)$ & $3.00$ & $7.44$ & $0.79$ & $0.44$ & $2.03$ & $40.38$ & $(0.04, 0.06)$ & $(0.43, 0.07)$ & $37.6$ & $1.3$ & $127$ \\
$87$ & $150$ & $(481, 629)$ & $5.00$ & $8.64$ & $0.86$ & $0.21$ & $1.37$ & $22.71$ & $(0.06, 0.08)$ & $(0.31, 0.05)$ & $19.7$ & $0.5$ & $112$ \\
$88$ & $150$ & $(481, 512)$ & $7.00$ & $9.12$ & $0.91$ & $0.03$ & $1.12$ & $10.18$ & $(0.07, 0.09)$ & $(0.18, 0.03)$ & $9.9$ & $0.1$ & $97$ \\
$89$ & $150$ & $(385, 629)$ & $8.00$ & $9.32$ & $0.93$ & $0.01$ & $1.05$ & $5.48$ & $(0.09, 0.11)$ & $(0.12, 0.02)$ & $6.2$ & $0.0$ & $90$ \\
$90$ & $150$ & $(481, 629)$ & $9.00$ & $9.56$ & $0.95$ & $0.00$ & $1.01$ & $2.26$ & $(0.17, 0.16)$ & $(0.08, 0.01)$ & $2.9$ & $0.0$ & $81$ \\
$91$ & $15$ & $(193, 320)$ & $5.00$ & $8.56$ & $0.86$ & $0.04$ & $1.18$ & $11.71$ & $(0.12, 0.15)$ & $(0.32, 0.05)$ & $7.7$ & $0.1$ & $59$ \\
$92$ & $15$ & $(193, 320)$ & $6.00$ & $8.77$ & $0.88$ & $0.02$ & $1.10$ & $8.18$ & $(0.14, 0.17)$ & $(0.26, 0.04)$ & $5.6$ & $0.0$ & $56$ \\
$93$ & $15$ & $(193, 320)$ & $7.00$ & $8.99$ & $0.90$ & $0.01$ & $1.06$ & $5.34$ & $(0.17, 0.19)$ & $(0.20, 0.03)$ & $3.9$ & $0.0$ & $53$ \\
$94$ & $15$ & $(193, 320)$ & $8.00$ & $9.24$ & $0.93$ & $0.01$ & $1.02$ & $3.13$ & $(0.22, 0.23)$ & $(0.16, 0.02)$ & $2.4$ & $0.0$ & $49$ \\
$95$ & $15$ & $(193, 320)$ & $9.00$ & $9.53$ & $0.95$ & $0.00$ & $1.01$ & $1.60$ & $(0.00, 0.00)$ & $(0.00, 0.00)$ & $1.1$ & $0.0$ & $43$ \\
$96$ & $1.5$ & $(97, 170)$ & $7.00$ & $8.94$ & $0.90$ & $0.01$ & $1.03$ & $3.03$ & $(0.39, 0.40)$ & $(0.26, 0.04)$ & $1.5$ & $0.0$ & $28$ \\
$97$ & $1.5$ & $(97, 170)$ & $8.00$ & $9.22$ & $0.92$ & $0.00$ & $1.01$ & $1.98$ & $(0.63, 0.54)$ & $(0.25, 0.03)$ & $1.0$ & $0.0$ & $26$ \\
$98$ & $1.5$ & $(97, 170)$ & $9.00$ & $9.51$ & $0.95$ & $0.00$ & $1.00$ & $1.27$ & $(0.00, 0.00)$ & $(0.00, 0.00)$ & $0.4$ & $0.0$ & $23$ \\
$99$ & $0.15$ & $(73, 106)$ & $1.50$ & $7.11$ & $0.74$ & $0.06$ & $1.34$ & $8.40$ & $(0.32, 0.39)$ & $(0.60, 0.10)$ & $4.2$ & $0.0$ & $21$ \\
\hline
\multicolumn{14}{c}{$Pr = 7, Le = 3$} \\ 
\hline
$100^{*\text{NS}}$ & $18300$ & $(1536, 1536)$ & $1.10$ & 
$1.56$ & $0.78$ & $13.58$ & $25.03$ & $102.9$ & $(0.01, 0.01)$ & $(0.30, 0.04)$ 
& $312.4$ & $119.4$ & $339$ \\
$101^{*\text{NS}}$ & $1830$ & $(641, 970)$ & $1.10$ & $1.74$ & $0.77$ & $18.29$ 
& $14.50$ & $58.78$ & $(0.02, 0.03)$ & $(0.35, 0.05)$ & $128.6$ & $58.6$ & $205$ 
\\
$102$ & $833.3$ & $(385, 469)$ & $2.50$ & $2.87$ & $0.95$ & $0.09$ & $1.89$ & 
$7.95$ & $(0.06, 0.08)$ & $(0.11, 0.02)$ & $15.5$ & $0.3$ & $131$ \\
$103^{*\text{NS}}$ & $732$ & $(481, 629)$ & $1.10$ & $1.80$ & $0.77$ & $18.19$ & 
$11.78$ & $47.10$ & $(0.03, 0.04)$ & $(0.37, 0.05)$ & $90.6$ & $38.9$ & $166$ \\
$104$ & $667$ & $(385, 469)$ & $2.00$ & $2.65$ & $0.90$ & $0.33$ & $3.70$ & $19.02$ & $(0.05, 0.06)$ & $(0.22, 0.03)$ & $30.7$ & $1.1$ & $149$ \\
$105^*$ & $500$ & $(385, 426)$ & $1.50$ & $2.20$ & $0.81$ & $11.48$ & $5.84$ & $27.61$ & $(0.04, 0.05)$ & $(0.29, 0.04)$ & $47.8$ & $16.6$ & $159$ \\
$106^*$ & $366$ & $(385, 469)$ & $1.001$ & $1.75$ & $0.77$ & $40.59$ & $11.82$ & 
$43.66$ & $(0.04, 0.04)$ & $(0.40, 0.05)$ & $80.0$ & $65.5$ & $133$ \\
$107^*$ & $366$ & $(385, 469)$ & $1.05$ & $1.81$ & $0.77$ & $41.38$ & $10.81$ & $41.40$ & $(0.04, 0.04)$ & $(0.39, 0.05)$ & $73.9$ & $61.5$ & $137$ \\
$108^*$ & $366$ & $(385, 426)$ & $1.10$ & $1.88$ & $0.78$ & $17.83$ & $10.18$ & $40.11$ & $(0.04, 0.04)$ & $(0.39, 0.05)$ & $68.9$ & $32.0$ & $142$ \\
$109^*$ & $366$ & $(385, 469)$ & $1.25$ & $2.03$ & $0.79$ & $14.36$ & $7.92$ & $33.96$ & $(0.04, 0.05)$ & $(0.35, 0.05)$ & $56.5$ & $22.7$ & $146$ \\
$110^*$ & $366$ & $(385, 469)$ & $1.50$ & $2.24$ & $0.82$ & $8.45$ & $5.55$ & $25.95$ & $(0.05, 0.06)$ & $(0.30, 0.04)$ & $42.4$ & $12.6$ & $146$ \\
$111$ & $366$ & $(385, 469)$ & $1.75$ & $2.51$ & $0.87$ & $0.54$ & $4.44$ & $21.85$ & $(0.05, 0.07)$ & $(0.27, 0.04)$ & $32.4$ & $1.9$ & $135$ \\
$112$ & $366$ & $(385, 469)$ & $2.00$ & $2.72$ & $0.91$ & $0.29$ & $3.40$ & $16.67$ & $(0.06, 0.08)$ & $(0.23, 0.04)$ & $24.3$ & $0.8$ & $124$ \\
$113$ & $366$ & $(385, 469)$ & $2.50$ & $2.86$ & $0.95$ & $0.04$ & $1.71$ & $6.39$ & $(0.08, 0.10)$ & $(0.12, 0.02)$ & $11.0$ & $0.1$ & $103$ \\
$114$ & $366$ & $(385, 469)$ & $2.75$ & $2.91$ & $0.96$ & $0.01$ & $1.19$ & $2.65$ & $(0.12, 0.13)$ & $(0.08, 0.01)$ & $5.3$ & $0.0$ & $94$ \\
$115$ & $366$ & $(385, 469)$ & $2.90$ & $2.95$ & $0.98$ & $0.00$ & $1.03$ & $1.29$ & $(0.00, 0.00)$ & $(0.00, 0.00)$ & $1.9$ & $0.0$ & $80$ \\
$116$ & $366$ & $(385, 469)$ & $2.95$ & $2.97$ & $0.99$ & $0.00$ & $1.01$ & $1.07$ & $(0.00, 0.00)$ & $(0.00, 0.00)$ & $0.9$ & $0.0$ & $70$ \\
$117^*$ & $350$ & $(385, 469)$ & $1.05$ & $1.81$ & $0.77$ & $39.71$ & $10.70$ & $40.97$ & $(0.04, 0.04)$ & $(0.39, 0.05)$ & $72.6$ & $58.3$ & $136$ \\
$118$ & $36.6$ & $(193, 256)$ & $1.10$ & $2.29$ & $0.82$ & $0.50$ & $6.50$ & $23.48$ & $(0.08, 0.10)$ & $(0.45, 0.07)$ & $27.7$ & $1.2$ & $75$ \\
$119$ & $36.6$ & $(193, 256)$ & $2.70$ & $2.89$ & $0.96$ & $0.01$ & $1.13$ & $2.13$ & $(0.27, 0.27)$ & $(0.13, 0.02)$ & $2.6$ & $0.0$ & $50$ \\
$120$ & $36.6$ & $(193, 256)$ & $2.79$ & $2.91$ & $0.97$ & $0.00$ & $1.06$ & $1.56$ & $(0.00, 0.00)$ & $(0.00, 0.00)$ & $1.7$ & $0.0$ & $48$ \\
$121$ & $36.6$ & $(193, 256)$ & $2.89$ & $2.95$ & $0.98$ & $0.00$ & $1.02$ & $1.16$ & $(0.00, 0.00)$ & $(0.00, 0.00)$ & $0.8$ & $0.0$ & $43$ \\
$122$ & $36.6$ & $(193, 256)$ & $2.94$ & $2.97$ & $0.99$ & $0.00$ & $1.00$ & $1.04$ & $(0.00, 0.00)$ & $(0.00, 0.00)$ & $0.4$ & $0.0$ & $39$ \\
$123$ & $3.66$ & $(129, 170)$ & $1.10$ & $2.55$ & $0.85$ & $0.11$ & $3.84$ & $12.37$ & $(0.17, 0.23)$ & $(0.49, 0.08)$ & $10.3$ & $0.1$ & $39$ \\
\end{longtable}
}

\section{Heat sources in spherical geometry}
\label{sec:app1}

This appendix demonstrates how the time-averaged buoyancy powers 
$\mathcal{P}_\xi$ and 
$\mathcal{P}_T$ can be related to $Ra_\xi(Sh-1)/Sc^2$ and $Ra_T(Nu-1)/Pr^2$ in 
spherical geometry. In contrast to the planar configuration where those 
quantities match to each other, gravity changes with 
radius and curvature prohibit such an exact relation in curvilinear 
geometries \citep[see the derivations by][]{Oruba2016}.
Here we detail the main steps involved in the approximation of 
$\mathcal{P}_\xi$ only, keeping in mind that the derivation $\mathcal{P}_T$ 
would be strictly the same with simple exchanges of $Ra_\xi$ by $Ra_T$, $Sh$ by 
$Nu$ and $Sc$ by $Pr$.

To provide an approximation of the time-averaged buoyancy power of chemical 
origin $\mathcal{P}_\xi$, we first start by noting that

\[
 \mathcal{P}_\xi = \dfrac{\rac}{Sc}\overline{\langle g u_r \xi 
\rangle_V}=\dfrac{3\rac}{Sc(r_o^3 
-r_i^3)}\int_{r_i}^{r_o}g \overline{\langle u_r \xi \rangle_S} 
\,r^2\mathrm{d}r\,.
\]
Using the definition of the Sherwood number at all radii given in 
Eq.~(\ref{eq:nu_sh}), we obtain

\[
  \mathcal{P}_\xi = \dfrac{3\rac}{Sc(r_o^3 
-r_i^3)}\left[-\dfrac{Sh}{Sc}\int_{r_i}^{r_o} 
g\dfrac{\mathrm{d}\xi_c}{\mathrm{d}r}\,r^2\mathrm{d} r
+\dfrac{1}{Sc}
\int_{r_i}^{r_o} 
g\dfrac{\mathrm{d}\Xi}{\mathrm{d}r}\,r^2\mathrm{d} r
\right]
\]
At this stage, it is already quite clear that only the peculiar configuration 
of  $g\propto r^{-2}$ would allow a closed form for the buoyancy 
power \citep[see][]{Gastine2015}.
Noting that the conducting background state reads $\mathrm{d} \xi_c/\mathrm{d} 
r=-r_i r_o/r^2$ and that here $g=r/r_o$, one gets

\[
  \mathcal{P}_\xi = \dfrac{3\rac}{Sc(r_o^3 
-r_i^3)}\left[\dfrac{1}{2}(r_o^2-r_i^2) r_i \dfrac{Sh}{Sc}
+\dfrac{1}{Sc}
\int_{r_i}^{r_o} 
g \dfrac{\mathrm{d}\Xi}{\mathrm{d}r}\,r^2\mathrm{d} r
\right]\,.
\]
Splitting the time-averaged radial profile of composition into the mean 
conducting state and a fluctuation such that $\Xi = \xi_c+\Xi'$ yields

\[
  \mathcal{P}_\xi = \dfrac{\rac}{Sc^2} \left[\dfrac{3}{2}
  \dfrac{r_i(r_o+r_i)}{r_o^2+ r_or_i+r_i^2}(Sh-1)
+\dfrac{3}{r_o(r_o^3-r_i^3)(Sh-1)}
\int_{r_i}^{r_o} 
\dfrac{\mathrm{d}\Xi'}{\mathrm{d}r}\,r^3\mathrm{d} r
\right]\,.
\]
The chemical composition being imposed at both boundaries 
$\Xi'(r_i)=\Xi'(r_o)=0$, an integration by part of the above expression 
yields

\begin{equation}
  \mathcal{P}_\xi = \dfrac{\rac}{Sc^2} \left[\dfrac{3}{2}
\dfrac{r_i(r_o+r_i)}{r_o^2+r_or_i+r_i^2}(Sh-1)
-\dfrac{3}{r_o(Sh-1)}\langle 
\Xi' \rangle_V \right]\,.
\label{eq:exact_dissip}
\end{equation}
The second term in the brackets is proportional to the volume and time averaged 
fluctuations of chemical composition. With the choice of imposed composition at 
both spherical shell boundaries, this quantity remains bounded within
$-1\leq \langle \Xi' \rangle_V \leq 1$. This second term is hence expected to 
play a negligible role when $Sh \gg 1$. A fair approximation of 
$\mathcal{P}_\xi$ in spherical shell when $g=r/r_o$ thus reads

\begin{equation}
   \mathcal{P}_\xi \approx \dfrac{3}{2}
\dfrac{r_i(r_o+r_i)}{r_o^2+r_or_i+r_i^2}\dfrac{\rac}{Sc^2} (Sh-1)
\approx \dfrac{4\pi}{V} r_i r_m\dfrac{\rac}{Sc^2} (Sh-1)
\,,
\label{eq:approx_dissip}
\end{equation}
where $V$ is the spherical shell volume and $r_m=(r_i+r_o)/2$ is the mid-shell 
radius. This approximation was already derived by \citet{Christensen2006} in 
the case of thermal convection.

\begin{figure}
  \centering
  \includegraphics[width=0.99\textwidth]{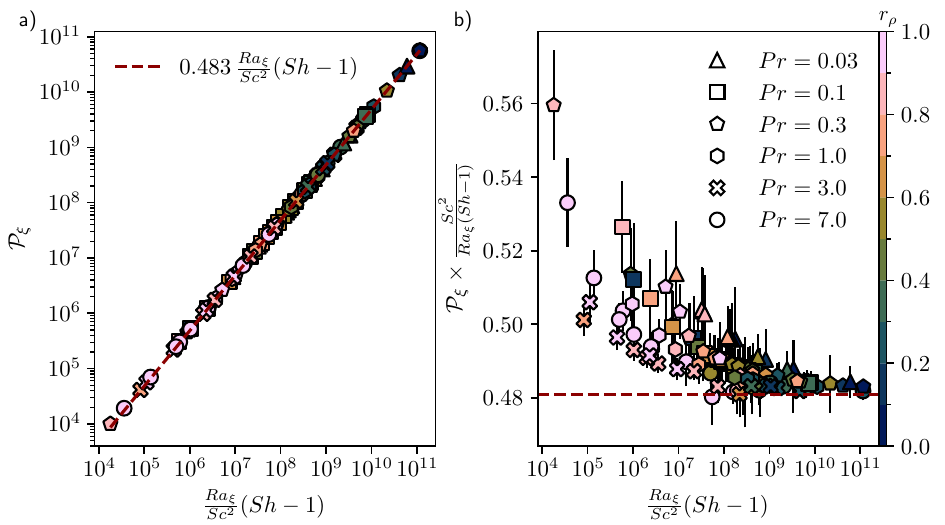}
  \caption{(\textit{a}) $\mathcal{P}_\xi$ as a function of $\rac Sc^{-2} 
(Sh - 1)$. The dashed line corresponds to a linear fit to the data.
  (\textit{b}) Compensated scaling of $\mathcal{P}_\xi \rac^{-1} 
Sc^{2} (Sh-1)^{-1}$ as a function of $\rac Sc^{-2} 
(Sh - 1)$. The horizontal line corresponds to the theoretical value derived in 
Eq.~\eqref{eq:approx_dissip} for spherical shells with $r_i/r_o=0.35$.}
  \label{fig:PC_Sh}
\end{figure}

Figure~\ref{fig:PC_Sh}(\textit{a}) shows
$\mathcal{P}_\xi$ as a function of $\rac Sc^{-2} (Sh - 1)$ for the simulations 
computed in this study. A numerical fit to the data yields 
\begin{equation}
\mathcal{P}_\xi = 0.483 \dfrac{\rac}{Sc^2} (Sh - 1)\,.
\label{eq:num_dissip}
\end{equation}
in excellent agreement with the approximated prefactor of $0.481$ obtained in 
Eq.~\eqref{eq:approx_dissip} for spherical shells with $r_i/r_o=0.35$. To 
highlight the deviations to this approximated scaling, 
Fig.~\ref{fig:PC_Sh}(\textit{b})
shows the compensated scaling of $\mathcal{P}_\xi \rac^{-1} Sc^{2} (Sh - 
1)^{-1}$ as a function of  $\rac Sc^{-2} (Sh - 1)$. As expected, the 
time-averaged buoyancy power gradually tends towards the scaling
\eqref{eq:approx_dissip} on increasing values of $\rac(Sh-1)$.

\section{A comparison with local unbounded simulations for $Pr<1$}
\label{sec:heuristic}

\begin{figure}
 \centering
 \includegraphics[width=.99\textwidth]{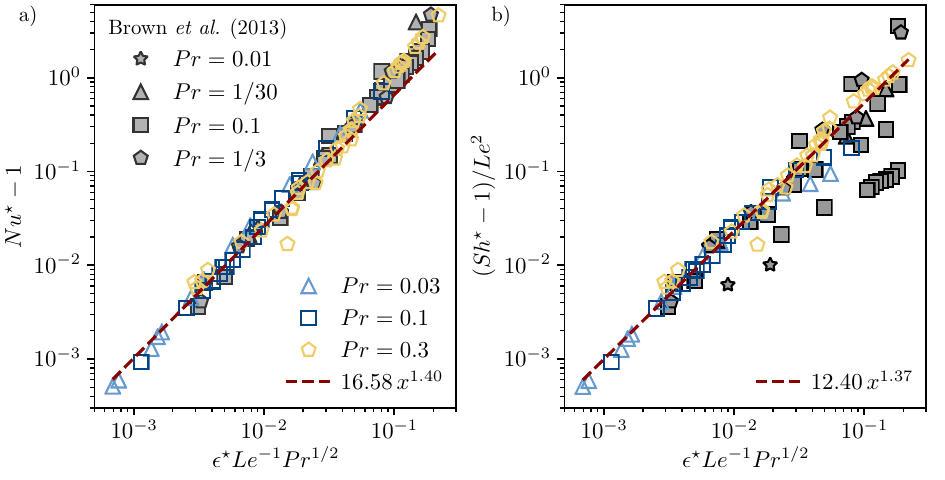}
 \caption{(\textit{a}) Convective heat transport $Nu^\star-1$ as a 
function of $\epsilon^\star Le^{-1} Pr^{1/2}$ for our simulations with $Pr <1$ 
alongside the unbounded Cartesian simulations from \cite{Brown2013}. 
(\textit{b}) Convective transport of chemical composition $(Sh^\star-1)/Le^2$ as 
a function of $\epsilon^\star Le^{-1} Pr^{1/2}$. The dashed lines correspond to 
best fits for the simulations with $\epsilon^\star/Le < 0.05$. Data from 
\citet{Brown2013} come from their Table~1.}
\label{fig:heuristic_eps}
\end{figure}

To compare our computations with local Cartesian unbounded models, we 
introduce adjusted diagnostics which take the modification of the background 
profiles into account. This practically defines effective quantities on the 
fluid bulk only
\begin{equation}
    \epsilon^\star = \dfrac{Le}{\Rp^\star}-1,\  Nu^\star - 1 = \dfrac{Nu - 
1}{{\mathrm d} \Theta / {\mathrm d}r (r_m)}, \   Sh^\star - 1 = \dfrac{Sh - 
1}{\left|{\mathrm d} \Xi / {\mathrm d}r (r_m)\right|}\,,
\end{equation}
where the background radial gradients of $\Theta$ and $\Xi$ are evaluated at the 
mid-shell radius. The introduction of these bulk gradients
 prevents the derivation of scaling laws that solely depend on 
control quantities, but this is the price to pay to make a comparison with local 
unbounded models possible. Assuming boldly that the convective and compositional heat 
transport $Nu^\star-1$ and $Sh^\star-1$ can be described by polynomial 
scaling laws of the form ${\epsilon^\star}^{\alpha_1} Pr^{\alpha_2} 
Le^{\alpha_3}$ and conducting a $3$-parameter least-squares fit on the exponents 
$\alpha_1$, $\alpha_2$ and $\alpha_3$ for all the simulations with $Pr < 1$ and 
$\epsilon^\star /Le < 0.05$ yields
\begin{equation}
Nu^\star - 1 \sim {\epsilon^\star}^{1.43} Pr^{0.68} Le^{-1.40},\quad
(Sh^\star - 1)/Le^2 \sim {\epsilon^\star}^{1.33} Pr^{0.69} Le^{-1.39}\,.
\end{equation}
These numerical fits are suggestive of a simpler parameter dependence of the 
form
\begin{equation}
 Nu^\star - 1 \sim  (Sh^\star - 1)/Le^2  \sim \left(\epsilon^\star Le^{-1}
Pr^{1/2}\right)^{\alpha_1}\,.
\label{eq:bold_scaling}
\end{equation}
To illustrate this parameter dependence, Fig.~\ref{fig:heuristic_eps} 
shows $Nu^\star-1$ and $(Sh^\star-1)/Le^2$ as a function of $\epsilon^\star 
Le^{-1}Pr^{1/2}$ for all our $68$ numerical simulations with $Pr < 1$ 
(coloured symbols) alongside $43$ local computations from \cite{Brown2013} 
(grey symbols). It is striking to note that our spherical shell data almost 
perfectly collapse with the local Cartesian unbounded computations. Weakly 
nonlinear models with the smallest $\epsilon^\star$ values are compatible with 
the simple power law \eqref{eq:bold_scaling} with a scaling exponent $\alpha_1 
\approx 1.40$, while a gradual steepening of the slope is observed on 
increasing supercriticalities. The scaling behaviour of $(Sh^\star-1)/Le^2$ also 
shows more scatter when $\epsilon^\star Le^{-1} Pr^{1/2} > 10^{-2}$. This 
likely comes from the approximation of the flux ratio $\gamma \approx 
\Rp^\star/Le$ being too crude far from onset. A quick look at 
Fig.~\ref{fig:Sh_ga}(\textit{b}) 
indeed reveals that another limit assuming a constant value of $\gamma$ instead 
could likely perform better on increasing $\epsilon^\star$.

Using the power balance \eqref{eq:Pe_gene} combined with 
Eq.~\eqref{eq:bold_scaling} 
allows to derive the corresponding scaling for $Pe$
\begin{equation}
 Pe \sim |\rat|^{1/4} {\epsilon^\star}^{\alpha_1/2+1/4} 
Pr^{\alpha_1/4}Le^{1-\alpha_1/2} \approx
|\rat|^{1/4} {\epsilon^\star}^{0.95} 
Pr^{0.17}Le^{0.30}\,,
\end{equation}
where $\alpha_1=1.4$ has been used.

Though we acknowledge the fact that Fig.~\ref{fig:heuristic_eps} merely 
provides empirical fits to the data, it is worthy to stress  that the simple 
polynomials fits considered here provide a better agreement with the actual 
data than the weakly-nonlinear theory derived in \S~\ref{sec:wnl_scalings}.

\bibliographystyle{jfm}


\end{document}